\documentclass[a4paper,11pt]{article}
\usepackage{amsmath}
\usepackage{amssymb}
\usepackage{amsthm}
\usepackage{amsfonts}
\usepackage[T1]{fontenc}
\usepackage[utf8]{inputenc}
\usepackage{graphicx}
\usepackage{float}
\usepackage{epstopdf}
\usepackage{centernot}
\usepackage{appendix}
\usepackage{caption}
\usepackage{booktabs}
\usepackage[nottoc]{tocbibind}
\usepackage{hyperref}
\usepackage{braket}
\usepackage[font+=footnotesize]{subcaption}
\usepackage[round]{natbib}
\usepackage{color}
\usepackage{bibunits}
\usepackage{url}
\usepackage{dcolumn}
\usepackage{booktabs}
\usepackage[top=1in, bottom=1.25in, left=0.9in, right=0.9in]{geometry}
\usepackage{wrapfig}
\usepackage[font={small}]{caption}
\usepackage{afterpage}

\newcommand{\mathds}{}
\newcommand{\Cov}{\text{Cov}}
\newcommand{\Cor}{\text{Cor}}
\newcommand{\tildey}{\tilde{\mathbf{y}}}
\newcommand{\boldy}{\mathbf{y}}
\newcommand{\boldw}{\mathbf{w}}

\theoremstyle{definition}

\definecolor{dark-gray}{gray}{0.3}
\definecolor{orange}{rgb}{1,0.5,0}

\newcommand{\proper}{\mathsf}

\newcommand{\mv}[1]{{\boldsymbol{\mathrm{#1}}}}
\DeclareMathOperator{\vep}{\varepsilon}
\DeclareMathOperator{\pd}{\partial}
\newcommand{\pN}{\proper{N}}
\DeclareMathOperator*{\argmax}{arg\,max}

\hypersetup{ 
	pdfborder={0 0 0}
}

\newcommand{\kronprod}{(4)}
\newcommand{\kroninverse}{(5)}
\newcommand{\noisemodel}{(1)}

\usepackage{fancyhdr}
\usepackage[switch, modulo]{lineno}

\title{\bf{A three-dimensional statistical model for imaged microstructures of porous polymer films}}
\author{\scshape{Sandra Barman and David Bolin\footnote{Correspondence to: David Bolin, Department of Mathematical Sciences, University of Gothenburg, 41296 Gothenburg, Sweden. Tel: +46 31 772 53 75; email: david.bolin@chalmers.se}}\\[1ex]
{Department of Mathematical Sciences}\\ {Chalmers University of Technology and the University of Gothenburg}}
\date{}

\begin{document}

\maketitle
\begin{center}\begin{minipage}{0.9\textwidth}
\noindent{\bf Abstract:}  A thresholded Gaussian random field model is developed for the microstructure of porous materials. Defining the random field as a solution to stochastic partial differential equation allows for flexible modelling of non-stationarities in the material and facilitates computationally efficient methods for simulation and model fitting.  A Markov Chain Monte Carlo algorithm is developed and used to fit the model to three-dimensional confocal laser scanning microscopy images. The methods are applied to study a porous ethylcellulose/hydroxypropylcellulose polymer blend that is used as a coating to control drug release from pharmaceutical tablets. The aim is to investigate how mass transport through the material depends on the microstructure. We derive a number of goodness-of-fit measures based on numerically calculated diffusion through the material. These are used in combination with measures that characterize the geometry of the pore structure to assess model fit. The model is found to fit stationary parts of the material well. 

\noindent \textit{Keywords:} Porous media, Gaussian field, Gaussian Markov random field, Markov Chain Monte Carlo, model validation.
\end{minipage}\end{center}

\section*{Introduction}

Studies of porous media have applications in many areas, ranging from geophysics, energy, electrical and chemical engineering, composite material design, and biomedical and pharmaceutical science \citep*{Vafai2015,Torquato2002}, to  the oil industry \citep*{Blunt2013}, and modelling of transport in polymer-electrolyte fuel cells \citep{Weber2014}.

The macroscopic properties---e.g.~electrical conductivity, heat or mass transfer---depend not only on the properties of the material, but also on the microscopic geometry of the pore structure. It is therefore important to characterize the microstructure, and determine how it influences the 
macroscopic properties.
We study the microstructure of a porous ethylcellulose/hydroxypropylcellulose (EC/HPC) polymer blend, which is used as a coating to control the drug release from pharmaceutical tablets. To create coatings with desired transport properties it is important to understand how the properties of the microstructure depend on the different manufacturing parameters, and how those properties influence mass transport---in this case diffusion---of the drug through the coating.

Statistical properties of the microstructure of porous media can be obtained from microscopy data, which then can be related to the macroscopic properties.  However, producing the porous material and doing the microscopy imaging can be time-consuming and costly. Formulating a stochastic model for the pore structure, which can be fitted to microscopy images of the material, is therefore useful. 
It is easier and cheaper to control and change different microscopic properties of the stochastic simulations from the model than it would be to produce samples of the material with these properties. By determining the macroscopic properties of each stochastic simulation, the simulations can be used to understand how the microscopic properties influence the macroscopic properties. 

Thresholded Gaussian random fields were among the first stochastic models used to reconstruct heterogeneous materials \citep*[p.\ 295]{Torquato2002}, and such models have been used to characterize the microstructure and its relation to macroscopic properties of porous media in e.g.\ \citet*{Adler1990,RobertsTeubner1995,Mukherjee2007}. Other examples of stochastic models and methods used to make inferences about macroscopic properties of porous media are random set models \citep*{Hermann2014}, network models \citep*{Blunt2002,Gaiselmann2014}, process-based models \citep*{Malek2007}, and simulated annealing-based reconstructions \citep*{Yeong1998,Kim2009}. These models are usually validated by comparing the statistical properties of the model with those of the imaged pore structures. Also macroscopic properties computed for the stochastic simulations from the model are often compared with those computed for the pore structure obtained from microscopy images, as well as with experimental results. These macroscopic properties can be estimated in different ways: by computing analytical bounds or by solving the governing equations approximately on simplified network models or in the pore structure (see e.g.\ \citet*[Chapter 21]{Torquato2002}; \citet*{Mukherjee2010}; \citet{Weber2014}).

In this work, we develop a new class of thresholded Gaussian random field models for the microstructure of porous materials. Instead of specifying the Gaussian random field through its mean and covariance function, we specify the field as a solution to a specific stochastic partial differential equation (SPDE). The reason for this is twofold: Firstly, the model can easily be extended to handle non-stationarities in the material by allowing for spatially varying parameters in the SPDE. Secondly, a computationally efficient representation of the model can be constructed by discretising the SPDE using a finite element method. Using this representation, we develop a Markov Chain Monte Carlo (MCMC) algorithm that can be used to fit the model to large three-dimensional confocal laser scanning microscopy images. 

We also construct a number of goodness-of-fit measures based on numerically calculated diffusion through the material. These can be combined with measures that characterize the geometry of the pore structure to test the goodness-of-fit when the model is used in practice. We use the methods to study the microstructure of EC/HPC polymer blends with two different weight ratios. The polymer blend is used as coating which controls drug release from pharmaceutical tablets and the aim is to investigate how mass transport properties of the material depend on the microstructure. The model is fitted to stationary but anisotropic parts of confocal laser scanning microscopy (CLSM) images of the EC/HPC films and is found to fit the data well. 

The article is structured as follows. In the next section, the data, model, estimation procedure, and validation methods are presented. After this, the results for the EC/HPC data are presented, and the article concludes with a discussion of the results and future work. The supplementary material for the article contains three appendices that gives additional details and results.

\section*{Materials and methods}\label{sec:methods}

\subsection*{CLSM images of porous EC/HPC polymer films}
\label{sec:micrdata}

EC/HPC polymer blends are used as coatings to control the release of drug from tablets. The HPC is water soluble while the EC is not, and so the HPC will be leached out when the tablet is immersed in water, creating a porous structure through which drug can be transported. The microstructure of the EC/HPC polymer blend is determined by the polymers' weight ratios and molecular weights, and by processing parameters such as the temperature and spraying rate used in the manufacturing of the coating \citep*{Marucci2009, Marucci2013, Andersson2013}. We analyse CLSM images of two EC/HPC free films---i.e.\ films that have been sprayed onto a rotating drum---one with $30\%$ and one with $40\%$ HPC weight ratio. More details about the preparation and imaging of these films can be found in \citet*{Habel2017}. The EC/HPC material has previously been studied experimentally in \citet*{Marucci2009, Marucci2013, Andersson2013} and characterized statistically in \citet*{Habel2016, Habel2017}. The mass transport properties of the material have been investigated using numerically calculated diffusion in \citet*{Geback2015}. 

\begin{figure}[t]
	\centering
	\subcaptionbox{}
	[0.24\textwidth]
	{\includegraphics[width=3.5cm]{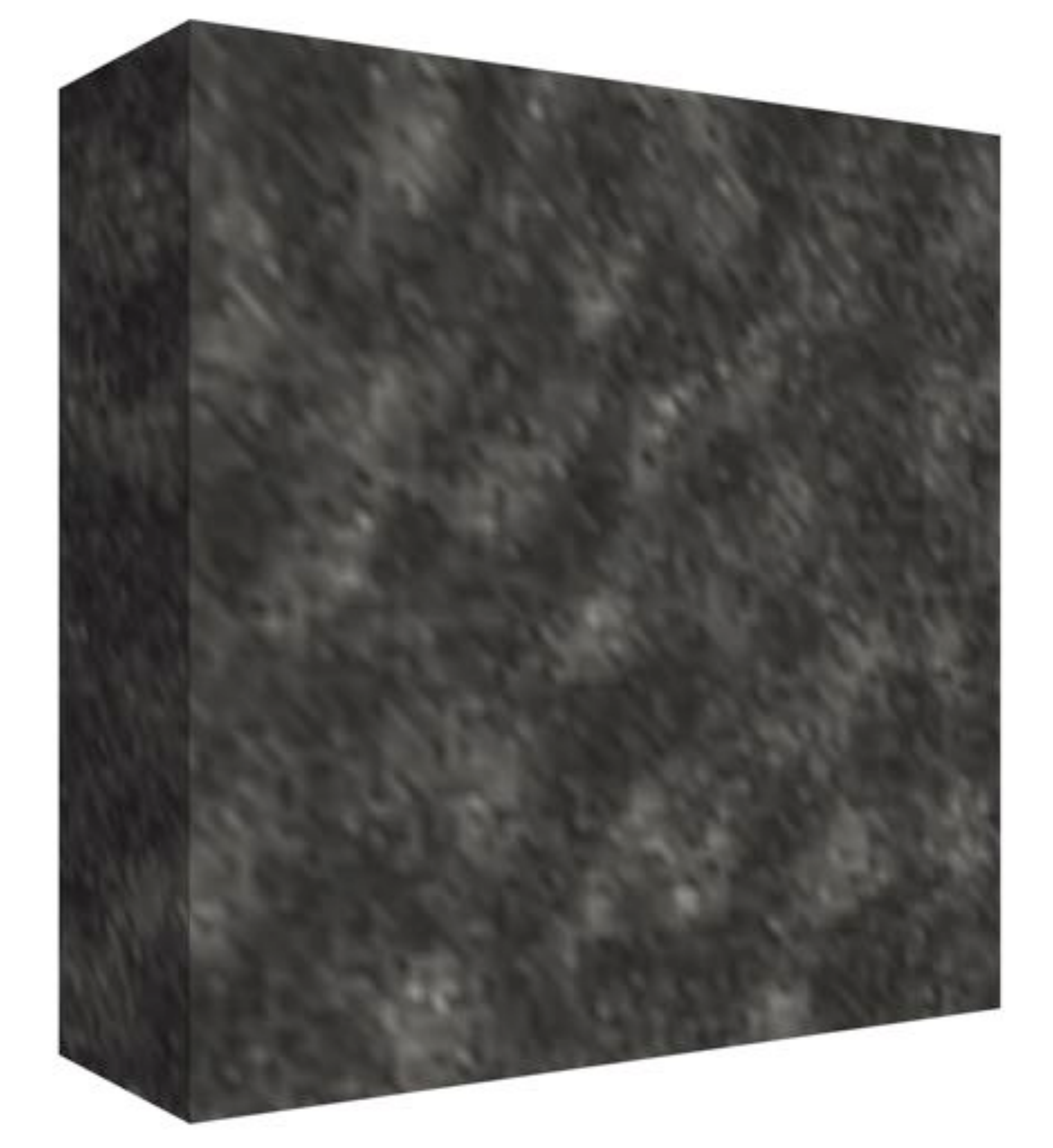}}
	\subcaptionbox{}
	[0.24\textwidth]
	{\includegraphics[width=3.5cm]{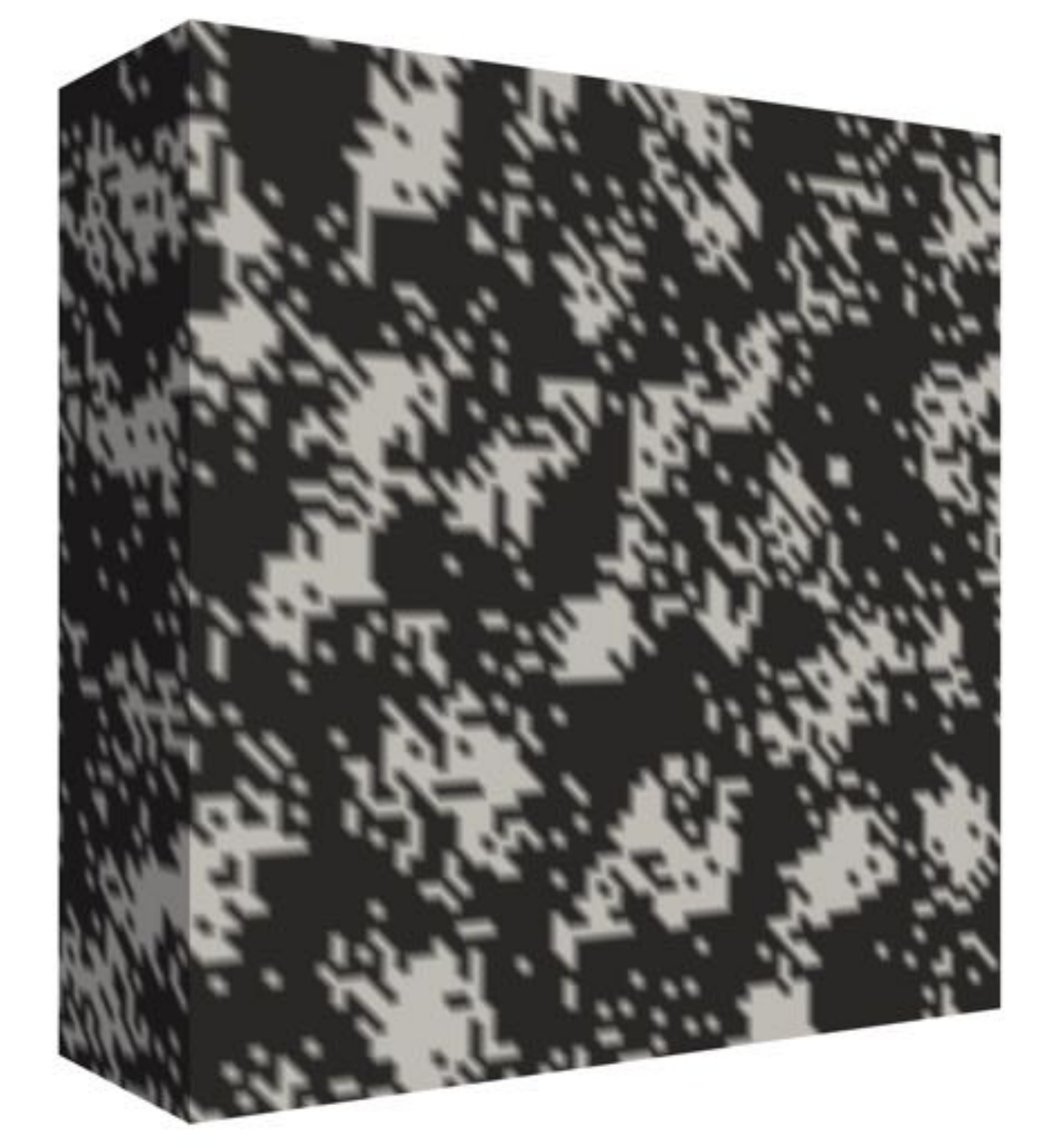}}
	\subcaptionbox{}
	[0.24\textwidth]
	{\includegraphics[width=3.5cm]{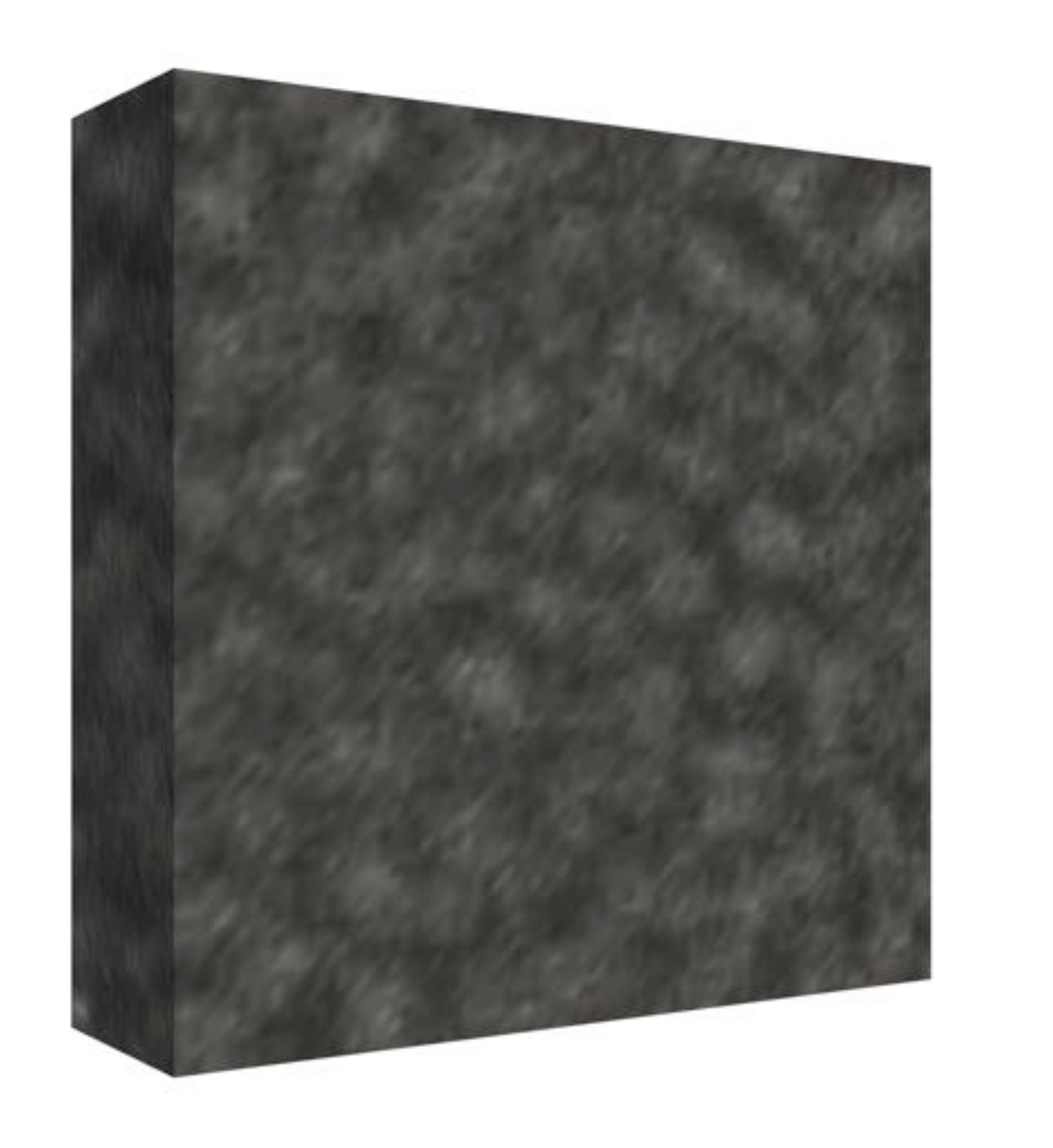}}
	\subcaptionbox{}
	[0.24\textwidth]
	{\includegraphics[width=3.5cm]{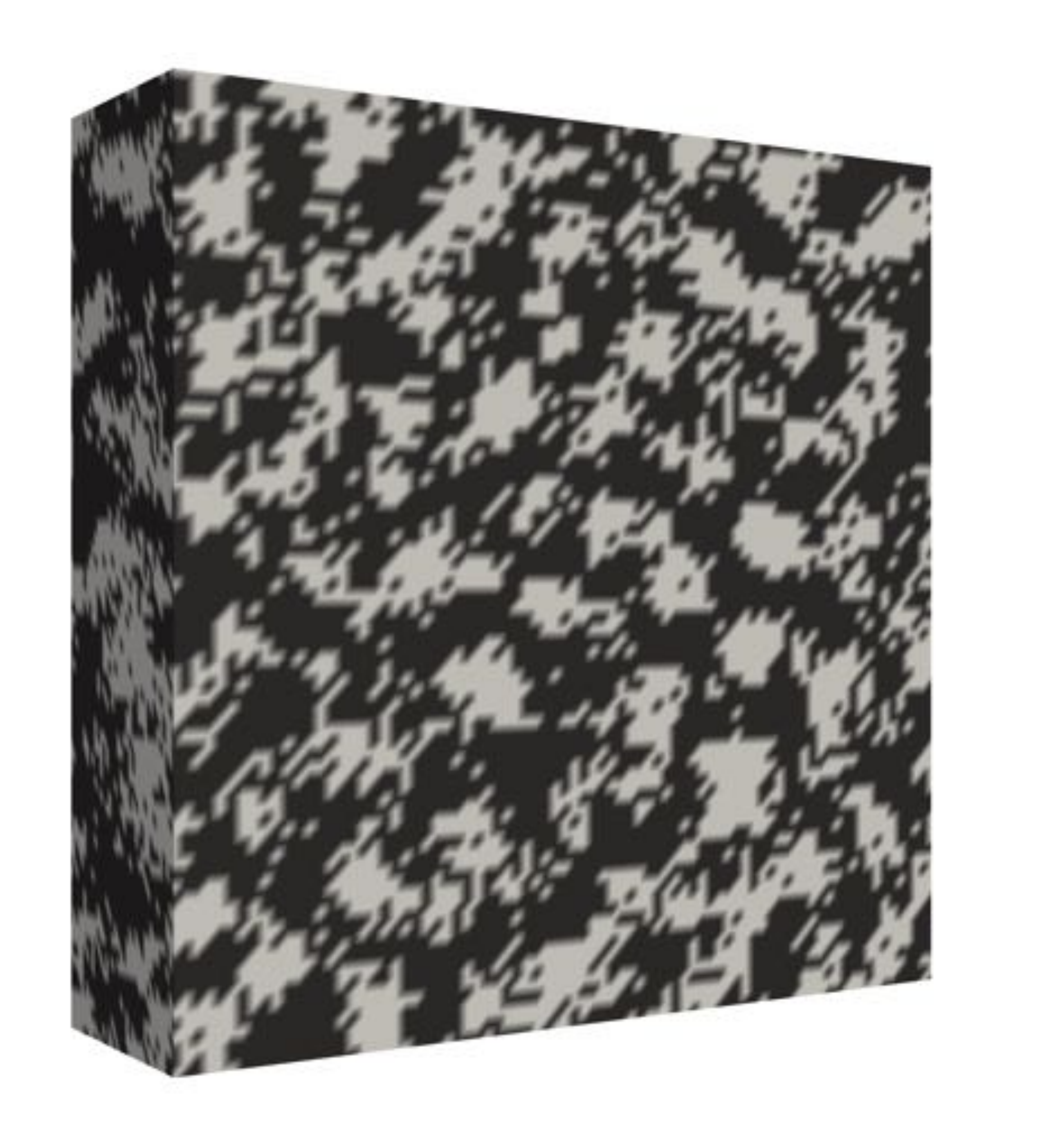}}
	\caption{(a) The CLSM images $HPC30_1$, (b) the binarized $HPC30_1$, (c) the CLSM images $HPC40_1$, and (c) the binarized $HPC40_1$. Sample $HPC30_1$ has $74 \times 74 \times 20$ voxels and the size $17.0 \times 17.0 \times 4$ $\mu$m$^3$, and sample $HPC40_1$ has $81 \times 81 \times 36$ voxels and the size $12.4 \times 12.4 \times 3.6$ $\mu$m$^3$.}
	\label{fig:CLSM}
\end{figure}

In the manufacturing of the films, the EC and HPC phase separate into regions rich in EC and regions rich in HPC. The phase separation continues until the solution has dried. Layers are sprayed onto the film one after the other. The drying of the film is slightly different at the top of the film compared to at the bottom, causing the average size of the pores to be smaller at the top than at the bottom. This is expected to have an effect on the pore shapes, and we observed that the pore shapes are slightly elongated in the direction perpendicular to the layers (depth), i.e.\ the pore shapes are anisotropic.

To fit our model, we selected parts of the CLSM image stacks which have approximately homogeneous pore sizes: four image samples $HPC30_1, \dots, HPC30_4$ from the HPC $30\%$ film, and four image samples $HPC40_1, \dots, HPC40_4$ from the HPC $40\%$ film. The eight sets of CLSM images were chosen to be representative of the variation in pore size in the middle part of the two films. 

The intensity of the signal decreases with depth. To account for this, we first thresholded (binarized) the CLSM images with a threshold that was a function of depth. See Figure \ref{fig:CLSM} for a comparison of binarized and original CLSM images. The threshold was determined by fitting a quadratic function to the $70\%$ and $60\%$ intensity value quantiles for the $30\%$ and $40\%$ HPC samples respectively, calculated for each layer individually. The resulting threshold decreases slowly with depth. 

\subsection*{The pore structure model}
\label{sec:themodel}
\begin{figure}[t]
	\centering
	\subcaptionbox{}
	 [0.24\textwidth]
	 {\includegraphics[width=3.5cm]{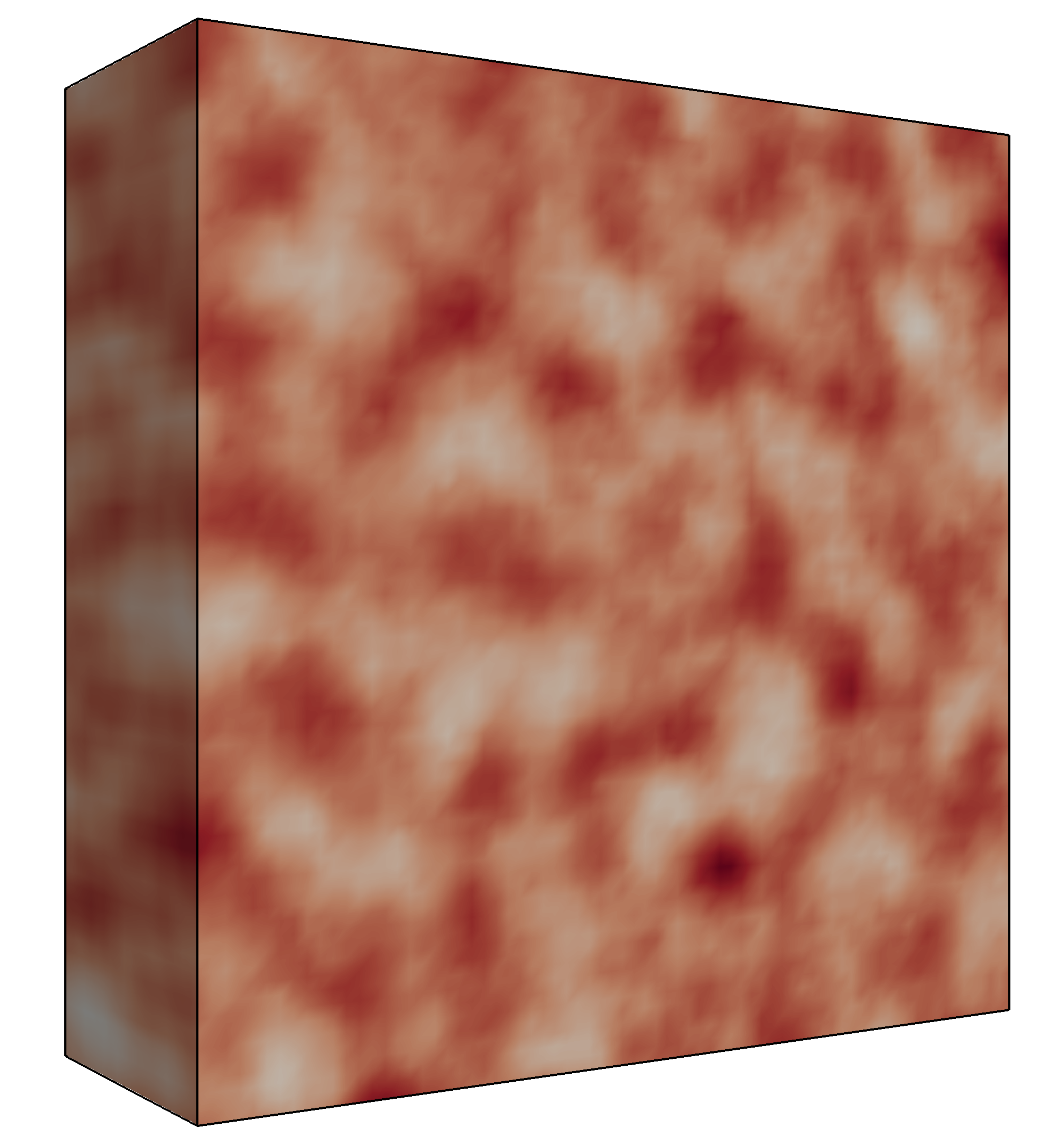}}
	 \subcaptionbox{}
	 [0.24\textwidth]
	 {\includegraphics[width=3.5cm]{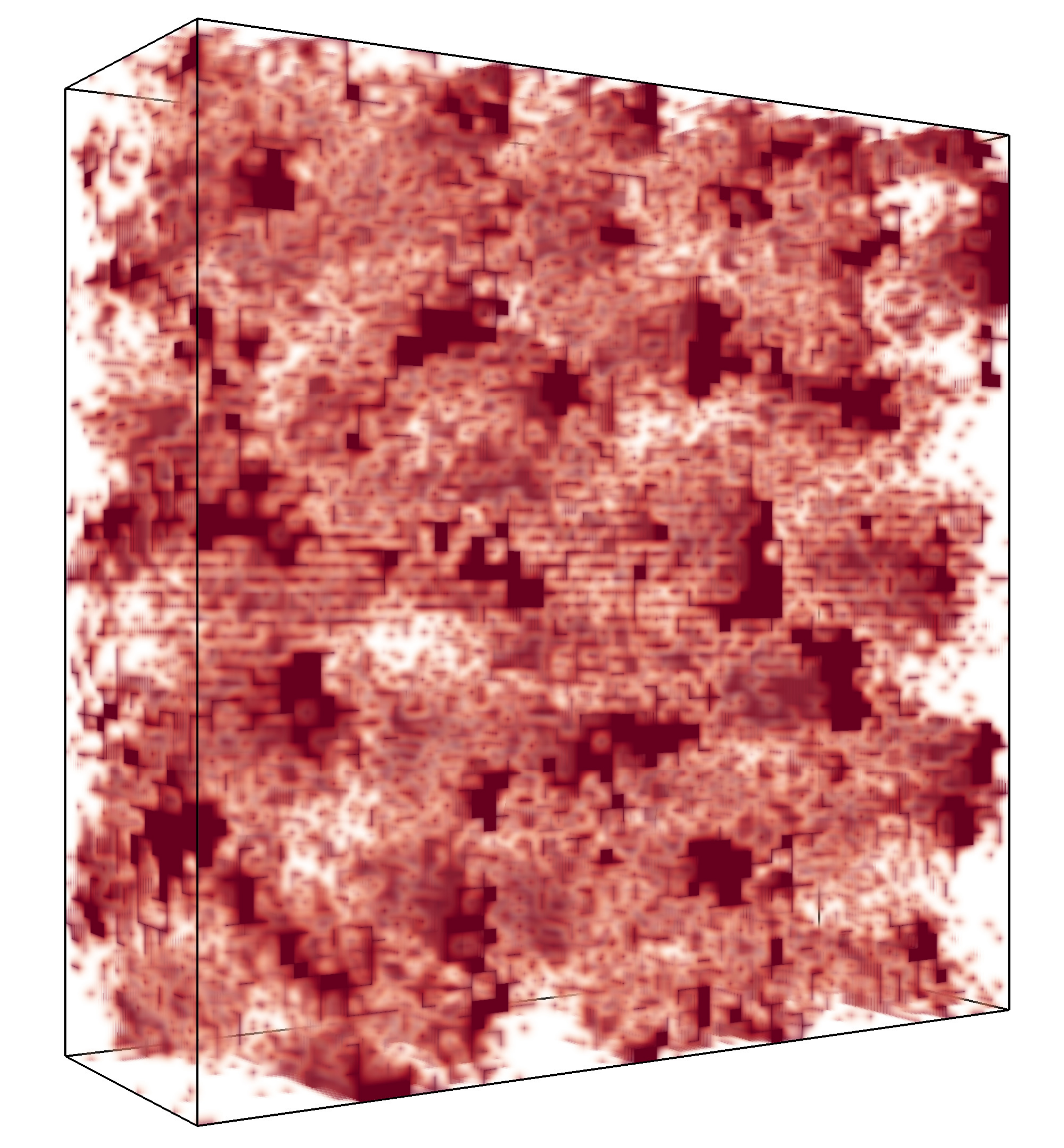}}
	\subcaptionbox{}
	[0.24\textwidth]
	{\includegraphics[width=3.5cm]{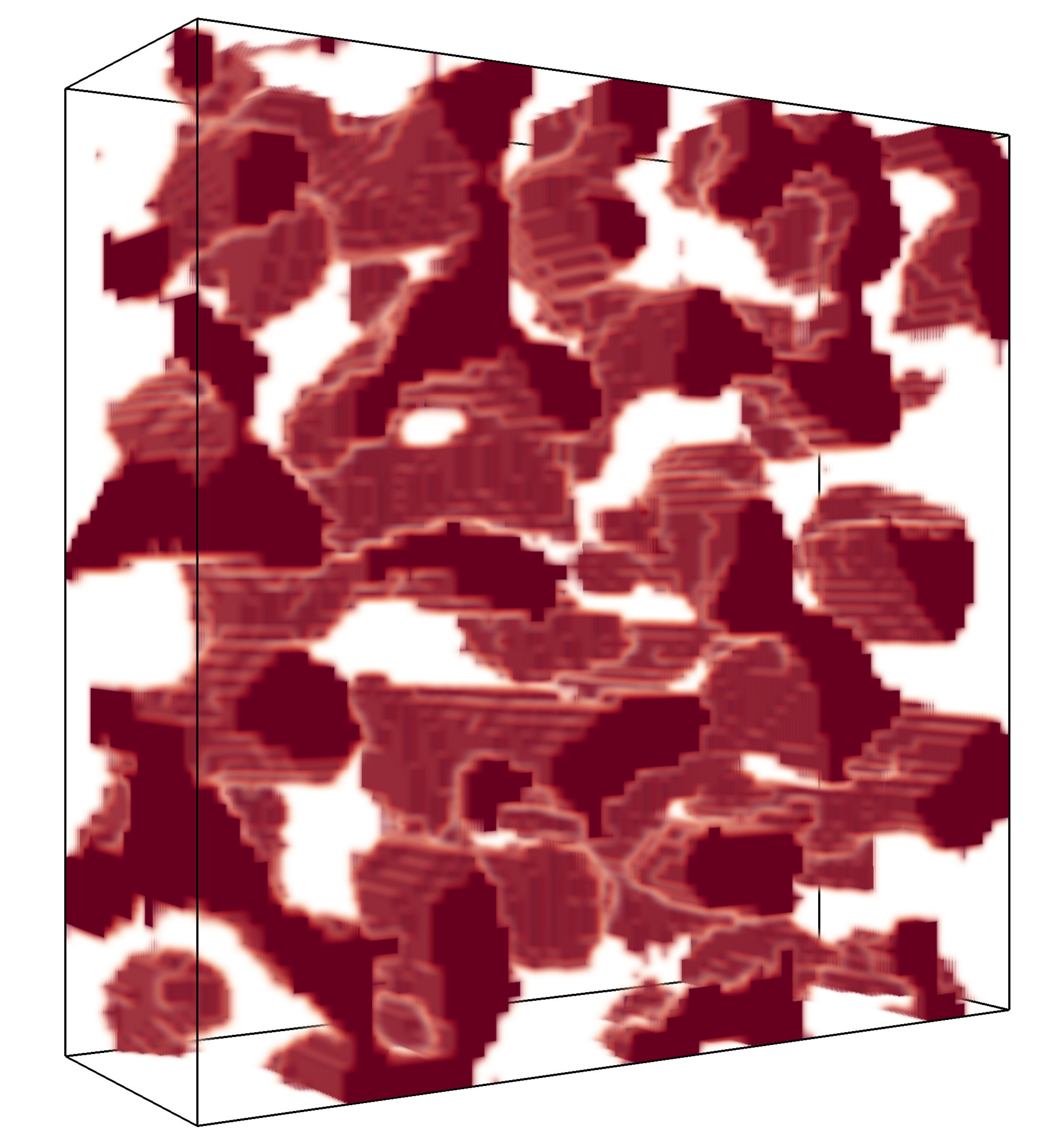}}
	 \caption{Stochastic simulations from the model, showing (a) the Gaussian field $X(\mv{s},z)$, (b) the noisy structure simulated from model \eqref{eq:noisemodelorig}, and (c) the pore structure simulated from model \eqref{eq:poremodel}. The size of the simulations are $74\times74\times20$ voxels.}
	 \label{fig:modelsim}
\end{figure}

In this section, we define the model for the pore structure. To simplify indexing, we define $\mv{s} = (x,y)$ so that a location $(x,y,z)$ in the CLSM image is written as $(\mv{s},z)$, where $z$ corresponds to the depth of the film. The model for the noisy binarized CLSM image $\boldy$ is obtained by taking a Gaussian field $X(\mv{s},z)$ with additive noise and thresholding it, as
\begin{align}
\label{eq:noisemodelorig}
y_i = \begin{cases}
1,&\text{ if $X(\mv{s}_i,z_i)+\vep_i \geq u$,} \\
0,&\text{ otherwise,}
\end{cases} \qquad	 i=1,\dots,m.
\end{align}
Here $m$ is the number of voxels in the sample, $u$ is the threshold, and $\vep_i$ are independent standard Gaussian variables that capture measurement noise. The model for the pore structure $\tildey$ is obtained by smoothing $\boldy$ with a filter $F$,
\begin{align}
\label{eq:poremodel}
\tildey = F(\boldy),
\end{align}
where $\tilde{y}_i = 1$ means that there is a pore at voxel $i$, and $\tilde{y}_i = 0$ that there is no pore. The filter that is used is a simple mean value filter, which takes the unweighted average in a 3-by-3-by-3 or a 5-by-5-by-5 neighbourhood depending on the pore sizes of the sample, combined with a global re-thresholding to obtain the correct volume fraction. See Figure \ref{fig:modelsim} for an example of a sample from $X(\mv{s},z)$ and the corresponding model structures, obtained using the model for $X(\mv{s},z)$ that is defined in the next section. 

\subsubsection*{Separable oscillating Gaussian Mat\'ern fields}\label{sec:theOscMGF}
To capture the different shapes of pores in the depth of the film (corresponding to the z-direction) compared to the within the layers (the $\mv{s}$-planes), the stochastic model should be anisotropic and allow for a different behaviour in the $z$-direction. We do this by using a Gaussian random field $X(\mv{s},z)$ with a separable covariance function given as $\Cov(X(\mv{s}_1,z_1),X(\mv{s}_2,z_2)) = \sigma^2\Cor_{\mv{s}}(\mv{s}_1,\mv{s}_2)\Cor_z(z_1,z_2)$, where $\sigma^2$ is the variance of the process. The anisotropy in the $z$-direction can then be controlled by letting the correlation function $\Cor_z$ be different from the correlation function $\Cor_{\mv{s}}$. 

\begin{figure}[t]
	\centering
	\subcaptionbox{} 
	{\includegraphics[width = 0.45\linewidth]{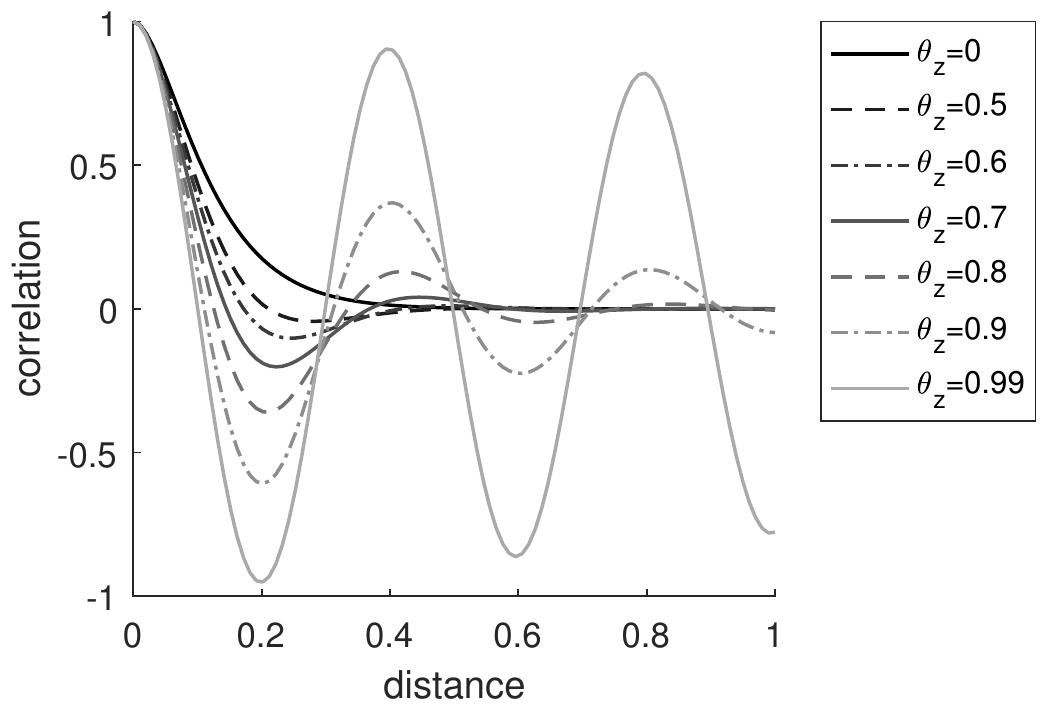}}
	\subcaptionbox{}
	{\includegraphics[width = 0.45\linewidth]{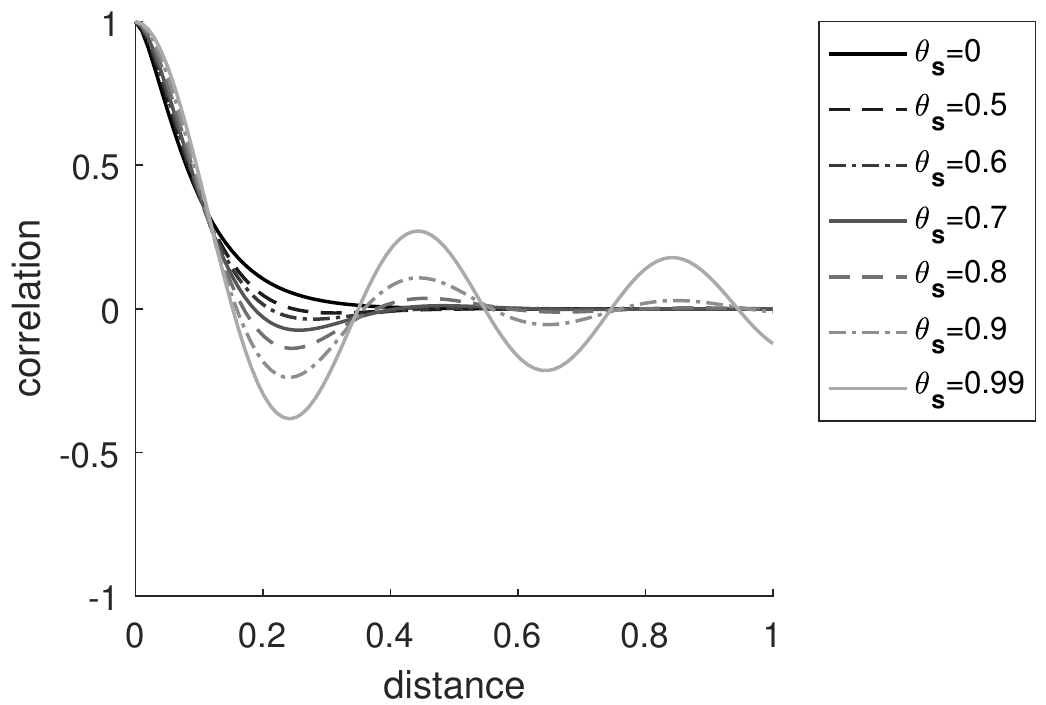}}
	\caption{Correlation functions with range parameters $250^{1/2}$, and different values of the oscillation parameters, for (a) the covariance $\Cov_z(z_1,z_2)$, shown as a function of the distance $|z_1-z_2|$, and (b) the covariance $\Cov_{\mv s}(\mv{s}_1,\mv{s}_2)$, shown as a function of the distance $\lVert \mv{s}_1-\mv{s}_2 \rVert$.}
	\label{fig:theta}
\end{figure}

\begin{figure}[t]
	\centering
	\subcaptionbox{} 
	[0.32\textwidth]
	{\includegraphics[height=4.5cm]{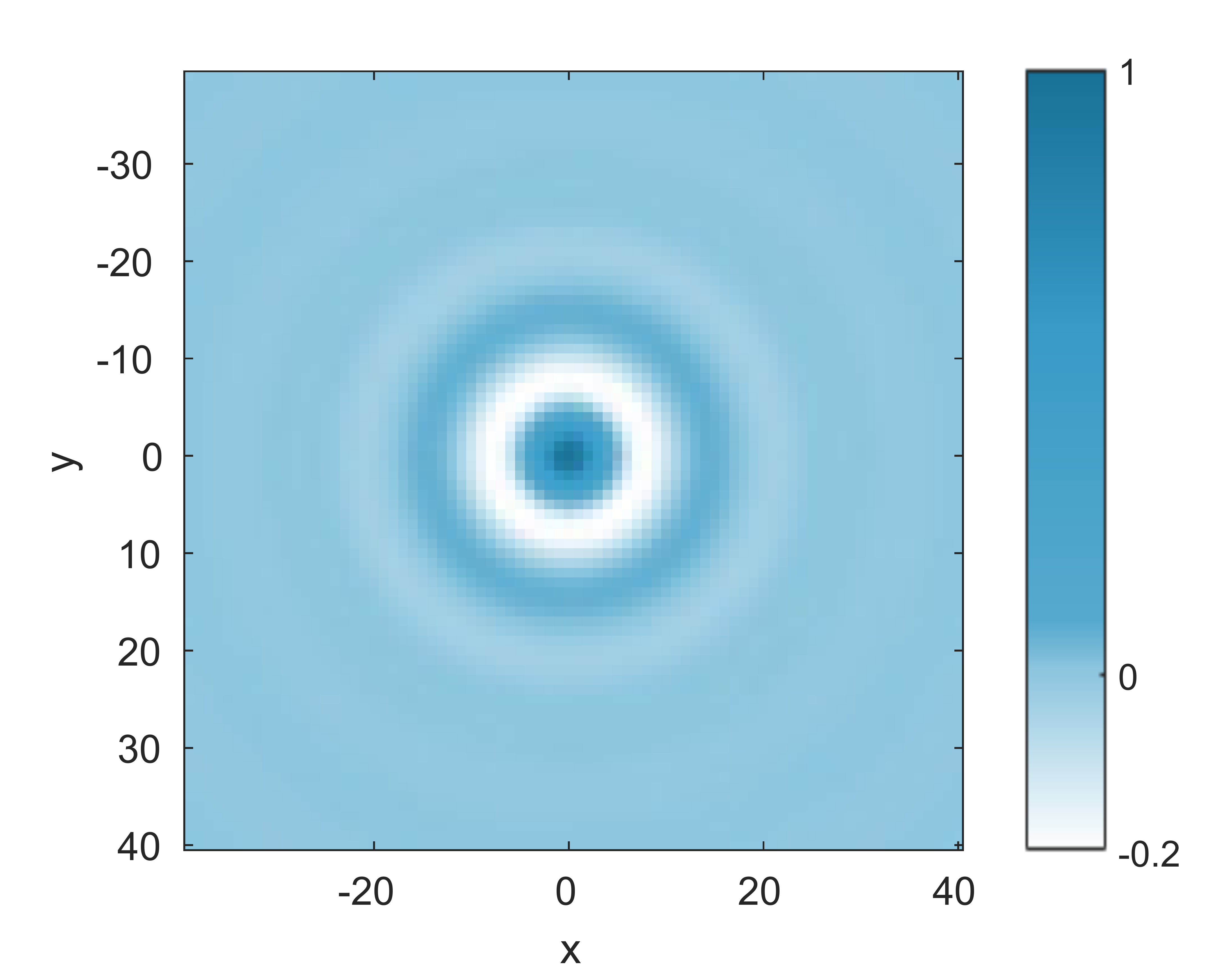}
	}
	\subcaptionbox{}
	[0.32\textwidth]
	{\includegraphics[height=4.5cm]{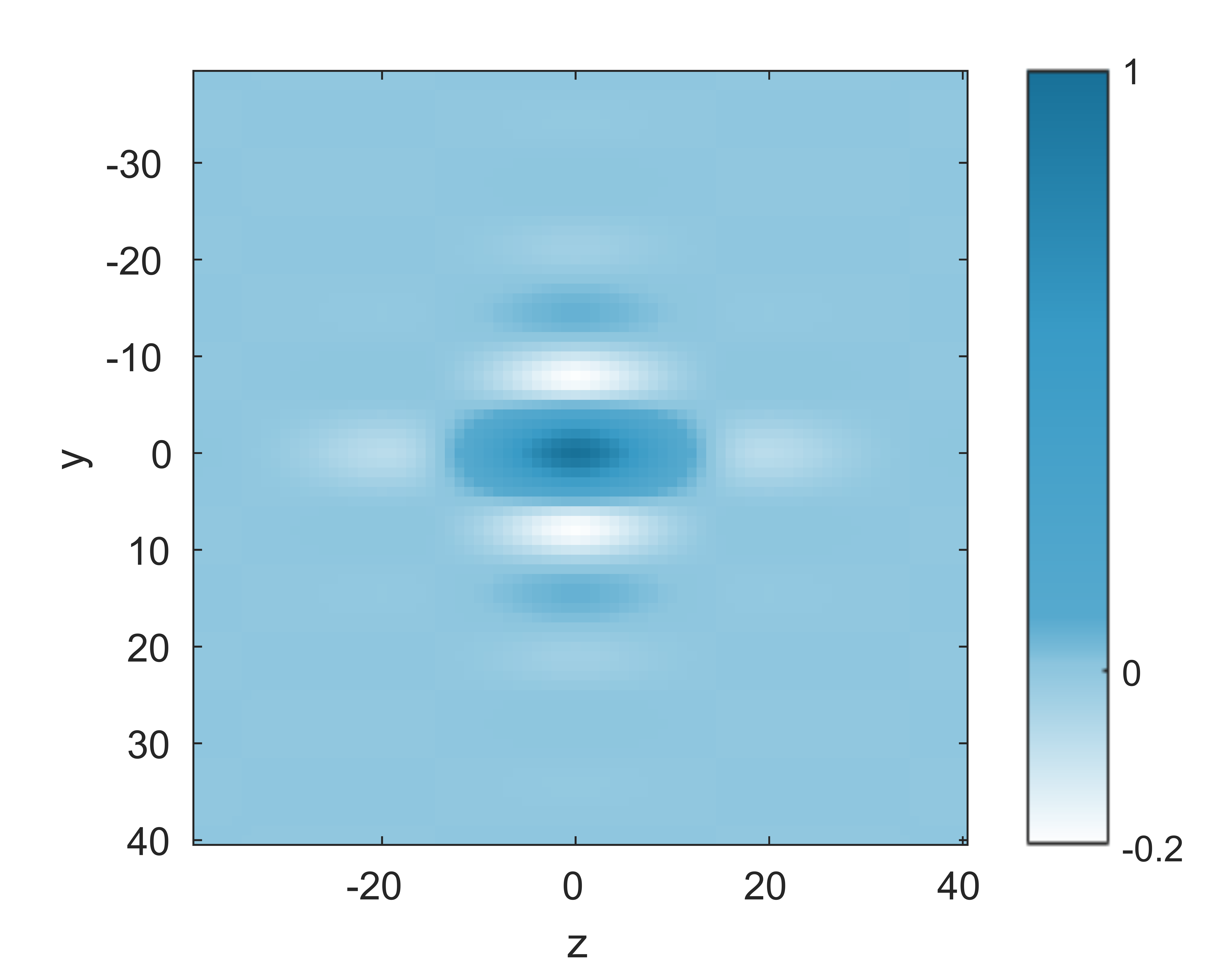}}
	\subcaptionbox{}
	[0.32\textwidth]
	{\includegraphics[height=4.5cm]{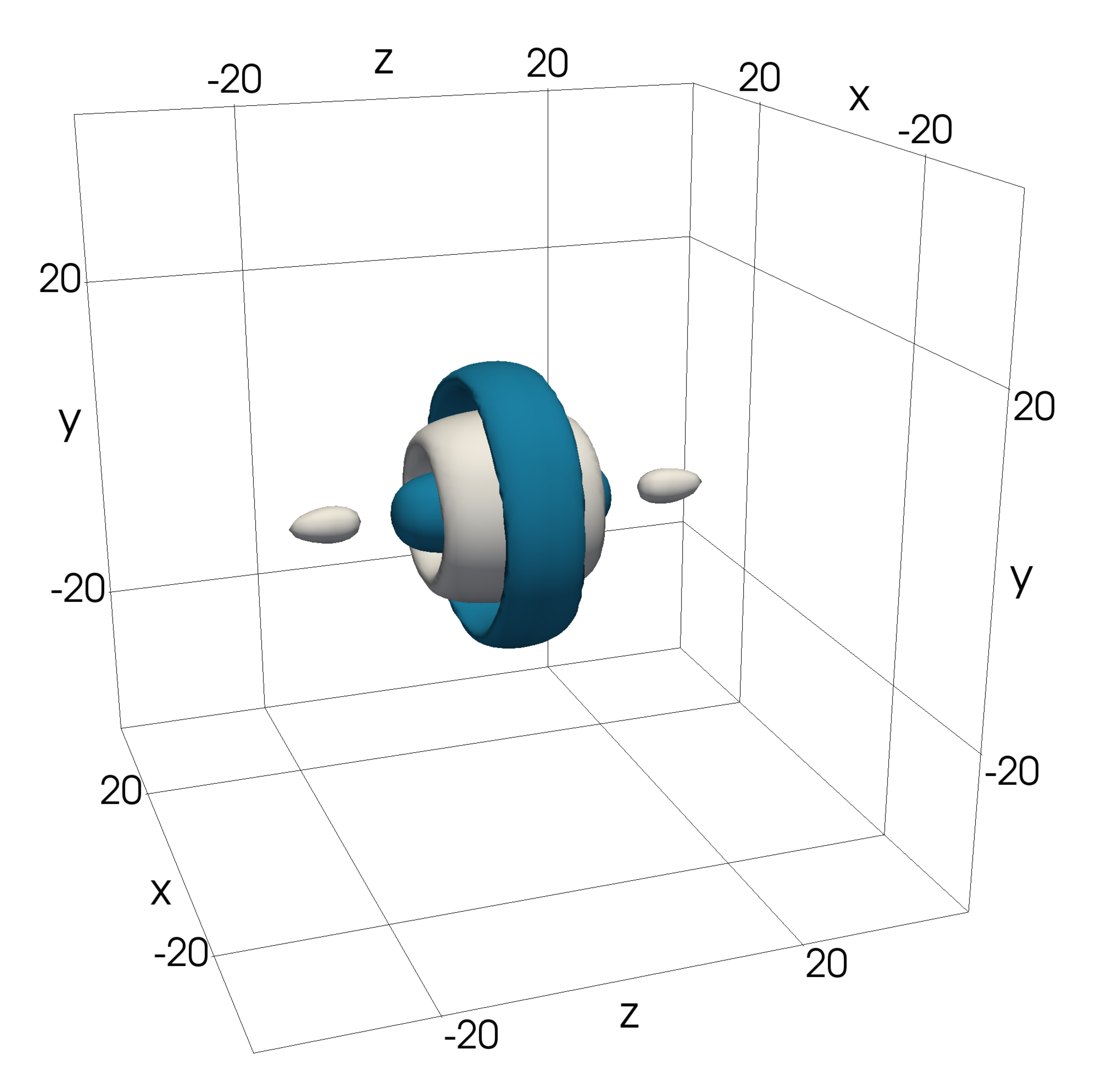}}
	\caption{Figures illustrating the covariance function of a separable oscillating Gaussian Mat\'ern field $X$, showing (a) the correlation of $X(\mv{0},0)$ with $X(\mv{s},0)$, (b) the correlation of $X(\mv{0},0)$ with $X(\mv{s}_0,z)$, where $\mv{s}_0 = (0,y)$, and (c) isosurfaces at positions where the correlation of $X(\mv{0},0)$ with $X(\mv{s},z)$ equals $0.05$ (blue) and $-0.05$ (white). The distance is given in voxels.} 
	\label{fig:modelcov}
\end{figure}

The imaged pore structures have regularly spaced pores, as can be seen from Figure \ref{fig:CLSM}. To obtain a Gaussian field that is regular we need to use an oscillating covariance function. Covariance functions with differing oscillating strengths should be allowed for, since covariance functions estimated from different binarized CLSM image samples are oscillating to various degrees. We use the family of oscillating Mat\'ern correlation functions 
\begin{align*}
\Cor_z(z_1,z_2) &= \frac{1}{\sin\left(\frac{\pi \theta_z}{2}\right)} \exp\left\{-\kappa_z \cos\left(\frac{\pi \theta_z}{2}\right) |z_1-z_2|\right\} \cdot \sin\left\{\frac{\pi \theta_z}{2} + \kappa_z \sin\left(\frac{\pi \theta_z}{2}\right) |z_1-z_2| \right\}, \\
\Cor_\mv{s}( \mv{s}_1,\mv{s}_2) &= \frac{1}{\pi \theta_{\mv s}i} \left[K_0\left\{\kappa_{\mv s} \lVert \mv{s}_1-\mv{s}_2 \rVert \exp\left(-\frac{i \pi \theta_{\mv s}}{2}\right) \right\} -  K_0\left\{\kappa_{\mv s} \lVert  \mv{s}_1-\mv{s}_2 \rVert \exp\left(\frac{i \pi \theta_{\mv s}}{2}\right) \right\}\right], 
\end{align*}
introduced by \citet{Lindgren2011}. Here $K_0$ is the modified Bessel function of the second kind and order zero. The positive parameters $\kappa_{*}>0$ scales the correlation functions, and so control the correlation ranges and the spacing between the correlation functions' peaks. The spacing determines the typical distance between neighbouring peaks in the Gaussian field. The parameters $\theta_{*}\in [0,1)$ control the amount of oscillation in the correlation functions. For $\theta_{*}=0$, the functions are equal to the well-known regular Mat\'ern correlation function, with smoothness parameter $\nu=1$ for $\mv{s}$ and $\nu = 3/2$ for $z$. The oscillation of the correlation functions is more pronounced the closer $\theta_{*}$ is to one, and the corresponding Gaussian field will therefore be more regular for higher values of $\theta_{*}$. Examples of the correlation functions $\Cor_z$ and $\Cor_\mv{s}$ are shown in Figure~\ref{fig:theta}. Figure \ref{fig:modelcov} shows correlations $\Cor_z\Cor_\mv{s}$ for the Gaussian field $X(\mv{s},z)$, with parameters taken from the model fitted to the dataset $HPC30_1$, where the oscillation parameters were $\theta_{\mv{s}}=0.86$ and $\theta_z=0.56$. A stochastic simulation of this Gaussian field is shown in Figure \ref{fig:modelsim}(a). The correlation is symmetric in the $\mv{s}$-plane since $\Cov_{\mv s}$ is isotropic, and has a different behaviour along the $z$-axis since the joint covariance is anisotropic.

\citet*{Lindgren2011} derived the oscillating Mat\'ern covariance function as the covariance function of the solution to a certain SPDE. By similar arguments, one can show that $X(\mv{s},z)$ in our case can be represented as the solution to the SPDE
\begin{equation}\label{eq:oscSPDE}
\left(\kappa_z^2 \exp(\text{i} \pi \theta_z) - \frac{\pd^2}{\pd z^2} \right)\left(\kappa_{s}^2 \exp(\text{i} \pi \theta_s) - \Delta \right)\{ \tau \left[X(\mv{s},z) + \text{i} X_2(\mv{s},z)\right] \}  = W_1(\mv{s},z) + \text{i} W_2(\mv{s},z), 
\end{equation}
where $\Delta = \frac{\pd^2}{\pd x^2} + \frac{\pd^2}{\pd y^2}$ is the Laplace operator,  $\tau$ is a parameter that controls the variance, and $W_1$ and $W_2$ are independent Gaussian white noise fields. A stationary weak solution $(X,X_2)$ to this SPDE has the property that the component fields $X$ and $X_2$ are real, independent, and have the same distribution. These fields are stationary and isotropic and we call them \textit{separable oscillating Gaussian Mat\'ern fields}. 

There are two important reasons for representing $X$ using the SPDE \eqref{eq:oscSPDE}. First, it allows us to define non-stationary extensions of the separable oscillating Gaussian Mat\'ern field by allowing the parameters to be spatially varying. This will be used in future work to analyse larger samples of EC/HPC films where the stationarity assumption is problematic. Secondly, the representation makes it possible to approximate the Gaussian field using a Gaussian Markov random field (GMRF). A GMRF is a multivariate Gaussian random vector with a sparse precision matrix (inverse covariance matrix). As we will see later, the sparsity greatly reduces the computational cost of fitting the model to data.

\subsubsection*{The Gaussian Markov random field approximation}\label{sec:theGMRFmodel}
A GMRF approximation of the separable oscillating Gaussian Mat\'ern field $X(\mv{s},z)$ is obtained by solving the SPDE \eqref{eq:oscSPDE} approximately using the finite element method, on a bounded domain $\Omega = \Omega_{\mv{s}}\times\Omega_z \subset \mathbb{R}^3$ with Neumann boundary conditions. 
The idea is to approximate $X(\mv{s},z)$ by a basis expansion $X_{\text{FEM}}(\mv{s},z) = \sum_{i=1}^n w_i \varphi_i(\mv{s},z) $, where $\{\varphi_i\}$ is a suitable basis and $\mv{w}$ are stochastic weights. Here a good choice of basis is the Kronecker basis $\{\varphi_{ij}(\mv{s},z) = \varphi_{\mv{s},i}(\mv{s})\varphi_{z,j}(z)\}$, where $ \{\varphi_{\mv{s},i}\}$ and $\{\varphi_{z,i}\}$ are piecewise linear and continuous basis functions obtained by a triangulation of $\Omega_{\mv{s}}$ and $\Omega_{z}$ respectively. The basis functions $\varphi_{\mv{s},i}$ and $\varphi_{z,i}$ take the value one in node $i$ and zero in all other nodes in their respective triangulations. The distribution of the stochastic weights $\mv{w}$ is derived using the Galerkin method \citep*{Lindgren2011}. Straightforward calculations give that $\mv{w}\sim N(0,\mv{Q}(\mv{\gamma})^{-1})$, where $\mv{\gamma} = (\theta_s,\kappa_s,\theta_z,\kappa_z,\tau)$ and $\mv{Q}(\mv{\gamma}) = \tau^2\mv{Q}_{\mv{s}}(\theta_s,\kappa_s)\otimes\mv{Q}_z(\theta_z,\kappa_z)$. The Kronecker structure of the precision matrix $\mv{Q}(\mv{\gamma})$ is a result of the separability of the covariance of $X(\mv{s},z)$, and of the choice of basis functions. The two matrices $\mv{Q}_{\mv{s}}$ and $\mv{Q}_{z}$ are given by 
\begin{equation*}
\label{eq:Qmatrices}
\mv{Q}_{\star}(\theta_{\star},\kappa_{\star}) = \kappa_{\star}^4 \mv{C}_{\star} + 2 \kappa_{\star}^2 \cos(\pi \theta_{\star}) \mv{G}_{\star} + \mv{G}_{\star}\mv{C}_{\star}^{-1} \mv{G}_{\star}, 
\end{equation*} 
for $\star= \mv{s}, z$, where $\mv{C}_{\star}$ is a diagonal matrix with diagonal elements $C_{\star,ii} = \int_{\Omega} \varphi_{\star,i}(t) dt$ and $\mv{G}_{\star}$ is a sparse matrix with elements $G_{\star,ij} = \int_{\Omega} \nabla \varphi_{\star,i}(t) \nabla \varphi_{\star,j}(t) dt$. Since the matrices $\mv{C}_{\star}$ and $\mv{G}_{\star}$ are sparse, so is $\mv{Q}_{\star}$, and the weights $\mv{w}$ is therefore a GMRF with the sparse precision matrix $\mv{Q}(\mv{\gamma})$. 

With this approach $X_{\text{FEM}}(\mv{s},z)$ is a piecewise linear approximation of $X(\mv{s},z)$. We refer to the stochastic weights $\mv{w}$ as the \textit{separable oscillating Mat\'ern GMRF}. The value of $X_{\text{FEM}}(\mv{s},z)$ at the node of a basis function $\varphi_{ij}(\mv{s},z)$ is the corresponding weight. When the triangulation used to construct the basis functions is refined, the approximating field $X_{\text{FEM}}$ converges weakly to $X$ \citep*{Lindgren2011} .

The domain $\Omega$ is chosen to be slightly larger than the domain of the CLSM image to reduce the effect of the boundary conditions imposed on the SPDE. We use one basis function for each voxel in the sample, with the node of the basis function at the center of the voxel, as well as basis functions outside the domain of the sample. For further details, see Appendix A.

\subsection*{Markov Chain Monte Carlo model fitting}\label{sec:MCMC}
To fit the model to the noisy binarized CLSM image sample $\boldy$, the parameters $\mv{\gamma}$ and the threshold $u$ have to be estimated. Evaluating the likelihood of the model is too computationally demanding, so we estimate the model using Markov Chain Monte Carlo (MCMC) in a Bayesian context instead of doing maximum likelihood estimation. 

Let $\mv{A} \in \mathbb{R}^{m \times n}$ be a matrix that maps the values of the GMRF $\mv{w}$ to the values of the piecewise linear approximation $X_{\text{FEM}}(\mv{s},z)$ at the voxels in the CLSM image, where as before $m$ is the number of voxels and $n$ is the number of stochastic weights. Specifically, element $ij$ in $\mv{A}$ is obtained by evaluating the $j$th basis function in the location of the $i$th voxel, so that $\mv{A}\mv{w}$ is a vector that contains the values for $X_{\text{FEM}}(\mv{s},z)$ at the voxels of the CLSM image. 

Approximating $X(\mv{s},z)$ with $X_{\text{FEM}}(\mv{s},z)$, we have from model \eqref{eq:noisemodelorig} that $y_i = 1$ if $\mv{A}_{i,\bullet}\mv{w} + \epsilon_i \geq u$, and $y_i = 0$ otherwise. This means that we can write the full model as 
\begin{align}
\label{eq:noisemodel} 
\begin{split}
y_i|\mv{w} &\sim  Be(\Phi(\mv{A}_{i,\bullet}\mv{w}-u)), \quad i=1,\dots,m,\\
\mv{w} &\sim \pN(0,\mv{Q}(\boldsymbol\gamma)^{-1}), \\
\boldsymbol\gamma &\sim \pi(\boldsymbol{\gamma}).
\end{split}
\end{align}
Here $Be$ denotes the Bernoulli distribution, $\Phi$ denotes the distribution function of a standard normal random variable, and $\pi(\mv{\gamma})$ is a prior distribution for the parameters. 

The MCMC algorithm we use to estimate the parameters is a blocked Metropolis-within-Gibbs sampler, see Appendix A for details. The algorithm is used to estimate the joint posterior density $\pi(\boldw, \boldsymbol\gamma, u \mid \boldy)$ and the posterior expectation $\hat{\mv{\gamma}} = E(\mv{\gamma} | \boldy)$ is taken as a point estimate of the parameters. When simulating from the fitted model, we draw parameters from the posterior distribution of the parameters.

Sampling from the posterior distribution of $\mv{w}$ is the most computationally demanding part of the MCMC algorithm. A method for reducing the complexity of the sampling is presented in Appendix A. The method takes advantage of the Kronecker structure of the precision matrix $\mv{Q}(\mv{\gamma}) = \tau^2\mv{Q}_{\mv{s}}\otimes\mv{Q}_z$ and its Cholesky factor, and the sparsity of these matrices. The most important implication of the Kronecker structure, which we use to improve the computational efficiency of the method, is that matrix-vector operations can be computed efficiently as follows. 

For a matrix $\mv{C} = \mv{C}_1 \otimes \mv{C}_2$, where $\mv{C}_1 \in \mathbb{R}^{n_{1} \times n_{1}}$ and $\mv{C}_2 \in \mathbb{R}^{n_{2} \times n_{2}}$, we have that
\begin{align}
\label{eq:kronprod}
\mv{C}\mv{v} = vec((\mv{C}_1(\mv{C}_2\mv{V})^T)^T),
\end{align} 
where $vec$ denotes the vectorization operator which maps a matrix to a vector by stacking the columns of the matrix, and $\mv{V}$ is an $n_{2}\times n_{1}$ matrix containing the elements of $\mv{v}$ with $\mv{V}_{ij} = \mv{v}_{(j-1)n_2+i}$ \citep*{BuisDyksen1996}. Using the same trick for the inverse we get
\begin{align}
\label{eq:kroninverse}
\mv{C}^{-1}\mv{v} = vec((\mv{C}_1^{-1}(\mv{C}_2^{-1}\mv{V})^T)^T),
\end{align} 
using that $\mv{C}^{-1} = \mv{C}_1^{-1} \otimes \mv{C}_2^{-1}$. 

For full matrices $\mv{C}_1$ and $\mv{C}_2$, computing $\mv{C}^{-1}\mv{v}$ without taking advantage of the Kronecker structure requires $\mathcal{O}((n_1 n_2)^3)$ operations, whereas the computation $vec((\mv{C}_1^{-1}(\mv{C}_2^{-1}\mv{V})^T)^T)$ only requires $\mathcal{O}(n_1^3 + n_2^3)$ operations, which is a considerable reduction. In the method for posterior sampling, the matrix-vector operations \eqref{eq:kronprod} and \eqref{eq:kroninverse} can be performed efficiently due to the sparsity of the Kronecker product matrices, further reducing the computational cost. Moreover, using \eqref{eq:kronprod} and \eqref{eq:kroninverse}, the MCMC algorithm can be implemented so that we never need to store full Kronecker products. These reductions in algorithmic complexity are crucial for making it possible to fit the model to images of realistic sizes. Further details are given in Appendix A.

\subsection*{Model validation}
To assess the model fit we need to measure the similarities between the binarized CLSM image and the pore structure obtained by filtering the binarized CLSM image, and the stochastic simulations from the corresponding fitted models \eqref{eq:noisemodel} and \eqref{eq:poremodel}. We have chosen the following measures: the empirical covariance function, size distributions with respect to linear and spherical shapes, and measures related to numerically calculated diffusion. Other popular summarizing functions that could be applied as measures of goodness-of-fit are, e.g., the contact distribution functions, empty space functions and chord length distribution functions \citep*[see][for a review]{Chiu2013ch7}. These are all related to the size distributions. Below we present the goodness-of-fit measures in more detail.

\subsubsection*{Covariance functions}
\label{sec:cov}
As a first assessment of the model fit, we look at marginal covariance functions for both models \eqref{eq:noisemodel} and \eqref{eq:poremodel}. To obtain the theoretical covariance of model \eqref{eq:noisemodel}, we use the following result for a thresholded Gaussian vector: If $(X,Y)$ is a zero-mean bivariate Gaussian vector with unit variances and correlation $\rho \in (-1,1)$, then the covariance between the elements of the vector thresholded at levels $u_X$ and $u_Y$ respectively is $\int_{0}^{\rho} \varphi(u_X,u_Y;z) dz$, where $\varphi(\cdot,\cdot;z)$ is the density of a bivariate Gaussian vector with unit variances and correlation $z$ \citep*[pp.\ 26--27]{CramerLeadbetter1967}. The covariances of the model are obtained using that $((\mv{A}_{i\cdot}\mv{w} + \epsilon_i)/\alpha_i)_{i=1}^n$ is a zero-mean Gaussian vector with unit variances and correlations $\rho_{i,j} = (\mv{Q}(\boldsymbol{\gamma})^{-1})_{i,j}/(\alpha_i \alpha_j)$, $i,j \in \{1,\dots,m\}$, where $\alpha_i^2 = \text{Var}(\mv{A}_{i\cdot}\mv{w} + \epsilon_i) = (\mv{Q}(\boldsymbol{\gamma})^{-1})_{i,i} + \sigma^2$.
Since the marginal distribution of the field evaluated in an $\mv{s}$-plane is stationary and isotropic except for boundary effects, the variances $\alpha_i^2$ will be approximately constant, and the correlations $\rho_{i,j}$ will depend only on the distance between the voxels $i$ and $j$ for voxels lying in the same $\mv{s}$-plane (as long as the voxels are far enough away from the boundary). Therefore we consider a single marginal covariance function $C_{\mv{s}}$ for voxels lying in the same $\mv{s}$-plane. Similarly, we consider a single marginal covariance function $C_z$ for voxels lying on the same $z$-line. 

We compare the empirical marginal covariance functions $\hat{C}_{\mv{s}}$ and $\hat{C}_{z}$ estimated from the binarized CLSM image $\boldy$, with envelopes estimated from stochastic simulations $\boldy_{\text{model}}$ from the fitted noise model \eqref{eq:noisemodel}, as well as with the theoretical covariance functions $C_{\mv{s}}$ and $C_z$ of this model. We also compare the empirical marginal covariance functions estimated from the filtered CLSM pore structure $\tildey$, with envelopes estimated from stochastic simulations $\tildey_{\text{model}}$ from the fitted pore model \eqref{eq:poremodel}. We use simultaneous $1-\alpha$-envelopes, defined as in \citet*{BolinarXiv} so that $1-\alpha$ percent of the covariance functions estimated from stochastic simulations lie completely within the envelopes.

\subsubsection*{Size distributions}
The morphological size distributions, or granulometries, were introduced in \citet*{Matheron1967} as a way of characterizing porous media and random closed sets, and is a common tool in image analysis \citep*{Soille2004}. 
Size distributions provide a measure of size for sets---such as porous structures---that do not have well-defined shapes. Local sizes are measured with respect to simpler sets, so called structuring elements, such as spheres and line segments, and the set of voxels in the three-dimensional pore space of a pore structure $\tildey$ is seen as a realization of a random set $\Xi_{\text{pores}} = \{i:\tilde{y}_i=1\}$, which consists of those voxels $i$ where there is a pore.

\begin{figure}[t]
	\centering
	\subcaptionbox{} 
	[0.3\textwidth]
	{\includegraphics[height=3.5cm]{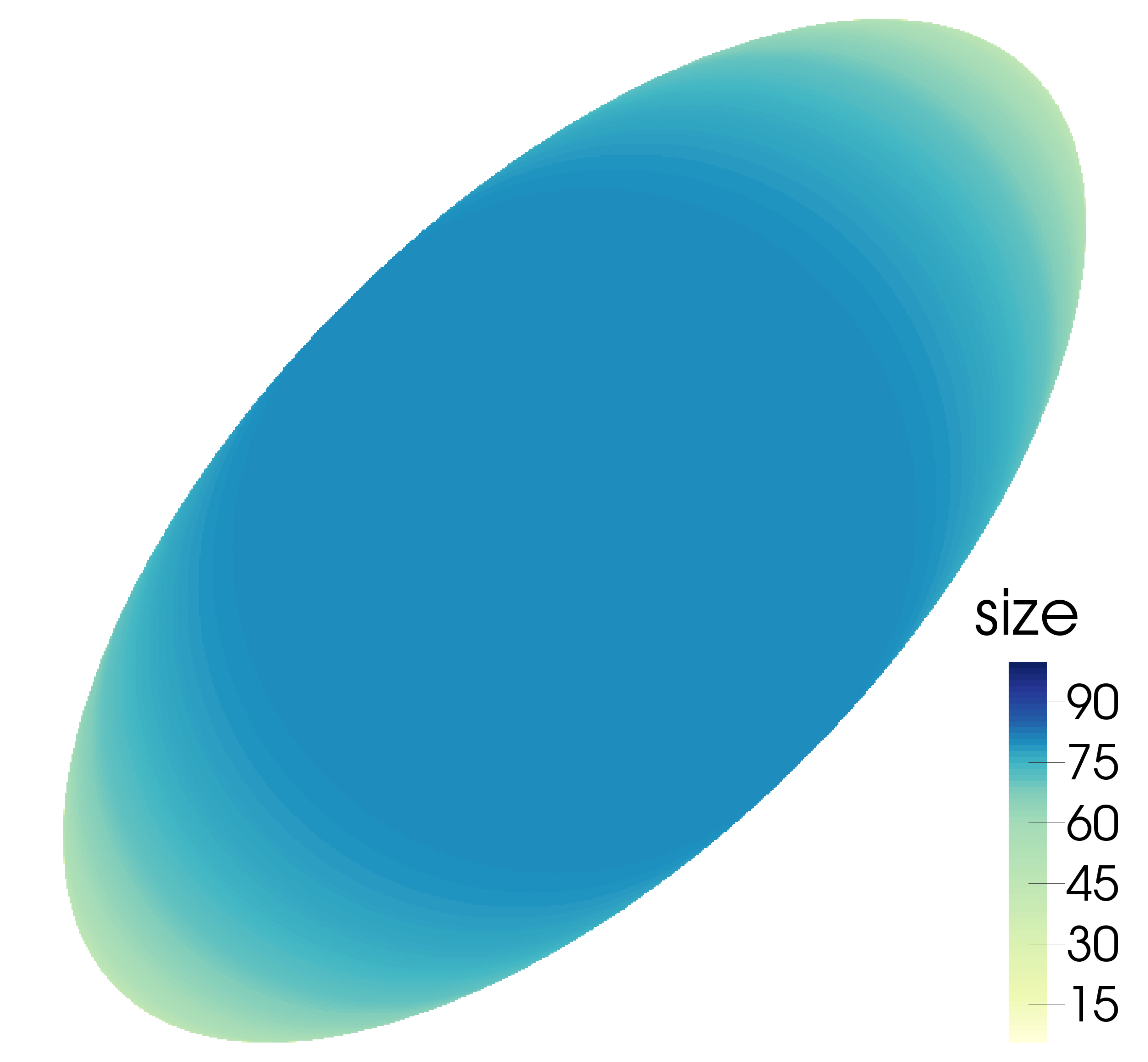}
	}
	\subcaptionbox{}
	[0.3\textwidth]
	{\includegraphics[height=3.5cm]{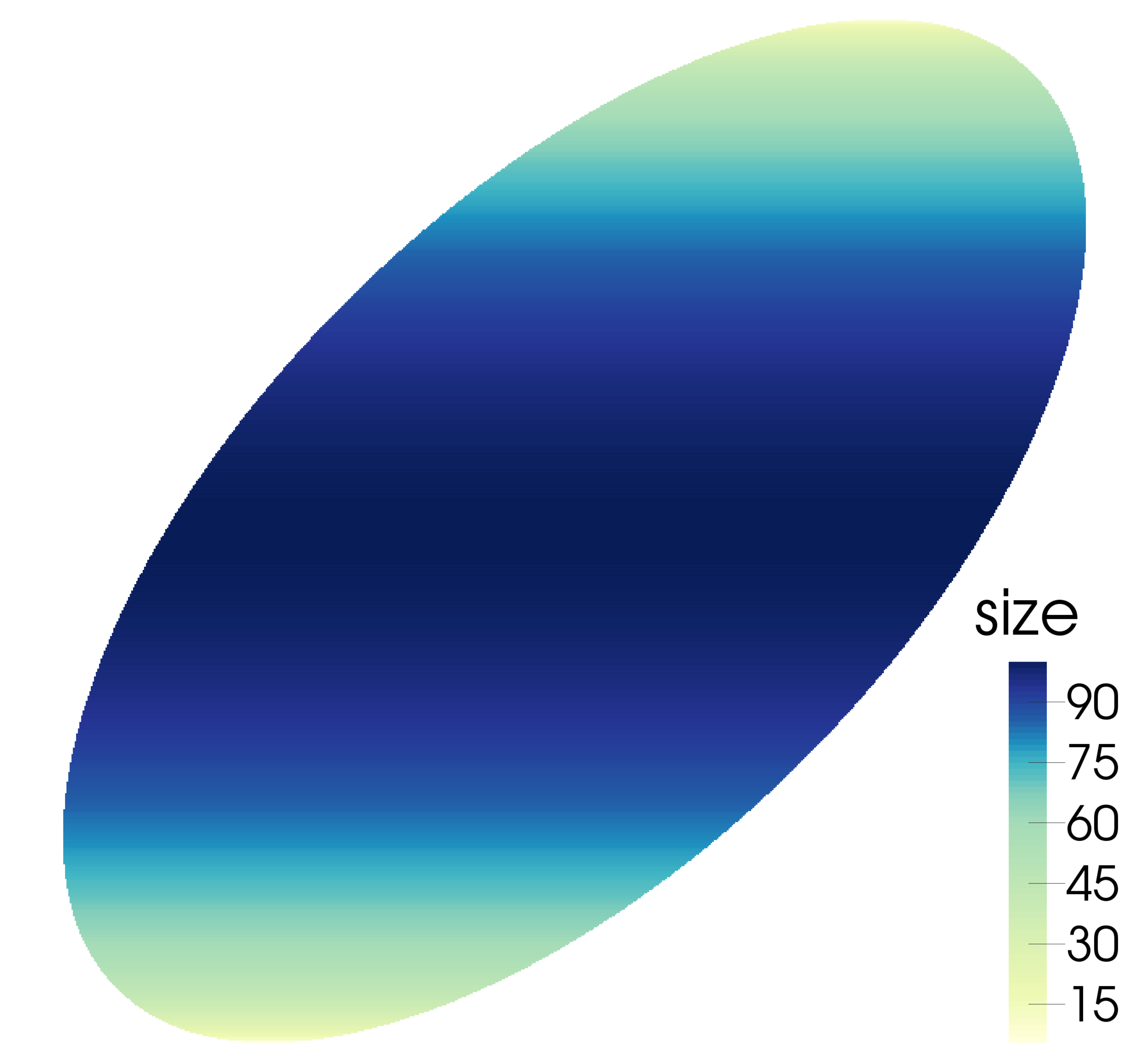}}
	\subcaptionbox{}
	[0.3\textwidth]
	{\includegraphics[height=3.5cm]{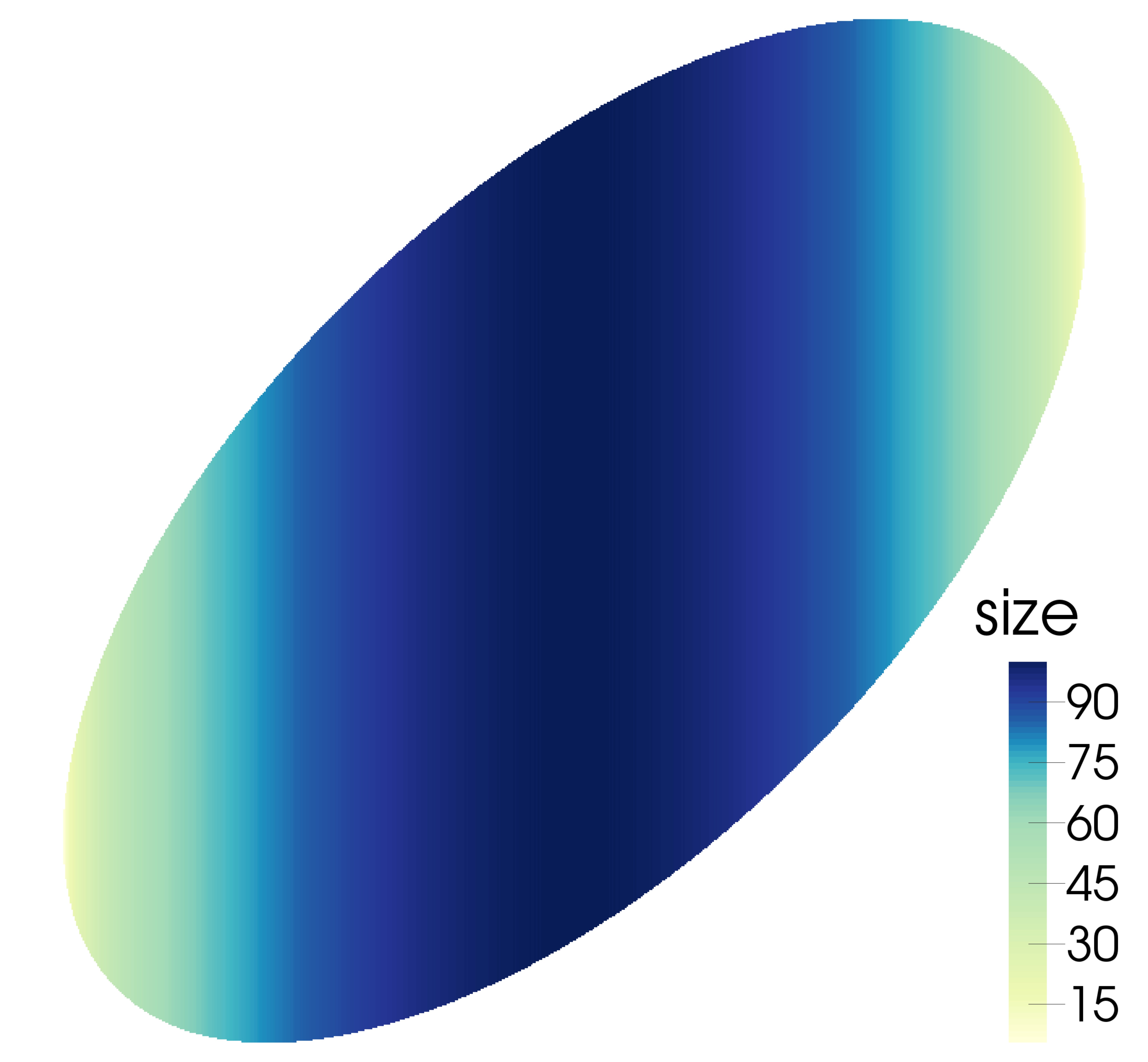}}
	\caption{A discretely indexed ellipse, where each pixel is marked with its size $h$ with respect to the following structuring elements: (a) the unit circle, (b) a line aligned with the x-axis, and (c) a line aligned with the y-axis. The line structuring elements have lengths two.}
	\label{fig:sizedistillustr}
\end{figure}

To define the size distribution of a random set $\Xi$ with respect to a structuring element $B$, first define the local size $h(x,\Xi,B)$ of a point $x \in \Xi$ with respect $B$ as the largest value for which the rescaled structuring element $B$ can be translated so that it fits within the set and covers the point $x$. Figure~\ref{fig:sizedistillustr} illustrates this local size concept, in this case for two-dimensional sets, showing local sizes with respect to three structuring elements $B$. The local sizes with respect to the circle (Panel (a)) are smaller in the middle of the ellipse compared to the local sizes with respect to the lines (Panels (b) and (c)), since it is possible to fit longer lines within the ellipse covering these pixels than it is possible to fit circles with those lines as radii. 

The size distribution $S(\lambda; x,\Xi,B) = P(h(x,\Xi,B) \geq \lambda \mid x \in \Xi)$ gives the probability that a point $x \in \Xi$ has a local size with respect to $B$ which is greater than or equal to $\lambda$. We let $s(\lambda; x,\Xi,B)$) denote the corresponding size density. The size distribution and density do not depend on the point $x$ if the random set is stationary.

For the discretely indexed random set $\Xi_{\text{pores}} = \{i:\tilde{y}_i=1\}$, we let the distance between two neighbouring voxels be one. We estimate stationary size distributions $S(\lambda; \Xi_{\text{pores}}, B)$, with lines aligned with the coordinate axes and the sphere as structuring elements. In addition, we also estimate these four size distributions for the random set $\Xi_{\text{matrix}} = \{i:\tilde{y}_i=0\}$, which consists of the voxels in the matrix (the solid part of the pore structure). The size distributions estimated from stochastic simulations $\tildey_{\text{model}}$ from the pore model \eqref{eq:poremodel} are compared to the size distributions estimated from the filtered CLSM pore structure $\tildey$, using simultaneous envelopes. For details about the estimation, see Appendix B.

\subsubsection*{Numerically calculated diffusion}

\begin{figure}[t]
	\centering
	\includegraphics[width=0.35\linewidth]{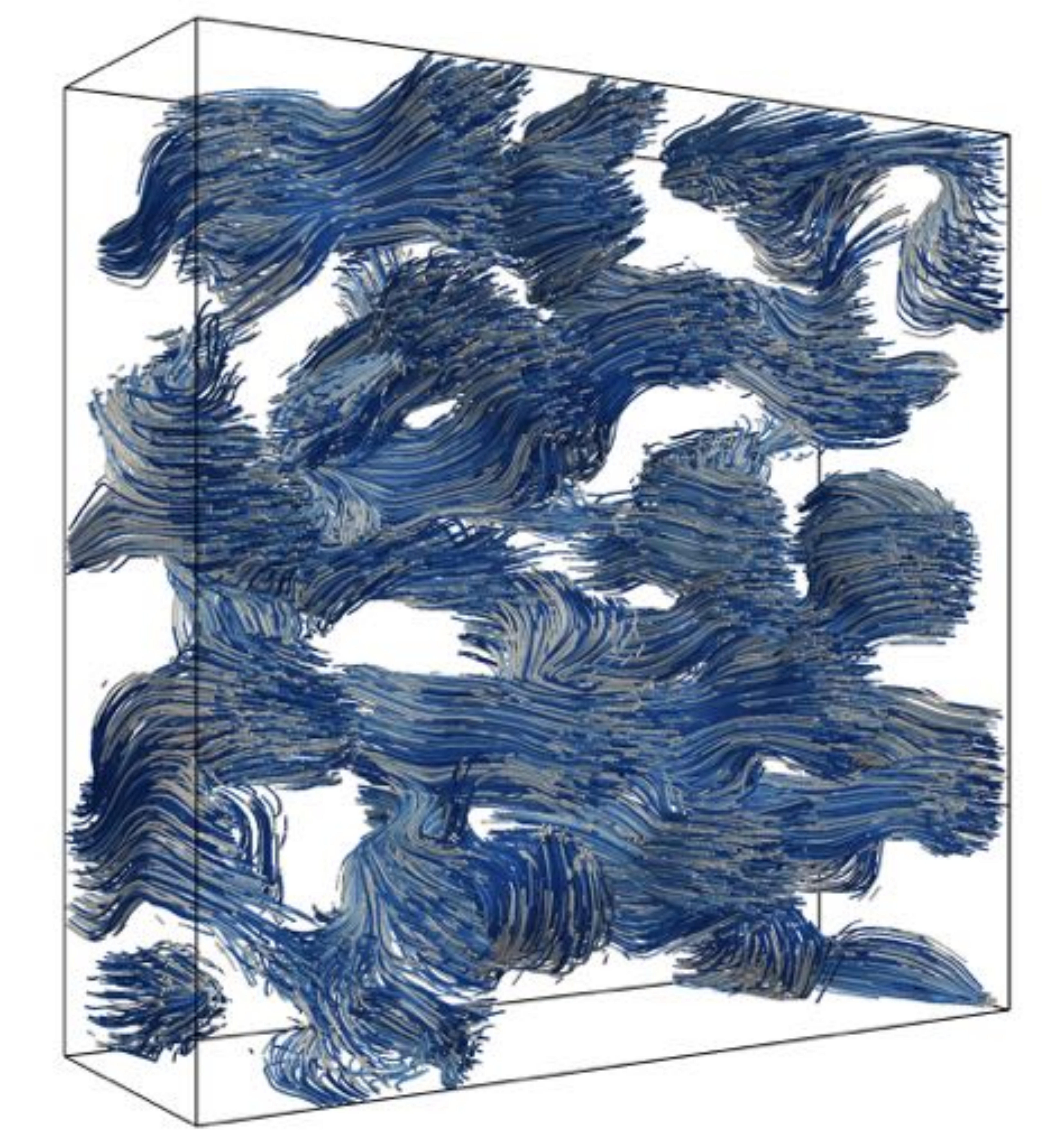}
	\caption{Streamlines following the diffusive flux calculated in the stochastic simulation shown in Figure~\ref{fig:modelsim}(c).}
	\label{fig:streamlines}
\end{figure}

The final aim is to use the developed model to analyze how mass transport---mainly diffusion---of drug through the EC/HPC coating depends on the microstructure. Because of this, we in this work use numerical calculations of diffusion to test the model fit. We compare the diffusion calculated in the filtered CLSM pore structure $\tildey$ with the diffusion calculated in stochastic simulations $\tildey_{\text{model}}$ from the fitted pore model \eqref{eq:poremodel}.

The diffusion calculations are done using the software Gesualdo \citep*{Geback2014}, which provides a diffusive flux vector $\mathbf{J}(i) = (J_x, J_y, J_z)(i)$ for each voxel $i$ in the pore space (see Figure \ref{fig:streamlines} for an illustration). The diffusive flux $J_x(i)$ gives the amount of particles transported in steady state in the $x$-direction, per unit square area and second, $J_y(i)$ the amount transported in the $y$-direction, and $J_z(i)$ the amount transported in the $z$-direction.
Since the $z$-direction is the direction of transport, the effective diffusion coefficient for the pore structure is defined as the average of $J_z$, taken over the whole structure including the matrix where $J_z$ is set to zero.

We compare both the effective diffusion coefficient and the diffusive flux. The comparison of the diffusive flux is done using 3D histograms of the diffusive flux vectors. Since a stochastic simulation $\tildey_{\text{model}}$ from the model is random, the corresponding flux histogram of the diffusive flux vectors $\{\mathbf{J}(i)\}_{i:\tilde{y}_{\text{model},i}=1}$ is a random histogram. To test if the random histograms for the simulated pore structures are similar to that from the CLSM pore structure, we test if the set where the model histogram values are usually high is included in the set where the CLSM histogram values are high, and if there are no high values in the CLSM histogram where the model histogram values are usually low. To do this formally we use the excursion set methodology by \citet*{BolinLindgren2015}. 

The set where the model histogram values are usually high is characterised by a positive excursion set. This is defined as a maximal set in which, with a given probability $1-\alpha$, all values of the model histogram are above a given value $u_J$:
\begin{align*}
E_{u_J,\alpha}^+ = 
\argmax_{D\subset\mathbb{R}^3} \{|D|: P(D \subseteq D_{u_J}^+) \geq 1-\alpha \}.
\end{align*}
Here $D_{u_J}^+ = \{b: H(b)>u_J\}$ is the random set of 3D bins where the model histogram $H$ exceeds the value $u_J$. 
Similarly, we define the negative excursion set as a maximal set where for most model histograms all values are below the given value $u_J$:
\begin{align*}
E_{u_J,\alpha}^- = 
\argmax_{D\subset\mathbb{R}^3} \{|D|: P(D \subseteq D_{u_J}^-) \geq 1-\alpha \},
\end{align*}
where $D_{u_J}^- = \{b: H(b)<u_J\}$. Set $E_{\text{CLSM},u_J}=\{b: H_{\text{CLSM}}(b)>u_J\}$, where $H_{\text{CLSM}}$ is the diffusive flux histogram of the CLSM pore structure. Selecting a high probability $1-\alpha$, we can then test if $E_{u_J,\alpha}^+ \subseteq E_{\text{CLSM},u_J} \subseteq (E_{u_J,\alpha}^-)^c$, where $E_{u_J,\alpha}^+$ and $E_{u_J,\alpha}^-$ are estimated from model simulations. If this inequality is satisfied, it implies that flux field through the observed CLSM pore structure is similar to flux fields obtained from model simulated pore structures.

\subsection*{Computation}

The MCMC algorithm was implemented in Matlab \citep*{Matlab2015b} and so was the estimation of the size distributions. The triangulation of the domain was done using the R package INLA \citep*{RINLA} and the 2D-covariance estimation was done using the R package RandomFields \citep*{RRF}. The excursion sets and simultaneous envelopes were estimated using the R package excursions \citep*{BolinarXiv}. The diffusive flux was calculated using the software Gesualdo \citep*{Geback2014}.

\section*{Results}
\label{sec:results}

\begin{figure}[t]
		\centering
		{\includegraphics[width=5.4cm]{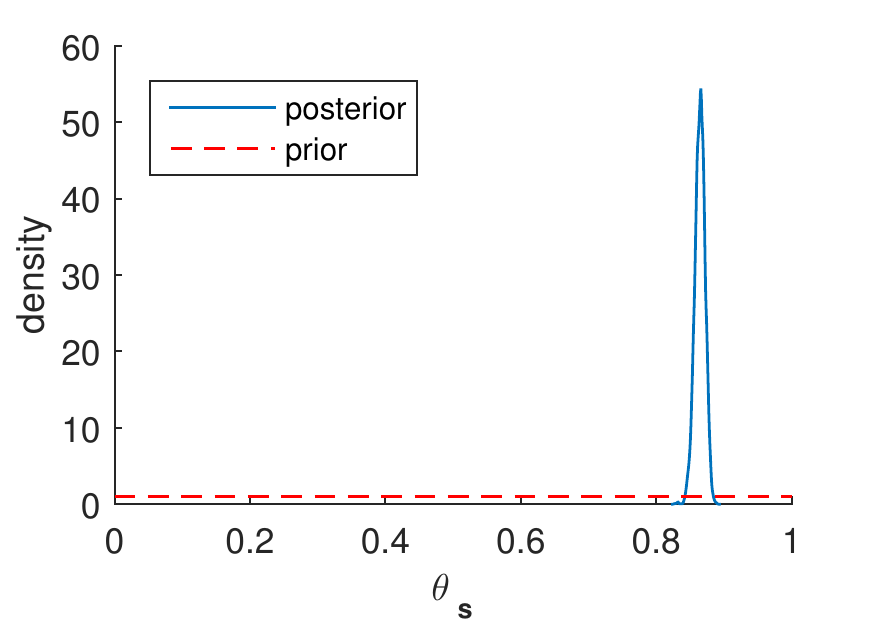}}
		{\includegraphics[width=5.4cm]{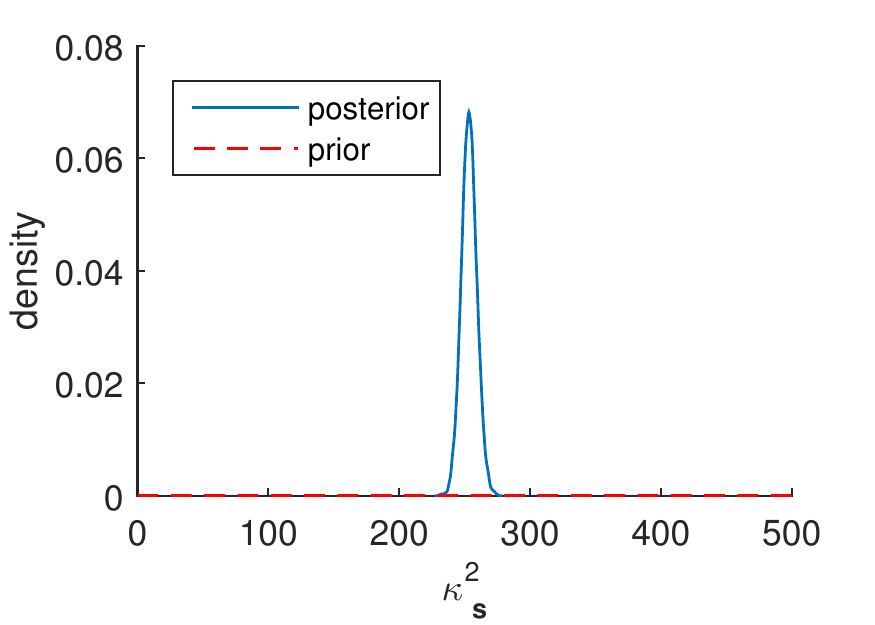}}
		{\includegraphics[width=5.4cm]{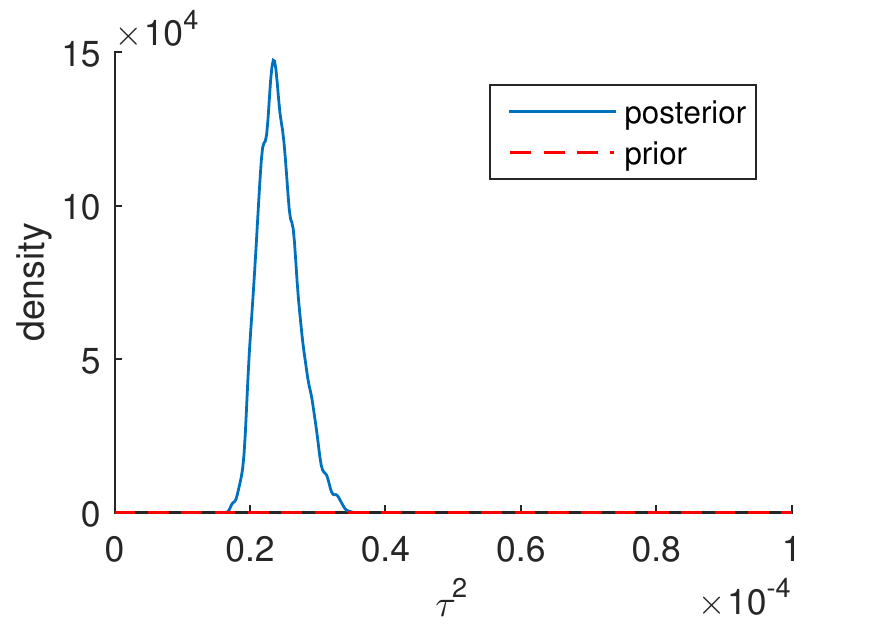}}
		\\
		{\includegraphics[width=5.4cm]{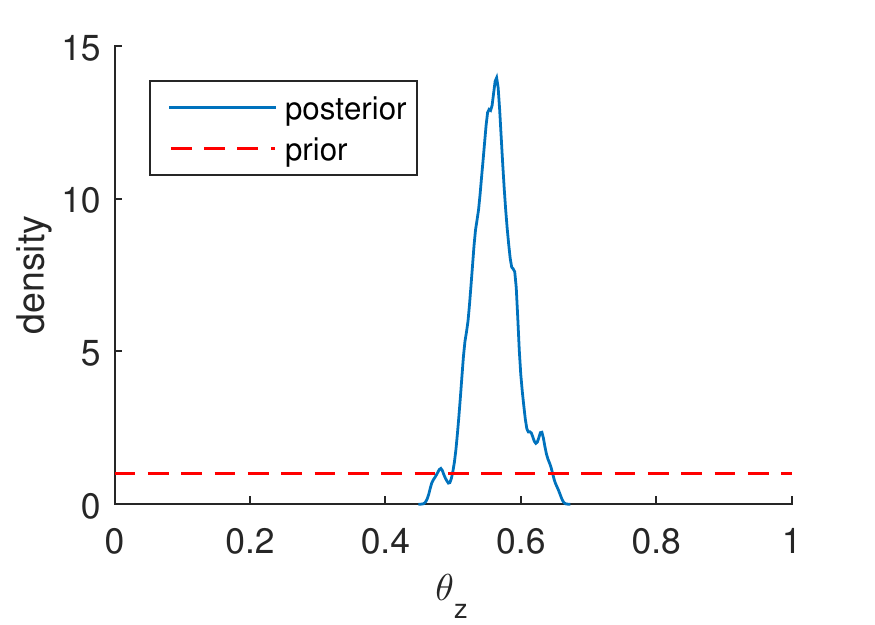}}
		{\includegraphics[width=5.4cm]{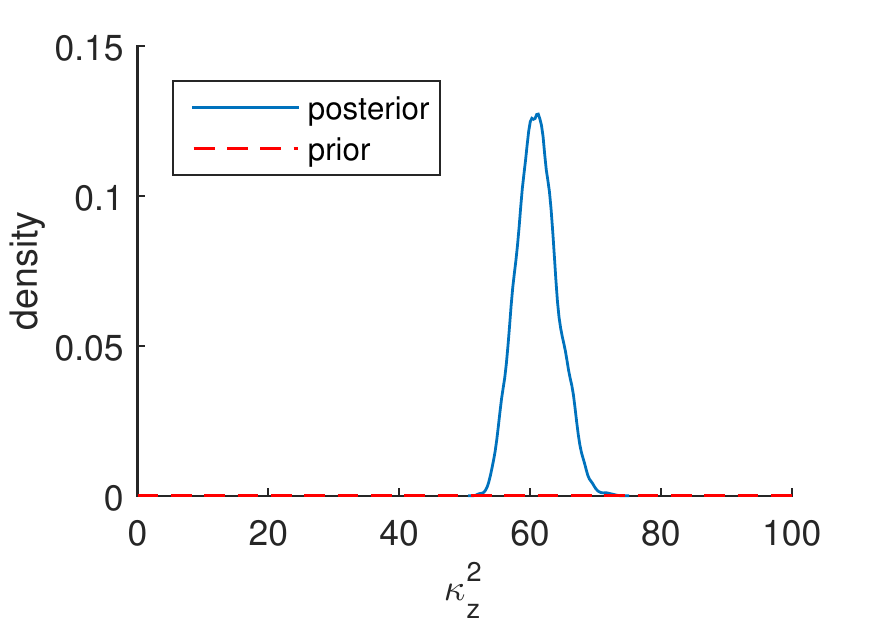}}
		{\includegraphics[width=5.4cm]{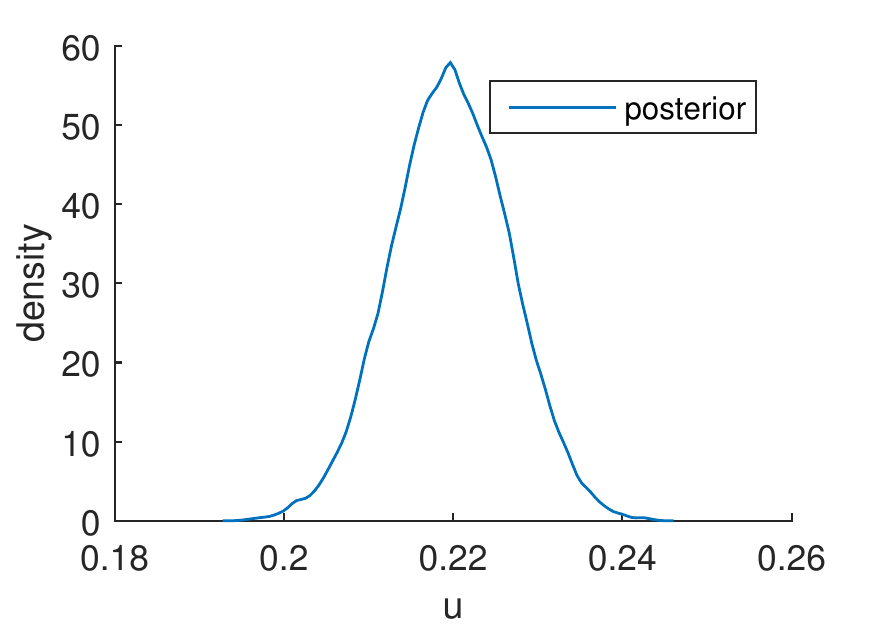}}
		\caption{Estimates of the posterior densities $\pi(\cdot | \boldy)$ (solid blue lines) for the GMRF parameters and the threshold $u$. The dashed red lines are the prior densities $\pi(\cdot)$ for the GMRF parameters. Since the prior for the threshold is improper it is not plotted.}
		\label{fig:MCMCdens}
	\end{figure}
	
	\begin{figure}[t]
		\centering
		\subcaptionbox{}
		[0.24\textwidth]
		{\includegraphics[width=3.5cm]{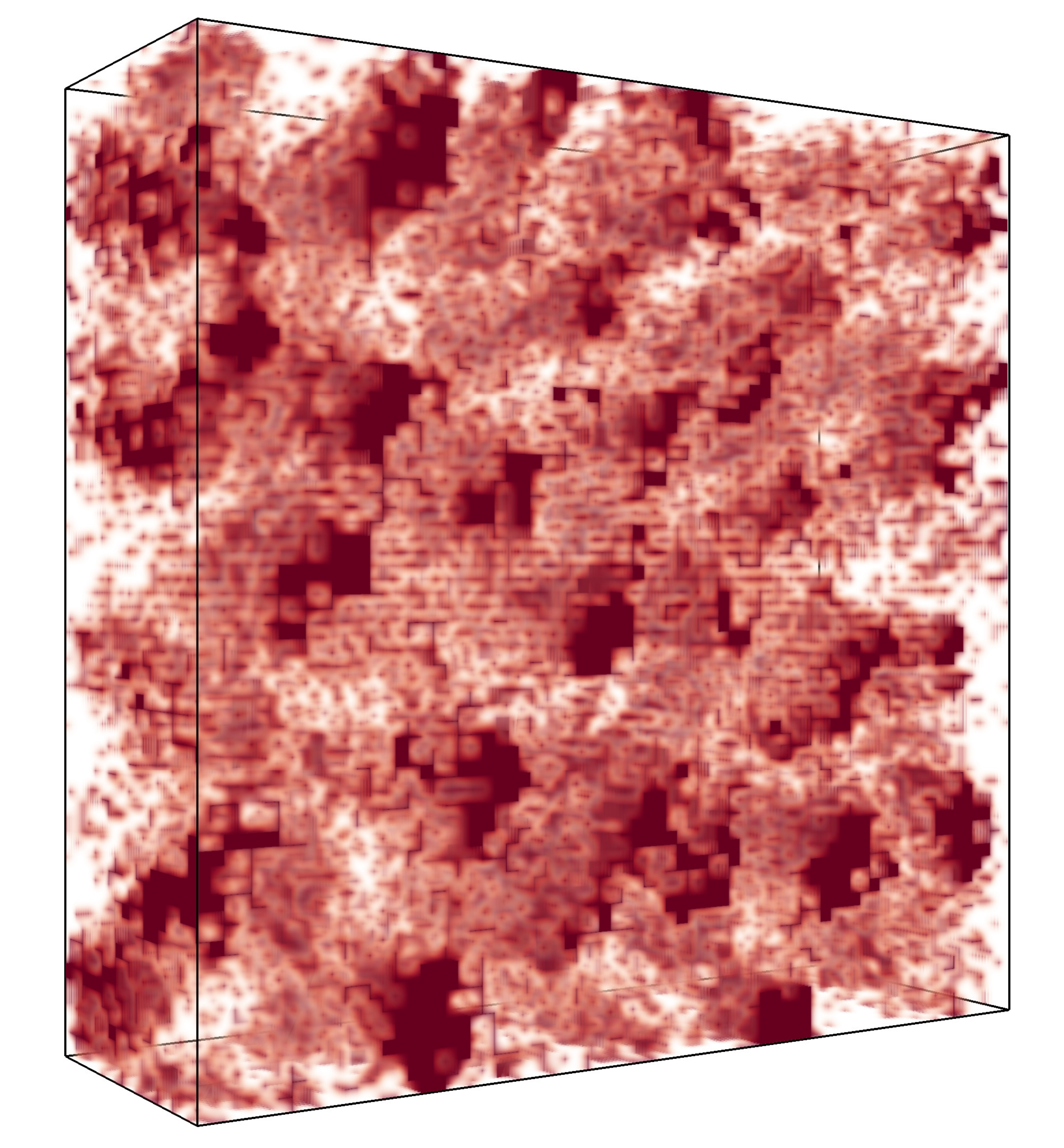}}
		\subcaptionbox{}
		[0.24\textwidth]
		{\includegraphics[width=3.5cm]{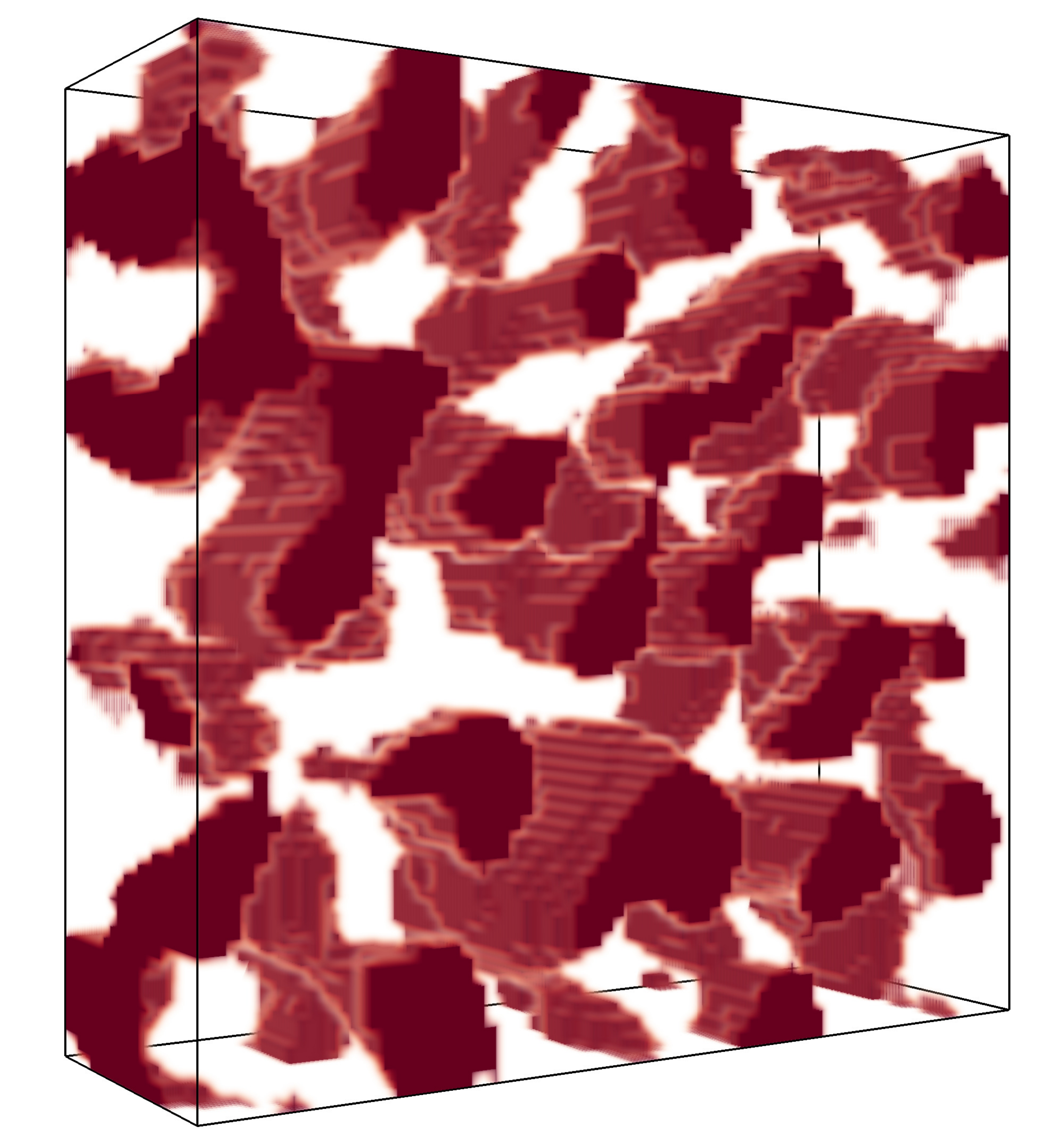}}
		\subcaptionbox{}
		[0.24\textwidth]
		{\includegraphics[width=3.5cm]{figs/comp_noise_rightp1_sim_darkred_highres-eps-converted-to.pdf}}
		\subcaptionbox{}
		[0.24\textwidth]
		{\includegraphics[width=3.5cm]{figs/comp_pores_rightp1_sim_darkred_highres-eps-converted-to.pdf}}
		\caption{(a) The binarized CLSM image, $\boldy$, (b) the filtered CLSM pore structure, $\tildey$, (c) a stochastic simulation from the fitted noise model \eqref{eq:noisemodel}, $\boldy_{\text{model}}$, and (d) the corresponding stochastic simulation from the fitted pore model \eqref{eq:poremodel}, $\tildey_{\text{model}}$. The binarized CLSM image is also shown in Figure \ref{fig:CLSM}(b), and the stochastic simulations are the same as those shown in Figure \ref{fig:modelsim}.}
		\label{fig:simmicrcomp}
	\end{figure}

In this section, detailed results for the model fitted to the CLSM image $HPC30_1$ are presented. The results for the other CLSM image samples are similar, see Appendix C. The MCMC algorithm was run for $2 \cdot 10^5$ iterations, making sure that the chains reach stationarity. Each iteration took around $1$--$2$ seconds. About $85$--$90\%$ of each iteration was spent sampling the posterior distribution of the GMRF. 

Regarding the prior densities for the parameters, we used uniform priors for the oscillation parameters, $\theta_{\mv{s}},\theta_z \sim U(0,1)$,  an improper uniform prior on the entire $\mathbb{R}$ for the threshold $u$, and gamma priors for the range  and variance parameters, $\kappa_{\mv{s}}^2,\kappa_z^2 \sim \Gamma(1,6\cdot10^{-5}), \tau^2~\sim~\Gamma(1,5\cdot10^{-3})$. Figure \ref{fig:MCMCdens} shows these densities as well as the estimated posterior densities. Because of the vast amount of data, the posterior distributions of the parameters are insensitive to the priors and have quite small variances. The posterior distribution for the oscillation parameter $\theta_z$ has a higher variance than the oscillation parameter $\theta_{\mv s}$, which is likely caused by the fact that there are fewer voxels and hence less information in the $z$-direction.

Figure \ref{fig:simmicrcomp} shows stochastic simulations from the models \eqref{eq:noisemodel} and \eqref{eq:poremodel} with parameters chosen as the posterior expectation $E(\mv{\gamma} | \boldy)$. Comparing these to the binarized CLSM image and the filtered CLSM pore structure, which are also shown in the figure, it seems like the model simulations are similar to the data.

\begin{figure}[t]
	\centering
	\subcaptionbox{}
	{\includegraphics[width=0.43\linewidth]{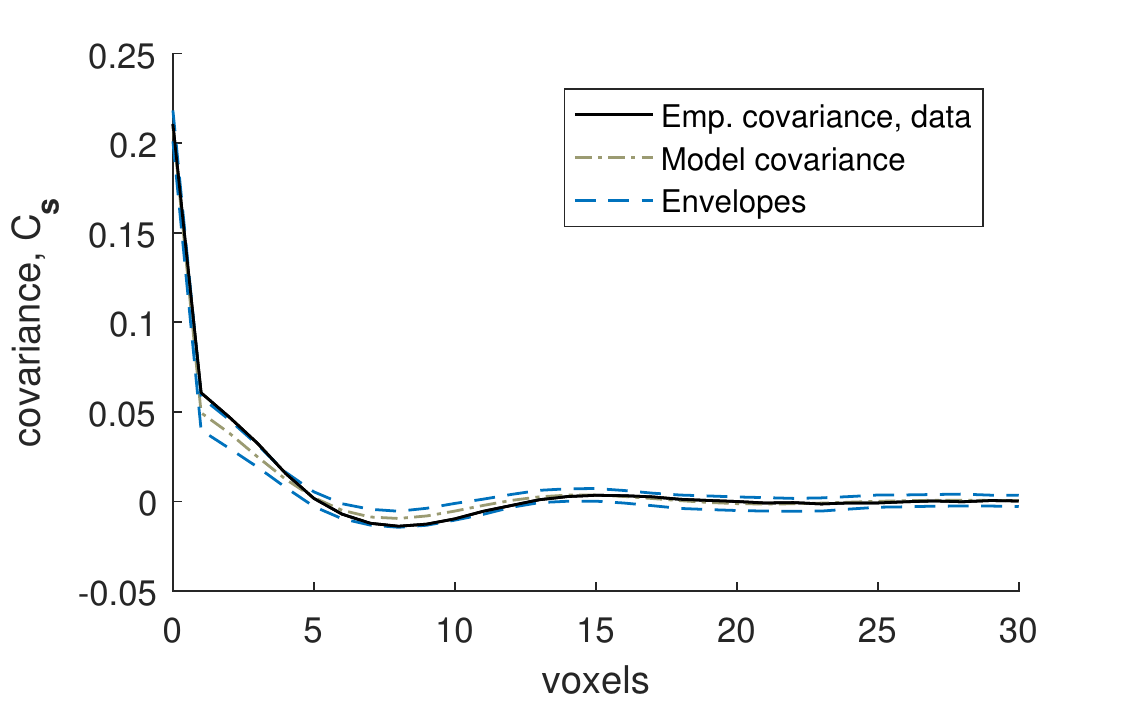}}
	\subcaptionbox{}
	{\includegraphics[width=0.43\linewidth]{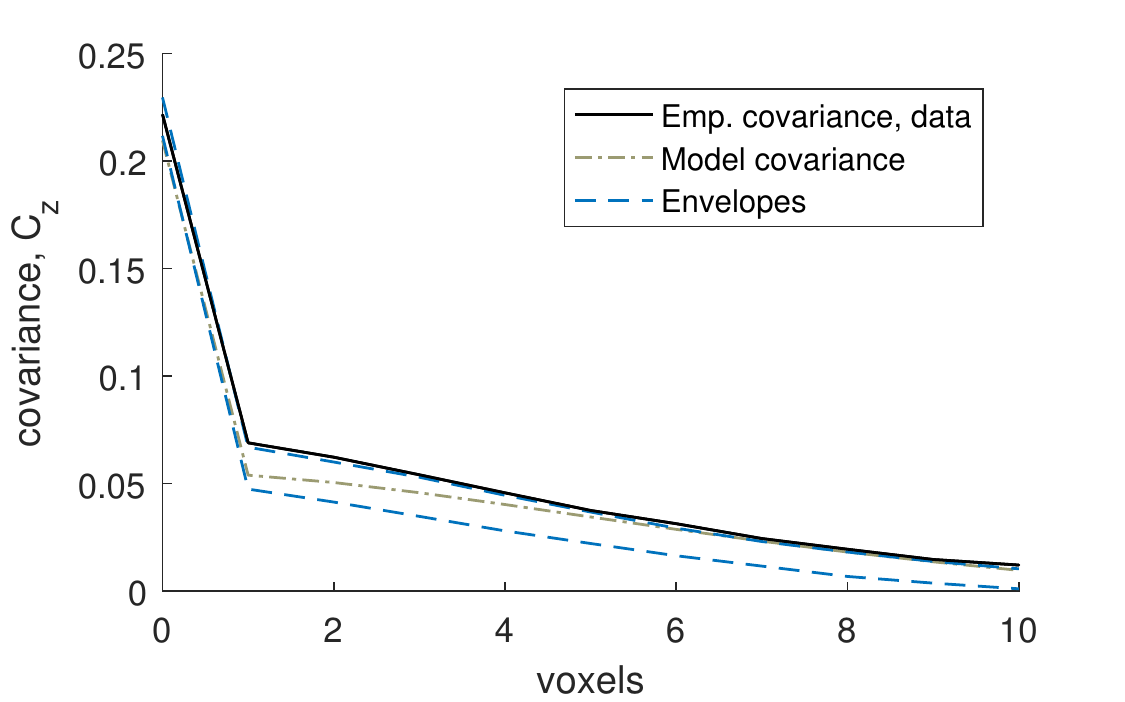}}
	\subcaptionbox{}
	{\includegraphics[width=0.43\linewidth]{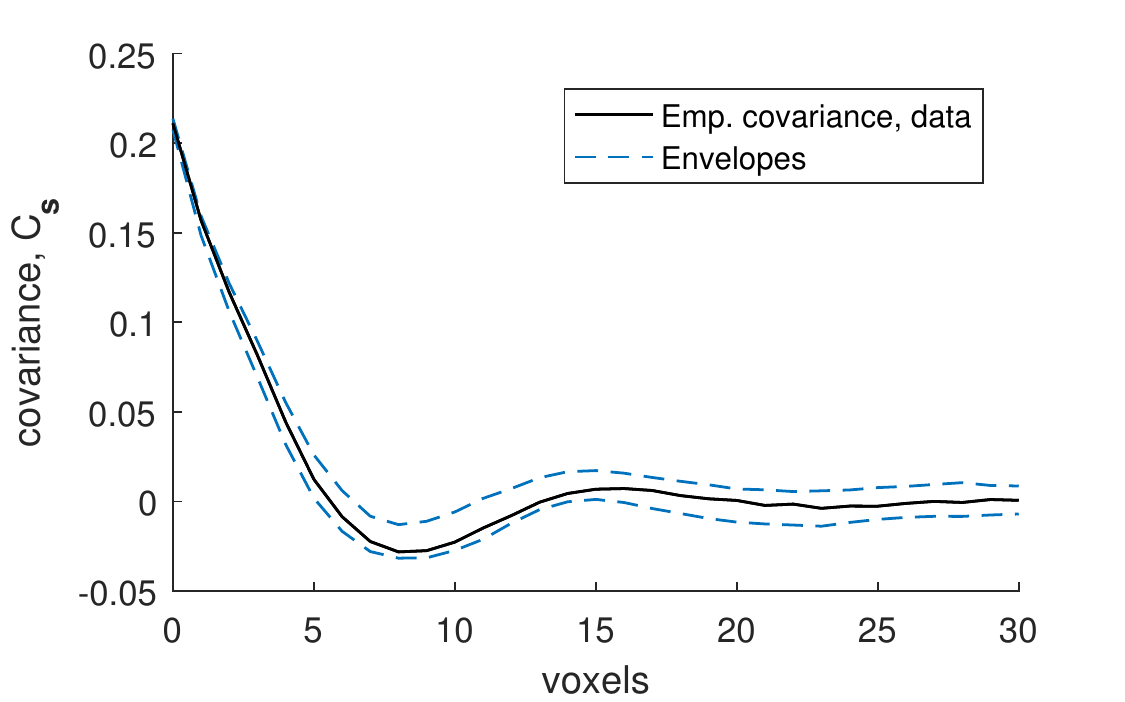}}
	\subcaptionbox{}
	{\includegraphics[width=0.43\linewidth]{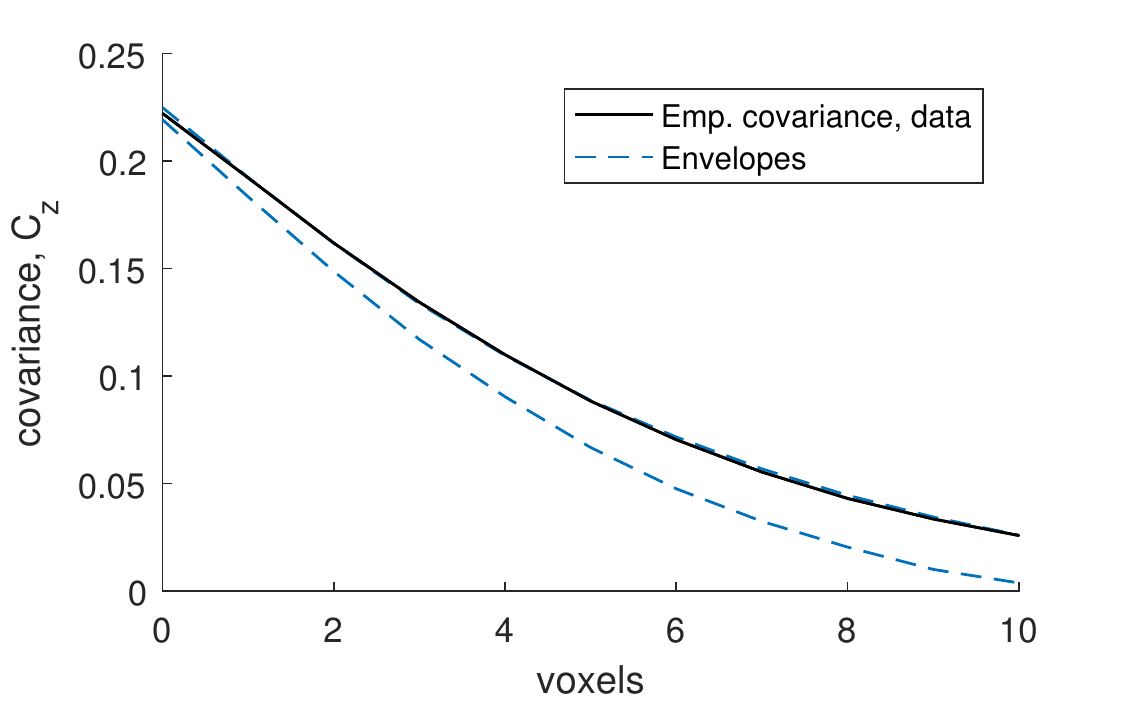}}
	\caption{Covariance functions for the noise model \eqref{eq:noisemodel} (top row), from which a stochastic simulation is shown in Figure \ref{fig:simmicrcomp}(c), and the pore model \eqref{eq:poremodel} (bottom row), from which a stochastic simulation is shown in Figure \ref{fig:simmicrcomp}(d). The marginal empirical covariance function estimated from the binarized CLSM image $\mv{y}$ and the CLSM pore structure $\tildey$ respectively (solid black), and a $95\%$ simultaneous envelope estimated from $500$ stochastic simulations from the corresponding model (dashed blue), plotted for (a), (c) the $\mv{s}$-plane, i.e.\ $C_{\mv{s}}$, and (b), (d) the $z$-line, i.e.\ $C_z$. For model \eqref{eq:noisemodel}, the model covariance functions using the point-estimates of the parameters (dash-dotted yellow) are also shown.}
	\label{fig:covenv}
\end{figure}

Figures \ref{fig:covenv}(a) and \ref{fig:covenv}(b) show that for the fitted noise model \eqref{eq:noisemodel}, the empirical covariances for the binarized CLSM image goes above the model envelopes at small distances. The model marginal covariance functions were computed using the posterior mean of the parameters. The variances in the model and in the empirical variance of the CLSM image are almost the same, and hence this is caused by the variance of the GMRF in the fitted model \eqref{eq:noisemodel} being too small compared to the variance of the noise. In other words, the signal to noise-ratio is underestimated in the fitted model. However, Figures \ref{fig:covenv}(c) and \ref{fig:covenv}(d) show that the empirical covariances of the CLSM pore structure falls within the model envelopes (although with a small margin for the $z$-line) for the fitted pore model \eqref{eq:poremodel}. Hence the underestimation of the signal to noise-ratio does not have a big effect on the pore model.

\begin{figure}[t]
	\centering
	\subcaptionbox{}
	[0.24\textwidth]
	{\includegraphics[width=4.5cm]{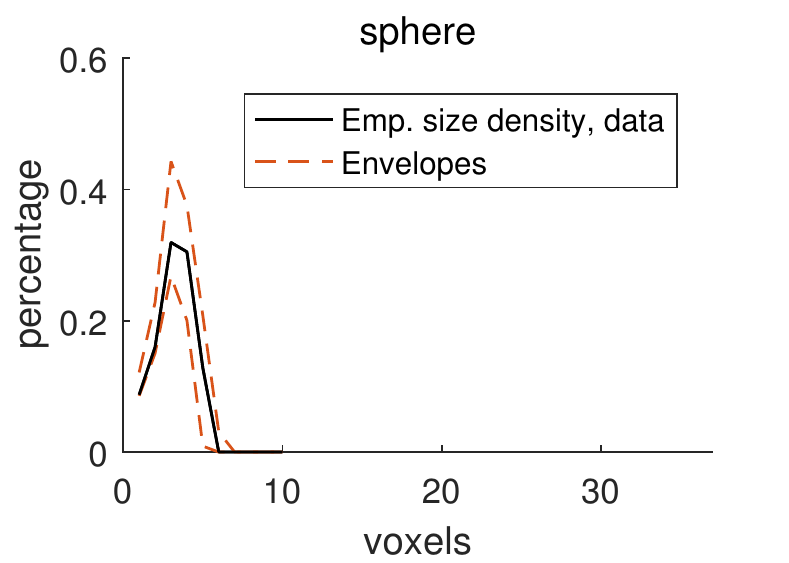}}
	\subcaptionbox{}
	[0.24\textwidth]
	{\includegraphics[width=4.5cm]{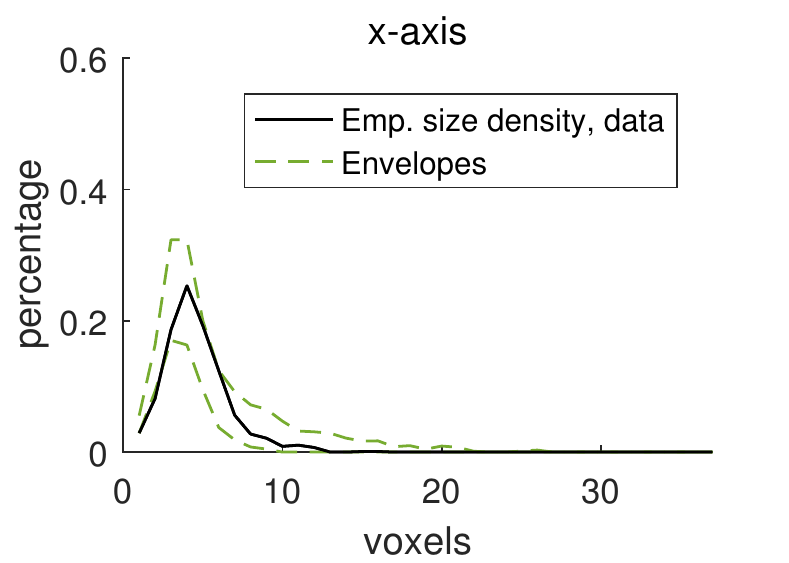}}
	\subcaptionbox{}
	[0.24\textwidth]
	{\includegraphics[width=4.5cm]{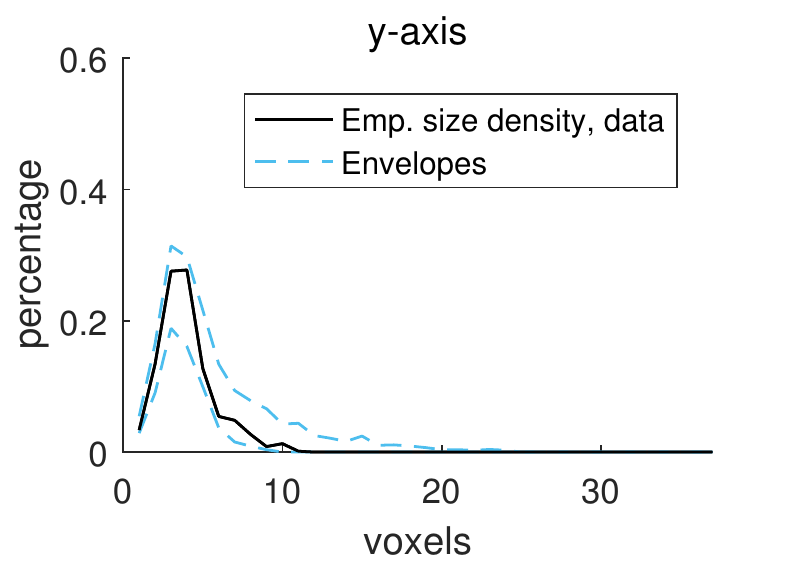}}
	\subcaptionbox{}
	[0.24\textwidth]
	{\includegraphics[width=4.5cm]{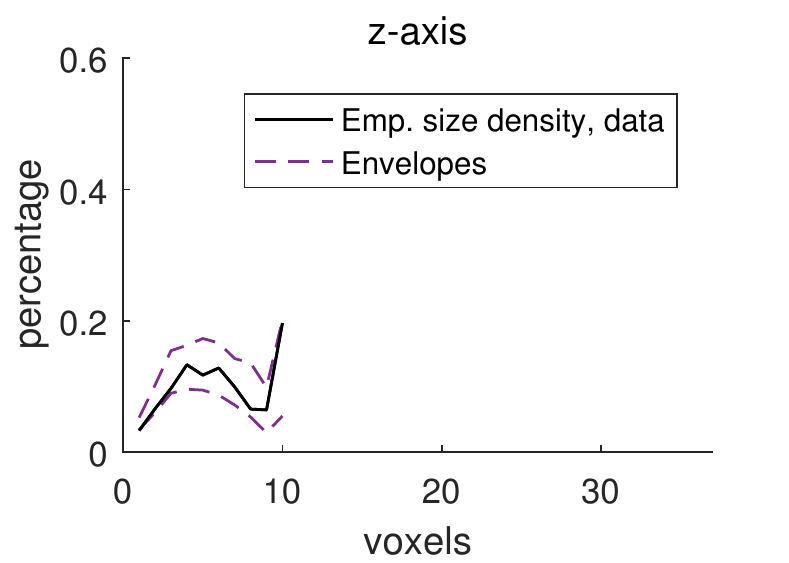}}
	\subcaptionbox{}
	[0.24\textwidth]
	{\includegraphics[width=4.5cm]{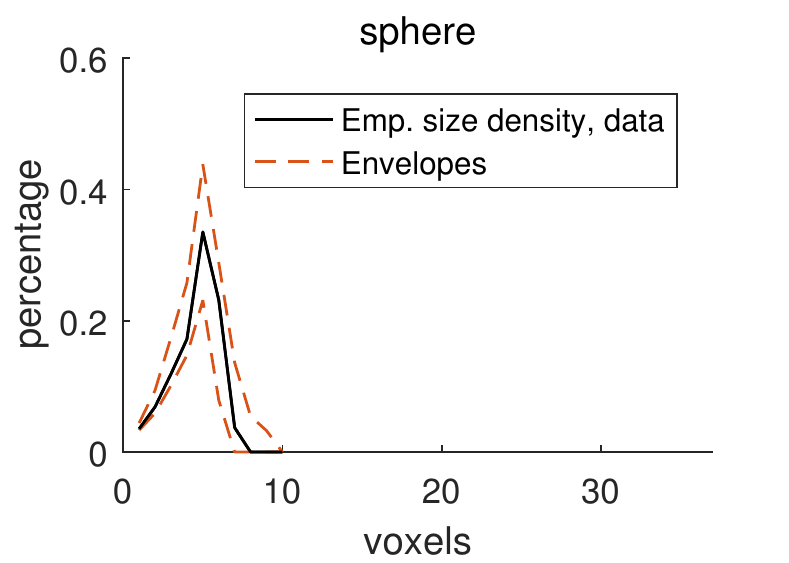}}
	\subcaptionbox{}
	[0.24\textwidth]
	{\includegraphics[width=4.5cm]{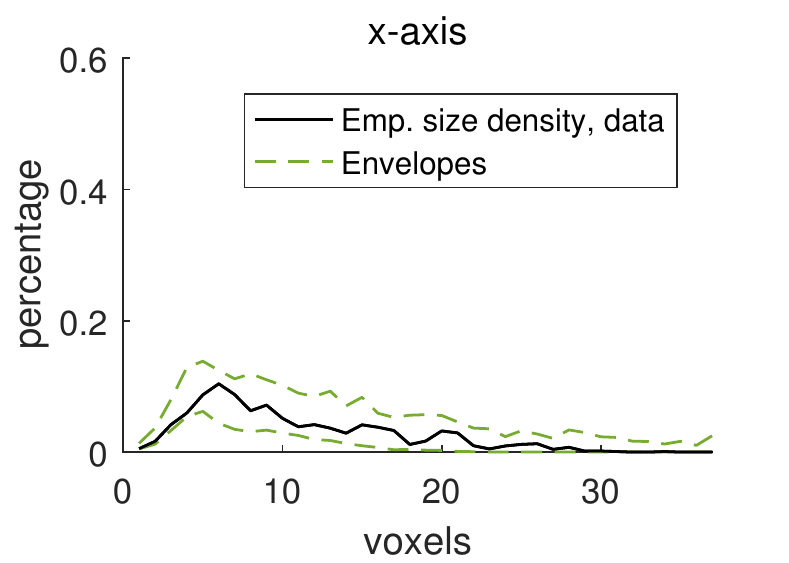}}
	\subcaptionbox{}
	[0.24\textwidth]
	{\includegraphics[width=4.5cm]{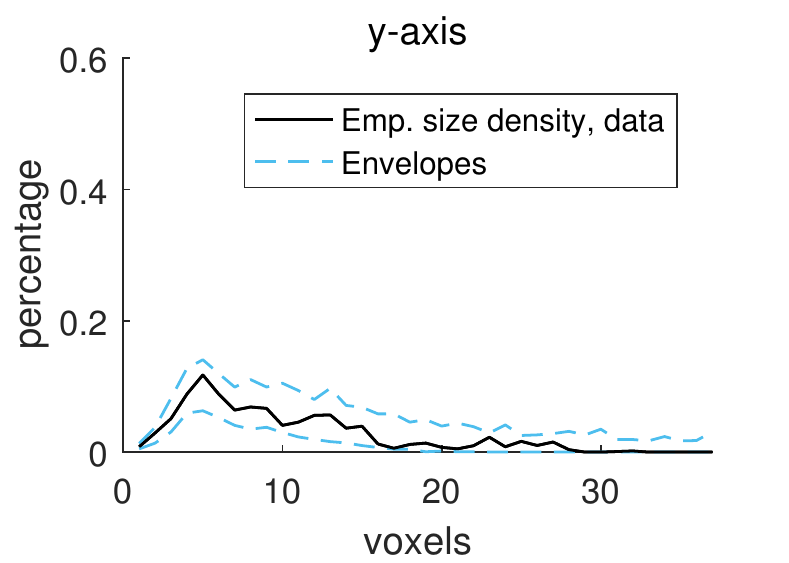}}
	\subcaptionbox{}
	[0.24\textwidth]
	{\includegraphics[width=4.5cm]{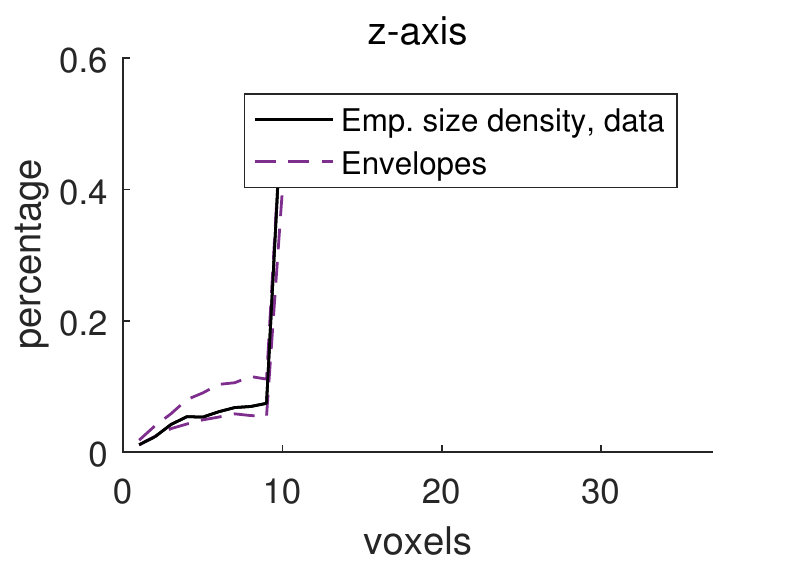}}
	\caption{Size densities estimated on the pore space (top row) and the pore matrix (bottom row). The estimate from the CLSM pore structure $\tildey$ (solid black) and a $95\%$ simultaneous envelope estimated from $500$ stochastic simulations from the pore model \eqref{eq:poremodel} (dashed colored), shown for the following structuring elements:  the unit sphere (a and e), and lines aligned with the $x$-axis (b and f), the $y$-axis (c and g), and the $z$-axis (d and h). The line structuring elements have lengths two.}
	\label{fig:sizedistenv}
\end{figure}

The empirical size densities in the pore space and in the matrix, shown in  Figure \ref{fig:sizedistenv}, tend to lie within the model envelopes, showing that the geometry of the CLSM pore structure and the stochastic simulations from model \eqref{eq:poremodel} are similar. Comparing the size densities with respect to the different structuring elements, 
there is room for longer lines along the $z$-axis than along the other two axes, and the size densities with respect to the lines are naturally larger than those for the sphere. Comparing the size densities of the pore space and the matrix, it is clear that we can fit larger structuring elements in the matrix than in the pore space. This is not surprising since we have a smaller volume fraction of pores than of matrix. The high values of the size density for the largest voxel in Panels (d) and (h) are caused by boundary effects. Since the size of a sample (the CLSM pore structure or stochastic simulations) is $74\times 74 \times 20$, local sizes with respect to the line aligned with the $z$-axis cannot be larger than $10$. But if we think of the sample as being part of a larger pore structure, we cannot tell if the local sizes that are found to be $10$ in fact are larger. For further discussion of this, see Appendix B.

\begin{figure}[t]
\centering
	\subcaptionbox{}
	{\includegraphics[width=0.32\linewidth]{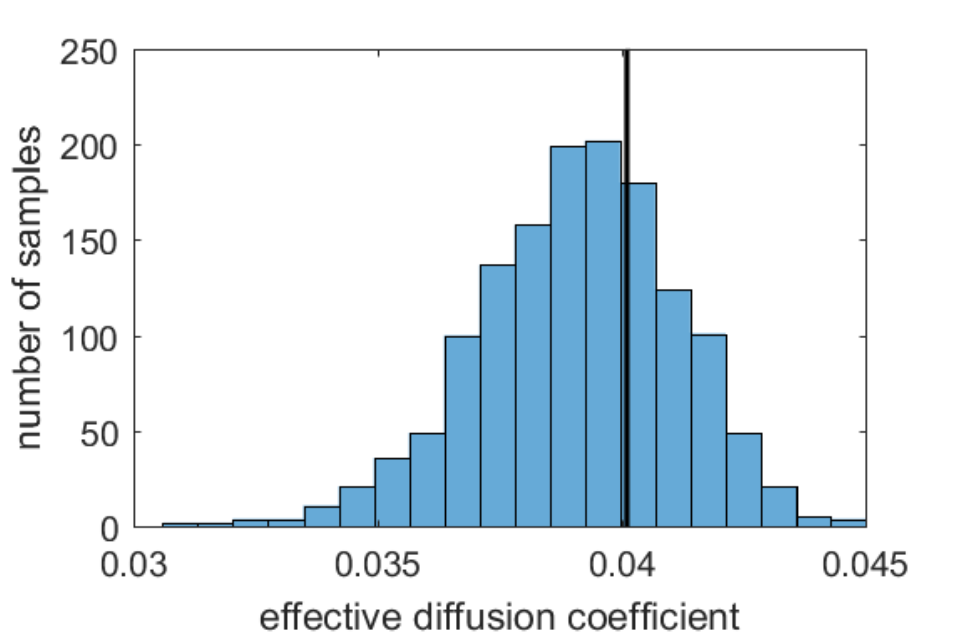}}
	\subcaptionbox{}
	{\includegraphics[width=0.21\linewidth]{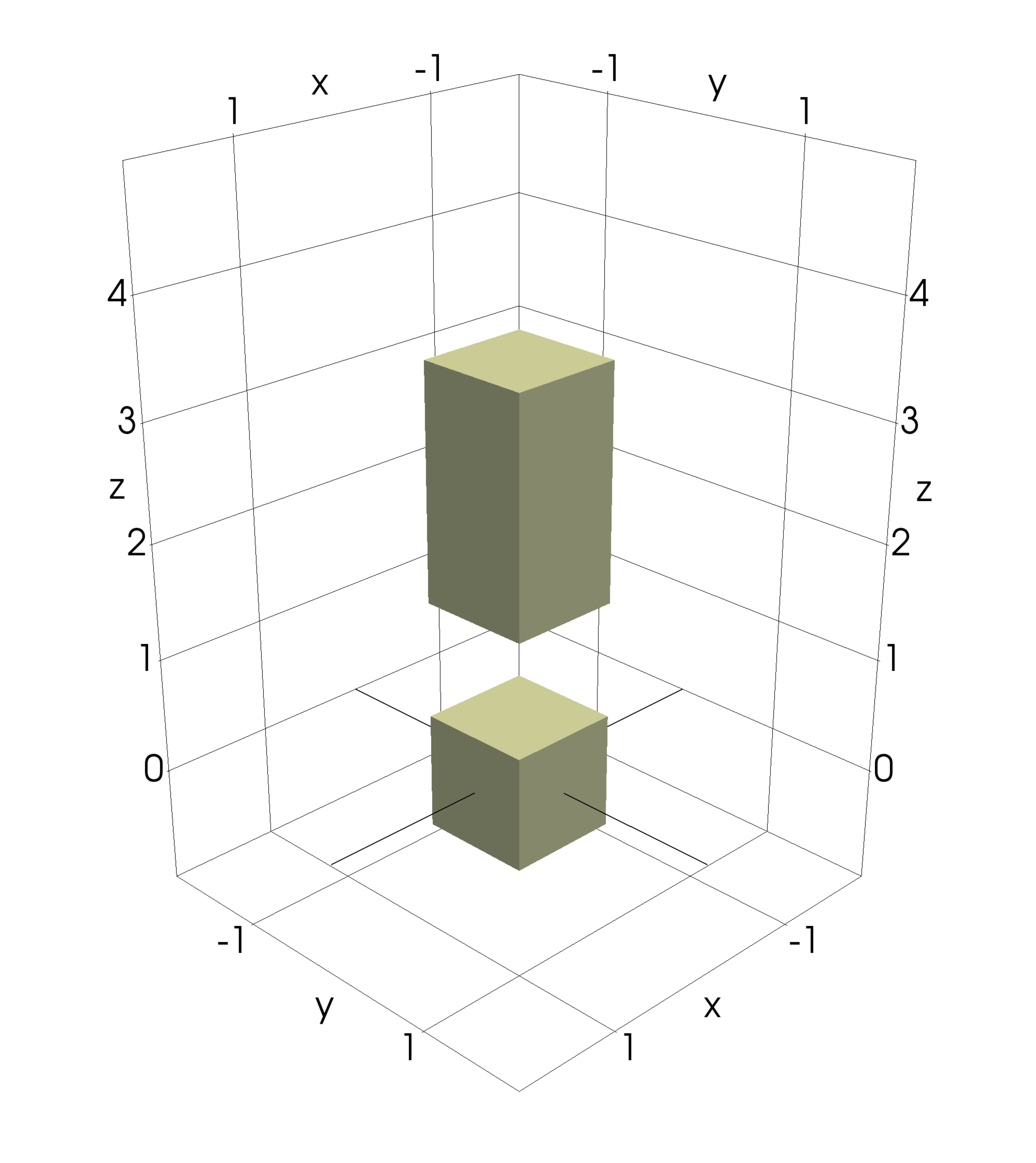}}
	\subcaptionbox{}
	{\includegraphics[width=0.21\linewidth]{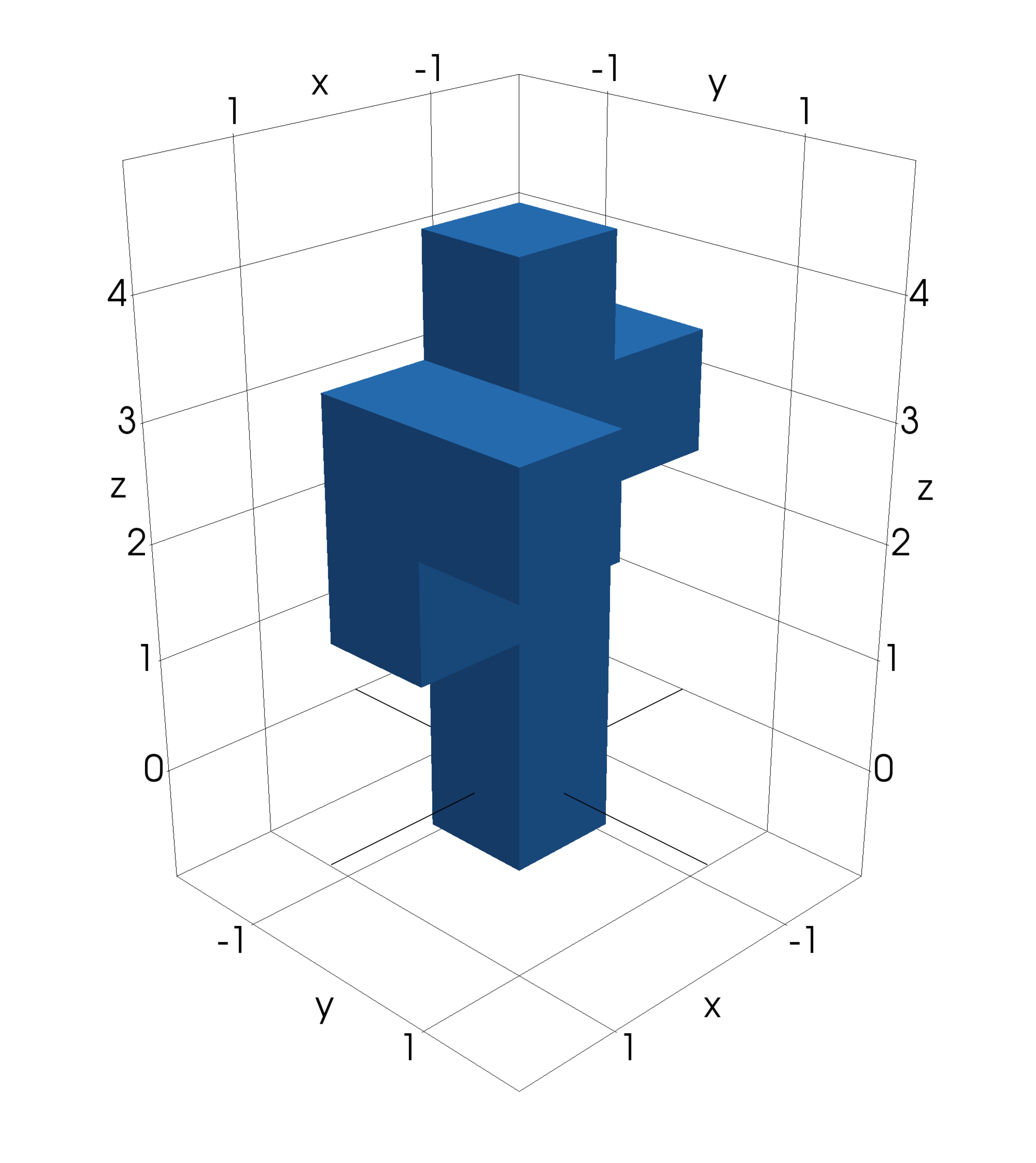}}
	\subcaptionbox{}
	{\includegraphics[width=0.21\linewidth]{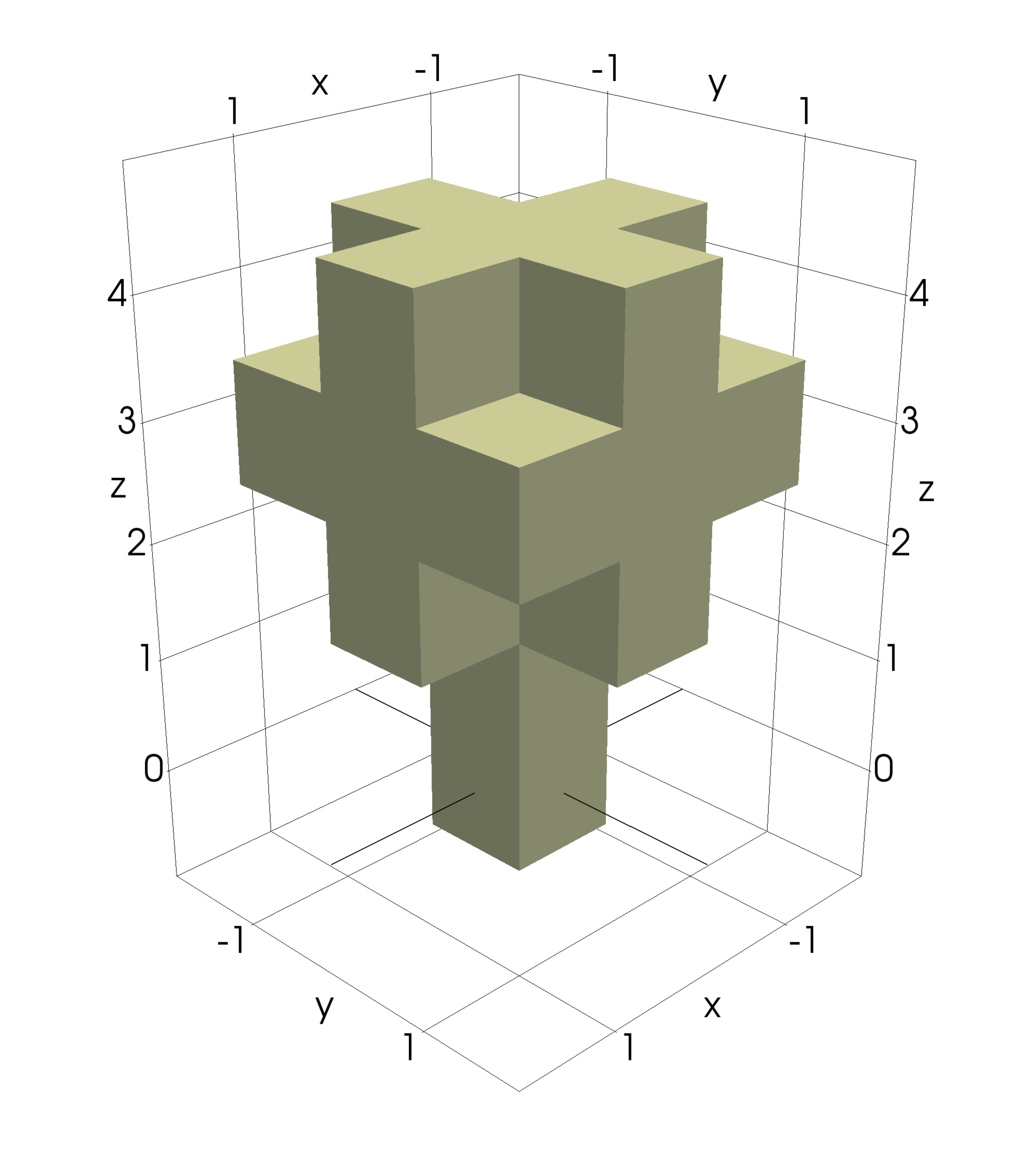}}
\caption{
Results based on $1400$ stochastic simulations from model \eqref{eq:poremodel}. Panel (a) shows a histogram of the  effective diffusion coefficient of each stochastic simulation and the effective diffusion coefficient of the CLSM pore structure ($0.0401$) as a vertical line. Also shown are sets corresponding to the 3D histograms of the diffusive flux of the simulations and the CLSM pore structure, where the histograms were standardized so that each 3D bin is of size $1\times 1 \times 1$, and the sums of the values over all bins are one:  (b) the estimated excursion set $E_{0.01,0.01}^+$,  (c) the set where the CLSM histogram takes values above $1\%$, $E_{\text{CLSM},0.01}$, and (d) the complement of the estimated excursion set $E_{0.01,0.01}^-$. The sets satisfy $E_{0.01,0.01}^+ \subset E_{\text{CLSM},0.01} \subset (E_{0.01,0.01}^-)^c$,which indicates that the diffusive flux in the CLSM pore structure is similar to the diffusive flux in the simulated pore structures.} 
	\label{fig:diffresults}
\end{figure}

Figure \ref{fig:diffresults} shows that the effective diffusion coefficient of the diffusion computed using Gesualdo in the CLSM pore structure lies well within the range of the effective diffusion coefficients computed in $1400$ stochastic simulations from model \eqref{eq:poremodel}. The figure also shows the estimated negative and positive excursion sets $E_{0.01,0.01}^-$ and $E_{0.01,0.01}^+$, which were estimated from the model diffusive flux histograms corresponding to the same $1400$ stochastic pore structures. In the figure one can see that the inclusions $E_{0.01,0.01}^+ \subset E_{\text{CLSM},0.01} \subset (E_{0.01,0.01}^-)^c$ hold, which means that the set where the CLSM histogram takes values above $1\%$ includes the set where for most model histograms all values are above $1\%$.  Also, no bins with values above $1\%$ in the CLSM histogram are included in the set where for most model histograms all values are below $1\%$.  This indicates that the diffusive flux computed in the CLSM pore structure is similar to the diffusive flux computed in the stochastic pore structures. 
One can also note that the estimated excursion sets for the model are symmetric in $x$ and $y$, which is explained by the fact that the model is isotropic in the $\mv s$-plane.

\section*{Discussion}\label{sec:discussion}

We have formulated a parametric stochastic model for the pore structure of films of EC/HPC polymer blends, with the aim to analyze the mass transport through the films. The model is based on a separable oscillating Gaussian Mat\'ern field, which is stationary and anisotropic. The field is obtained as the solution to an SPDE, which is solved using the finite element method giving an approximation of the Gaussian field by a GMRF. The model was fitted to CLSM images using an MCMC algorithm which takes advantage of the sparsity of the GMRF approximation and uses the Kronecker structure of its precision matrix in a, for this type of problem, new way to reduce the complexity of the algorithm. The computational efficiency of this model fitting algorithm allowed fitting the model to CLSM images of realistic sizes. We characterized the stochastic simulations from the model using the covariance and pore size distributions, and developed a measure using excursion sets to characterize the diffusive flux computed in the stochastic simulations. Using these measures we concluded that the model fits the CLSM images well.

A strength of the GMRF approximation is that it is not constrained to the triangulation that was used to estimate the parameters in the GMRF. Instead a finer triangulation can be used for the stochastic simulation from the model. As long as the same parameters are used, the model will be a linear approximation of the same continuously indexed Gaussian field. This is useful, since it is more computationally demanding to fit the model than to simulate from the model. 

We use a simple filter to obtain the pore structure of model \eqref{eq:poremodel}. A better alternative could be to use the posterior mean of the pore structure given the data, obtained from the MCMC algorithm, or to use a filter that e.g.\ takes the point spread function into account. However, the filter has no effect on the model fit since the model is fitted to model \eqref{eq:noisemodel}. Also, since the binarized CLSM data and stochastic simulations from model \eqref{eq:noisemodel} are filtered in the same way, the filter should not have a large effect on the model validation either.

An important aspect to consider is what size of the observation window is sufficient to get good parameter estimates, i.e.\ how big the so called representative volume element is \citep*[pp. 236--237]{Chiu2013ch7}. In our case, the biggest concern is whether we have enough data in the $z$-direction, since we use rather thin samples to ensure stationarity.  
The range of the correlation in the $z$-direction is large compared to the size of the samples---as can be seen by comparing the correlation for the fitted model in the $z$-direction, with the size of the CLSM image in the $z$-direction, which is $20$ voxels. To see if the parameter estimates are sensitive to the size of the CLSM image sample in the $z$-direction, we fitted the model to the combination of two in the $z$-direction consecutive samples, both of size $20$ voxels in the $z$-direction, where the two samples have different pore sizes. With this larger combined sample, the point-estimates of all the parameters---and particularly the range parameters $\tau_{\mv s}$ and $\tau_{z}$ which control the size of the pores in the $\mv{s}$-plane and $z$-direction respectively---turned out to lie between the point-estimates for the model fitted to the two samples individually. Thus the estimate of the size of the pores in the combined sample lies somewhere between the estimates of the individual samples, which indicates that the observation windows in the individual samples are large enough for us to obtain valid estimates of the range parameters.

Despite the different techniques that were used to decrease the computational effort, the model fitting method is still computationally demanding. Other less demanding methods, such as the minimum contrast method, where the parameters are chosen to minimize the distance between a summarizing function and the estimate of the function computed from the data, and the pairwise likelihood have previously been used for this type of problems, see \citet*{NottWilson2001}. An advantage with the likelihood-based MCMC algorithm is that it uses more of the information in the data than the less computationally demanding methods. It would be interesting to perform a comparison of the parameter estimates from the MCMC algorithm with estimates from other model fitting methods, similar to what was done in \citet*{Siden2017}.

Following \citet*{Lindgren2011}, the model can be made non-stationary by letting the parameters of \eqref{eq:oscSPDE} be slowly varying functions, which is useful for capturing non-stationarities in the EC/HPC films. The model can also be extended using nested SPDEs \citep*{Bolin2011} to capture other types of anisotropies in the data. These extensions are interesting for future work where the model will be used to investigate how the mass transport depends on the microstructure of the material.

\section*{Acknowledgements}

This work is funded by the Swedish Foundation for Strategic Research (SSF grant AM13-0066), the Knut and Alice Wallenberg foundation (KAW grant 20012.0067), and the Swedish Research Council (grant 2016-04187). We are grateful to Holger Rootz\'en for modelling ideas and valuable suggestions. We would also like to thank Henrike H{\"a}bel, Mariagrazia Marucci and Christine Boissier for supplying the data, and Jonas Wallin for sharing code that was used in the estimation method. Finally, we thank Aila S{\"a}rkk{\"a}, Christian von Corswant, and everyone involved in the SSF project for valuable discussions. 

\bibliography{3Dmodelref}

\begin{appendix}
\section{Details of the model and the model fitting algorithm}
\label{supp:supplementarydetails}
\subsection*{The triangulation used in the GMRF approximation}
\label{app:boundary}
\begin{figure}[t]
	\centering
	\includegraphics[width=7cm]{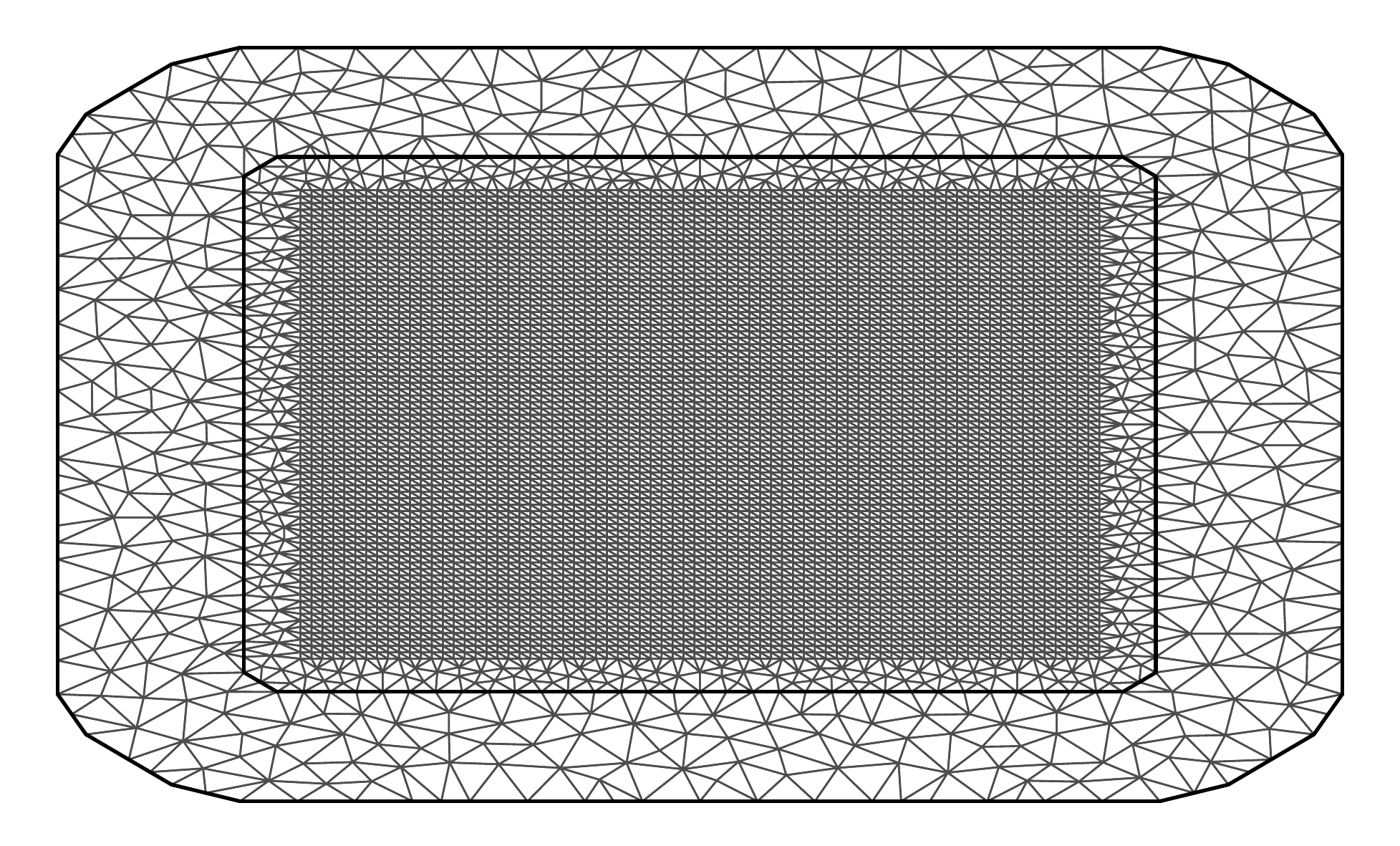}
	\caption{Triangulation of the domain $\Omega_{\mv s}$, for the model fitted to dataset $HPC30_1$.}
	\label{fig:triangulation}
\end{figure}

The triangulations of the domains $\Omega_{\mv s}$ and $\Omega_{z}$ have a node in each voxel of the CLSM image. The domains are larger than the domain of the CLSM voxels, to reduce the effect of the boundary conditions imposed on the SPDE \citep*[see][]{Lindgren2011}. 
For an example, see Figure \ref{fig:triangulation}. The interior nodes in the dense subset of the triangulation of $\Omega_{\mv s}$ are the microscopy data voxels. The exterior nodes are less densely spaced for computational efficiency. Because $\Omega_z$ is an interval, the $z$-triangulation is simply a division of the domain into subintervals. 
However, instead of using the model on the full domain $\Omega_z$, we consider the marginal distribution of only the interior nodes. This allows us to reduce the effect of the boundary conditions at a lower computational cost. In practice, this means that we use the precision matrix $\mv{Q}_{z,\text{marg}} = \mv{Q}_{z,II} - \mv{Q}_{z,IE} \mv{Q}_{z,EE}^{-1} \mv{Q}_{z,EI}$, instead of $\mv{Q}_z$. Here,
\begin{align*}
\mv{Q}_z = 
\begin{bmatrix}
\mv{Q}_{z,EI} & \mv{Q}_{z,EE} \\
\mv{Q}_{z,II} & \mv{Q}_{z,IE} \\
\end{bmatrix},
\end{align*}
where $I$ denotes the indices that correspond to the interior nodes, and $E$ denotes the indices that correspond to the exterior nodes. 
Finding the marginal precision matrix $\mv{Q}_{z,\text{marg}}$ can be done efficiently since $\mv{Q}_{z,EE}$ is low-dimensional, but is too computationally demanding for $\mv{Q}_{\mv s}$. 

\subsection*{The MCMC algorithm}
\label{app:modelfitting}
The MCMC algorithm used to to estimate the model parameters is explained in more detail here.

To simplify the sampling of $\boldw$, we introduce the auxiliary variables $\mathbf{s} = \mv{A}\boldw + \boldsymbol{\epsilon}$. Then the distribution of $\boldw \mid \mathbf{s}, \boldsymbol\gamma, u, \boldy$ is multivariate Gaussian, which is easy to simulate from. The MCMC algorithm uses a Metropolis-within-Gibbs sampler, from which we obtain a Markov chain $\{(\boldw^{(i)}, \mathbf{s}^{(i)}, \boldsymbol\gamma^{(i)}, u^{(i)})\}$ which has the desired posterior density as stationary distribution. The variables and parameters are updated in three separate blocks, which makes simulation of the vectors $\boldw$ and $\mathbf{s}$ easier while maintaining a low autocorrelation of the Markov Chain. A similar Gibbs sampler for binary data can be found in \citet*{AlbertChib1993}. 

We start by selecting starting values $\{(\boldw^{(0)}, \mathbf{s}^{(0)}, \boldsymbol\gamma^{(0)}, u^{(0)})\}$, and then repeat the following three steps for $i=1,2,\ldots$.
	
\begin{enumerate}
\item Sample $\boldw^{(i+1)}\sim \pi(\boldw \mid \mathbf{s}^{(i)}, \boldsymbol\gamma^{(i)}, u^{(i)},\boldy)$ for $\boldw \mid \mathbf{s}, \boldsymbol\gamma, u, \boldy \sim \pN\left(\hat{\mv{Q}}^{-1}\mv{A}^\top \mathbf{s}/\sigma^2,\hat{\mv{Q}}^{-1} \right)$ and $\hat{\mv{Q}} = \mv{Q}(\boldsymbol\gamma) + \mv{A}^\top \mv{A}/\sigma^2$. This step requires sampling a multivariate GMRF, which is the most computationally expensive step of the estimation procedure. How this is done is explained in the next section.  
\item Sample $\mathbf{s}^{(i+1)},u^{(i+1)} \sim \pi(\mathbf{s},u \mid \boldw^{(i+1)} \boldsymbol\gamma^{(i)},\boldy)$ using a Metropolis Hastings step. For this, a new threshold $u'$ is proposed using a random walk proposal, and given the proposed threshold the proposal density for $\mathbf{s}$ is
$$
\pi(\mathbf{s} \mid \mathbf{w}, \boldsymbol\gamma, u', \mathbf{y}) \propto \varphi_{(\mv{A}\mathbf{w},\sigma^2\mv{I})} (\mathbf{s}) \prod_{i: y_i=0} \mathds{1}_{(-\infty,u')} (s_i) \prod_{i: y_i=1} \mathds{1}_{[u',\infty)} (s_i),
$$
which is a product of independent truncated Gaussians. Here 
$\mathds{1}_{I} (s_i)=1$ if $s_i \in I$, and $\mathds{1}_{I} (s_i)=0$ otherwise.
Finally $\mv{s}$ and $u'$ are accepted or rejected jointly.
\item Sample $\boldsymbol\gamma^{(i+1)}\sim \pi(\boldsymbol\gamma \mid \boldw^{(i+1)},\mathbf{s}^{(i+1)}, u^{(i+1)},\boldy)$ using a Metropolis Hastings step where each parameter is proposed separately. Constrained random walk proposals are used for the parameters $\theta_{\mv{s}}$ and $\theta_z$, and log-normal proposals for the parameters $\kappa_{\mv{s}}$ and $\kappa_z$.
With the proposed parameters $(\theta_{\mv{s}}',\theta_{z}',\kappa_{\mv{s}}',\kappa_{z}')$, the gamma conditional density 
		$$\pi(\tau \mid \mathbf{w}, \mathbf{s}, \theta_{\mv{s}}',\theta_{z}',\kappa_{\mv{s}}',\kappa_{z}', u, \mathbf{y})
		\propto \pi(\tau) \varphi_{(\mathbf{0},\mv{Q}^{-1}(\boldsymbol\gamma))} (\mathbf{w})$$ 
is then used as the proposal density for $\tau$. The parameters are accepted or rejected jointly.
\end{enumerate}
		
\subsection*{Reducing the computational complexity of the MCMC algorithm}
\label{app:redmemorysim}
The most computationally demanding part of the MCMC algorithm is to sample from $\boldw \mid \mathbf{s}, \boldsymbol\gamma, u, \boldy$. We use the method introduced in \citet*{PapandreouYuille2010}, and used in \citet*{Siden2017} to sample large GMRFs without having to compute the full Cholesky factor of $\hat{\mv{Q}}$, with the following two steps:
\begin{enumerate}
\item Generate $\boldsymbol\xi= \tau (\mv{R}_{z} \otimes \mv{R}_{\mv{s}})^\top \mathbf{z}_1 + \mv{A}^\top \mathbf{z}_2/\sigma + \mv{A}^\top \mathbf{s}/\sigma^2$, where $\mathbf{z}_1, \mathbf{z}_2 \sim \pN(\mathbf{0},\mv{I})$ are independent, and $\mv{R}_z$ and $\mv{R}_{\mv s}$ are the upper triangular Cholesky factors of $\mv{Q}_z$ and $\mv{Q}_{\mv s}$ respectively. By the properties of the Kronecker product, $\tau \mv{R}_z \otimes \mv{R}_{\mv s}$ is the Cholesky factor of $\mv{Q}(\boldsymbol{\gamma})$. All three matrix-vector products can be computed without evaluating the full matrices. The first product is computed as in \kronprod{}, and the other two are computed using that $\mv{A}$ is a binary matrix with one non-zero element per row.
\item Compute $\mathbf{w}$ by solving $\hat{\mv{Q}}\mathbf{w} = \boldsymbol\xi$ approximately using a preconditioned conjugate gradient (PCG) method \citep*{Demmel1997}. 
\end{enumerate}
The PCG-method in the second step solves the system $\mv{P}^{-1} \hat{\mv{Q}} \mv{w} = \mv{P}^{-1}\boldsymbol{\xi}$ approximately where $\mv{P} = \mv{R}_{\mv{P}}\mv{R}_{\mv{P}}^T$ is a positive definite and symmetric preconditioner. A good preconditioner approximates $\hat{\mv{Q}}$ in such a way that $\mv{P}^{-1} \hat{\mv{Q}}$ is well-conditioned and $\mv{P}^{-1} \mv{v}$ can be computed efficiently. \citet*{Siden2017} used the preconditioner with $\mv{R}_{\mv{P}} = ichol(\hat{\mv{Q}})$, where $ichol$ denotes an incomplete Cholesky factorization. We instead use the preconditioner with Cholesky factor $ichol(\mv{Q}_{z}) \otimes ichol(\mv{Q}_{\mv{s}})$, and use the structure of the preconditioner and of $\hat{\mv{Q}}$ to compute $\mv{P}^{-1}(\hat{\mv{Q}} \mv{w})$ using \kronprod{} and \kroninverse{} without evaluating the full matrices. Table \ref{tab:precond} compares these two choices with some other natural choices of preconditioners for sampling $\boldw$, and shows that using $ichol(\mv{Q}_{z})$ and $ichol(\mv{Q}_{\mv{s}})$ is better than using $\mv{R}_z$ and $\mv{R}_{\mv s}$ in this case, since it leads to a more sparse preconditioner. From the results in the table, it is also clear that it is better to avoid evaluating the full Cholesky factor of the preconditioner by using its Kronecker structure, if possible. 

Using \kronprod{} and \kroninverse{} reduces the amount of elements we need to store in the MCMC algorithm from $\mathcal{O}((n_z n_{\mv s})^2)$ to $\mathcal{O}(n_z^2 + n_{\mv s}^2)$, where $\mv{Q}_{z} \in \mathbb{R}^{n_{z} \times n_{z}}$ and $\mv{Q}_{\mv s} \in \mathbb{R}^{n_{\mv s} \times n_{\mv s}}$---not taking the sparsity of $\mv{Q}_z$ and $\mv{Q}_{\mv s}$ into account---since we never store full Kronecker products. 
As can be seen in Table~\ref{tab:precond}, the gain from using \kroninverse{} can be substantial for the PCG step, which is the most computationally demanding part of the algorithm. Another potentially demanding part is the computation of the Cholesky factor of $\mv{Q}(\boldsymbol{\gamma})$, which is done to evaluate $\varphi_{(\mathbf{0},\mv{Q}^{-1}(\boldsymbol\gamma))} (\mathbf{w})$ when sampling the parameters $\boldsymbol{\gamma}$.
The complexity of the Cholesky factorization of the sparse $\mv{Q}_z$ is $\mathcal{O}(n_z)$, and the complexity of the factorization of the sparse $\mv{Q}_{\mv s}$ with an optimal reordering is $\mathcal{O}(n_{\mv s}^{3/2})$ \citep*[Ch.\ 2.4]{Rue2005}. Compare this with the cost of factorizing a dense $n \times n$ matrix, which is $\mathcal{O}(n^3)$. Performing the Cholesky factorization on the full matrix $\mv{Q}(\boldsymbol{\gamma})$ would therefore take $\mathcal{O}((n_z n_{\mv s})^3)$ operations if $\mv{Q}(\boldsymbol{\gamma})$ were dense, whereas the lower bound using the Kronecker structure of its Cholesky factor $\tau \mv{R}_z \otimes \mv{R}_{\mv s}$ is $\mathcal{O}(n_z + n_{\mv s}^{3/2})$ operations, taking into account the sparsity. 

Thus it is clear that the algorithmic complexity of the MCMC algorithm is greatly reduced by using the sparsity and Kronecker product structure of the precision matrix $\mv{Q}(\boldsymbol{\gamma})$ and its Cholesky factor. This allows us to fit the model to much larger data sets. 
Using \kroninverse{} we can also generate samples from the GMRF $\mv{w}$ efficiently.

\begin{table}[t]
	\footnotesize
	\begin{center}
		\begin{tabular}{lcrrrr}  
			\toprule
			& & \multicolumn{2}{c}{CPU-time} \\
			\cline{3-4}
			Preconditioner $\mv{P} = \mv{R}_{\mv{P}}^\top \mv{R}_{\mv{P}}$ & \parbox[r]{2.5cm}{Full Cholesky\\factor evaluated} & PCG & Total & \parbox[r]{1.5cm}{PCG\\iterations} & \parbox[r]{1.5cm}{Proportion\\non-zero elements} \\
			\midrule
$\mv{R}_{\mv{P}} = ichol(\mv{Q}_z) \otimes ichol(\mv{Q}_{\mv s})$ & (yes) no & (7.3) 0.9 & (8.4) 1.0 & (42) 42 & 0.003 \\ 
$\mv{R}_{\mv{P}} = \mv{R}_z \otimes \mv{R}_{\mv s}$ & (yes) no & (32.8) 2.9 & (35.4) 2.9 & (70) 70 & 0.004\ \\ 
$\mv{P} = diag(\hat{\mv{Q}})$ & - & 9.3 & 9.3 & 2555 & $8\cdot 10^{-6}$ \\ 
no preconditioner & - & 19.4 & 19.4 & 5967 & - \\ 
$\mv{R}_{\mv{P}} = ichol(\mv{Q}_z \otimes \mv{Q}_{\mv s})$ & yes & 27.3 & 30.2 & 135 & 0.007 \\ 
$\mv{R}_{\mv{P}} = ichol(\hat{\mv{Q}})$ & yes & 27.8 & 30.7 & 138 & 0.007 \\
			\bottomrule
		\end{tabular}
	\end{center}
	\caption{Differences in performance for simulating from $\boldw \mid \mathbf{s}, \boldsymbol\gamma, u, \boldy$ by using a PCG method with different preconditioners.Here $diag(\hat{\mv{Q}})$ is a diagonal matrix with the diagonal elements of $\hat{\mv{Q}}$. The results shown were computed from $100$ iterations of the MCMC algorithm for dataset $HPC30_1$, for which $\mv{Q}_z \in \mathbb{R}^{20 \times 20}$ and $\mv{Q}_{\mv s} \in \mathbb{R}^{6305\times 6305}$, where one sample was generated in each iteration. When possible, the performance when using \kroninverse{} to compute $\mv{P}^{-1} \mv{v} = \mv{R}_{\mv{P}}^{-1} (\mv{R}_{\mv{P}}^{-1})^\top \mv{v}$ was compared with the performance when the full Cholesky factor $\mv{R}_{\mv{P}}$ was evaluated. The PCG CPU-time is the mean time that was spent in the PCG-routine simulating one vector from $\boldw \mid \mathbf{s}, \boldsymbol\gamma, u, \boldy$; the total CPU-time is the mean total time it took to simulate one vector, which includes the time spent in the PCG-routine, as well as the time it took to calculate the preconditioner and to generate $\boldsymbol{\xi}$. The PCG-iterations are the mean number of iterations of the PCG-routine. The proportion of non-zero elements of the preconditioner is also indicated. A approximate minumum degree reordering \citep*{amestoy1996approximate} was performed when it led to a reduction in computation time.}
	\label{tab:precond}
\end{table}

\section{Estimation of the size distributions}
\label{app:sizedistdetails}

The local size in a point in a (random) set $\Xi$ with respect to a structuring element $B$ can be defined using the morphological operation opening, denoted $\circ$. The opening of $\Xi$ by $B$ consists of an erosion followed by a Minkowski addition with $B$. The opened set can be written 
$$
\Xi \circ B = \{x \in \Xi: \exists z \text{ such that } x \in z + B \text{ and } z + B \subseteq \Xi \}.
$$
 The local size $h$ is defined as $h(x,\Xi,B) = \sup \{\lambda: x \in \Xi \circ \lambda B \}$, where $\lambda B= \{\lambda x: x\in B\}$. 
It follows that, if the supremum can be attained, the local size in a point $x \in \Xi$ is the largest value $\lambda$ for which the rescaled structuring element $\lambda B$ can be translated so that it fits within the set $\Xi$ and covers the point $x$. 

We define the size distribution for a random set 
$\Xi$ with respect to a convex 
structuring element $B$ as 
$$S(\lambda; x,\Xi,B) = P(h(x,\Xi,B) \geq \lambda \mid x \in \Xi),$$ 
i.e.\ the size distribution $S(\lambda; x,\Xi,B)$ gives the probability that the local size with respect to $B$ in a point $x$ in $\Xi$ is greater than or equal to $\lambda$. Since the structuring element is convex, $h(x,\Xi,B) \geq \lambda$ is equivalent to $x \in \Xi \circ \lambda B$, and so the size distribution can be written as $S(\lambda; x,\Xi,B) = P(x \in \Xi \circ \lambda B \mid x \in \Xi) = P(x \in \Xi \circ \lambda B)/P(x \in \Xi)$ \citep*{Matheron1975,Delfiner1972}. 

When estimating the size distribution of a stationary random set $\Xi$ defined on $\Omega$, which is only observed in a bounded window $W \subset \Omega$, we have to consider the boundary effects to get an unbiased estimate. If we do not have any information about the set outside of $W$, we can in general only observe the opened set $\Xi \circ \lambda B$ in the smaller window $\tilde{W}(\lambda B) =  (W \ominus \lambda B) \ominus -\lambda B \subset \lambda B$ \citep*{OhserSchladitz2009}. This means that we cannot determine the local size in a point close to the boundary of the observation window $W$, since we need information about the surrounding of the point. To account for this, we can use a minus-sampling estimator of the size distribution based on the unbiased estimate $v_d(\Xi \circ \lambda B \cap \tilde{W}(\lambda B))/v_d(\tilde{W}(\lambda B))$ of $P(x \in \Xi \circ \lambda B)$, where $v_d(\cdot)$ denotes the volume of the set. Properties of minus-sampling estimators for size distributions for random closed sets can be found in \citet*{Moore1991}, and properties for discretely indexed random sets in \citet*{SivakumarGoutsias1997}.

Because we select microscopy data samples that have homogeneous pore sizes, our observation windows are small in the $z$-direction. Therefore, we do not use the unbiased minus-sampling estimator since that would reduce the observation window of the sample further. Instead we approximate $\Xi \circ \lambda B \cap W$ with the opened set $(\Xi \cap W) \circ \lambda B$, which is equivalent to assuming that there are no pores outside of the observation window $W$. This gives us a biased estimate $v_d((\Xi \cap W) \circ \lambda B)/v_d(W)$ of $P(x \in \Xi \circ \lambda B)$. However, the bias has the same effect for the filtered CLSM pore structure and the stochastic simulations from the model, since the observation windows are of the same size, and hence will not affect comparisons.

The size distribution for a structuring element $B$ is thus estimated as $$\hat{S}(\lambda) =  \frac{\hat{P}(x \in \Xi \circ \lambda B)}{\hat{P}(x \in \Xi)} = \frac{v_d((\Xi \cap W) \circ \lambda B)/v_d(W)}{v_d(\Xi \cap W)/v_d(W)},$$ for $\lambda = 1,2,\dots,N$, where the maximal local size $N$ is determined by the observation window $W$ in which the set $\Xi$ is observed. The size density is estimated as $\hat{s}(\lambda) = \hat{S}(\lambda) - \hat{S}(\lambda+1)$. 

\section{Results for all CLSM samples}
\label{supp:supplementaryresults}

Here we present results for the model fitted to all CLSM samples. More results for sample $HPC30_1$ can be found in Section ``Results''. 
For some samples, only every second or third voxel in the CLSM image was kept, depending on the average size of the pores. There is also a resolution difference between the $30\%$ and $40\%$ CLSM images. Thus, the distances between voxels are not the same in all samples (see Table \ref{tab:samplesizes}). However, we chose the size of the domains $\Omega$ so that the fitted parameters are comparable between the samples. The estimated oscillation parameters are fairly similar for all model fits (Table \ref{tab:sampleresults}). 

\begin{table}[t]
	\footnotesize
	\begin{center}
		\begin{tabular}{lcc}  
			\toprule
			Sample & Voxel size ($\mu$m$^3$) & Number of voxels \\
			\midrule
			$HPC30_1$ & $0.23 \times 0.23 \times 0.20$ & $74 \times 74 \times 20$ \\
			$HPC30_2$ & $0.23 \times 0.23 \times 0.20$ & $74 \times 74 \times 20$ \\
			$HPC30_3$ & $0.15 \times 0.15 \times 0.20$ & $101 \times 76 \times 25$ \\
			$HPC30_4$ & $0.15 \times 0.15 \times 0.20$ & $101 \times 76 \times 26$ \\
			$HPC40_1$ & $0.15 \times 0.15 \times 0.10$ & $81 \times 81 \times 36$ \\
			$HPC40_2$ & $0.15 \times 0.15 \times 0.10$ & $81 \times 81 \times 31$ \\
			$HPC40_3$ & $0.15 \times 0.15 \times 0.10$ & $91 \times 91 \times 26$ \\
			$HPC40_4$ & $0.15 \times 0.15 \times 0.10$ & $91 \times 91 \times 21$ \\
			\bottomrule
		\end{tabular}
	\end{center}
	\caption{Voxel sizes and number of voxels for all the samples.}
	\label{tab:samplesizes}
\end{table}

\begin{table}[t]
	\footnotesize
	\begin{center}
		\begin{tabular}{lcrrrrr}  
			\toprule
			& \multicolumn{2}{c}{Parameter estimates} & \multicolumn{2}{c}{Volume fraction} & \multicolumn{2}{c}{Surface area} \\
			\cline{2-3} \cline{4-5} \cline{6-7}
			Sample & ($\theta_{\mv{s}}$, $\theta_z$) & ($\kappa^2_{\mv{s}}$, $\kappa^2_z$) & Sample & Fitted model & Sample & Fitted model \\
			\midrule
			$HPC30_1$ & $(0.86,0.56)$ & $(254,61)$ & 0.697 & 0.698 (0.004) & 0.161 & 0.173 (0.004)\\
			$HPC30_2$ & $(0.84,0.51)$ & $(219,78)$ & 0.694 & 0.694 (0.004) & 0.150 & 0.170 (0.004) \\
			$HPC30_3$ & $(0.86,0.58)$ & $(369,87)$ & 0.691 & 0.692 (0.004) & 0.148 & 0.159 (0.003) \\
			$HPC30_4$ & $(0.88,0.56)$ & $(801,112)$ & 0.684 & 0.685 (0.006) & 0.210 & 0.217 (0.004)\\
			$HPC40_1$ & $(0.89,0.56)$ & $(1088,271)$ & 0.610 & 0.609 (0.007) & 0.254 & 0.271 (0.003) \\
			$HPC40_2$ & $(0.90,0.57)$ & $(1077,272)$ & 0.615 & 0.621 (0.007) & 0.248 & 0.269 (0.003)\\
			$HPC40_3$ & $(0.90,0.54)$ & $(1532,319)$ & 0.618 & 0.616 (0.005) & 0.291 & 0.306 (0.003)\\
			$HPC40_4$ & $(0.87,0.50)$ & $(1825,384)$ & 0.619 & 0.606 (0.006) & 0.345 & 0.358 (0.003)\\
			\bottomrule
		\end{tabular}
	\end{center}
	\caption{The oscillation parameters and range parameters for the model fitted to each sample. The volume fractions and surface areas for the pore structures obtained from the CLSM samples, as well as the mean values of $500$ stochastic simulations from each fitted pore model with standard deviations in parentheses.}
	\label{tab:sampleresults}
\end{table}

A larger range parameter for the model corresponds to a smaller covariance range and hence smaller pores. The samples for the two CLSM images were ordered so that the first two samples had smaller pore sizes than the last two samples for each image, which can be seen in the estimated range parameters. One can also note that the estimated range parameters are larger for the samples with $40\%$ HPC than for the samples with $30\%$ HPC. 

The differences between the sample and fitted model volume fractions are no more than two percent, and the differences between the sample and fitted model surface areas are $3$--$13\%$, where the surface areas were calculated using the algorithm provided by \citet*{Legland2011}. The volume fractions of the samples differ from the theoretical HPC weight ratios. The depth dependent threshold was always determined using a larger part of the CLSM image than the sample itself, and there may e.g.\ be edge effects or dust particles outside of the sample contributing to the volume fraction discrepancies.

\begin{figure}[H]
	\centering
	\subcaptionbox{}
	[0.45\textwidth]
	{\includegraphics[width=7cm]{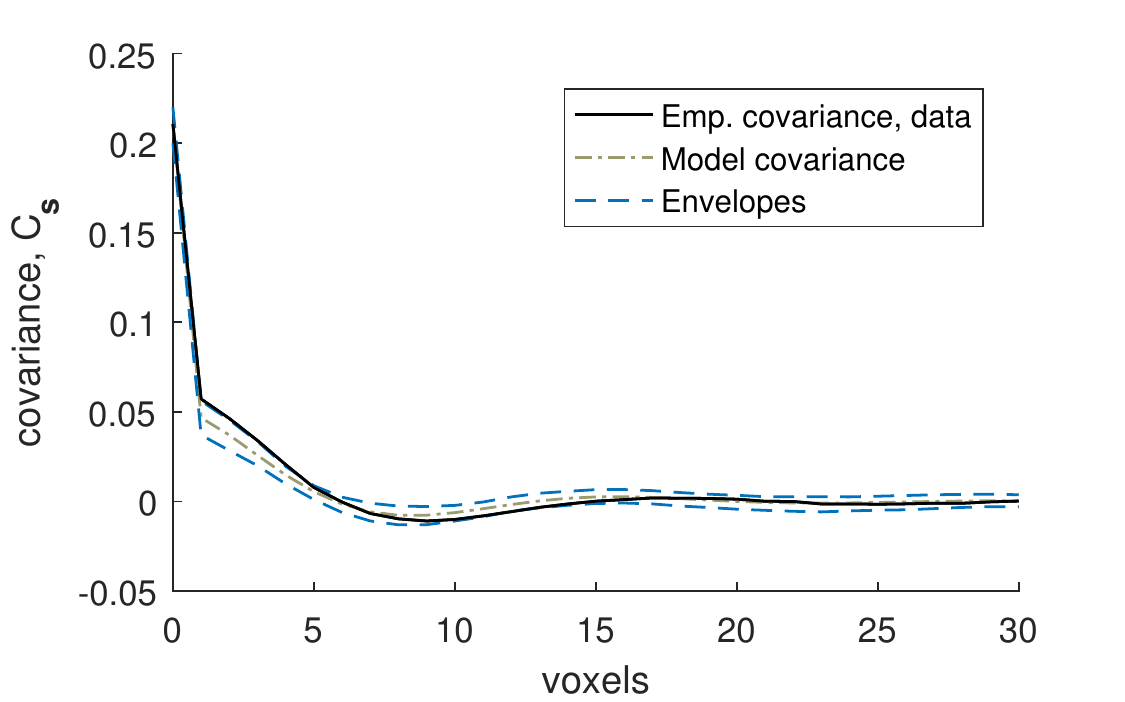}}
	\subcaptionbox{}
	[0.45\textwidth]
	{\includegraphics[width=7cm]{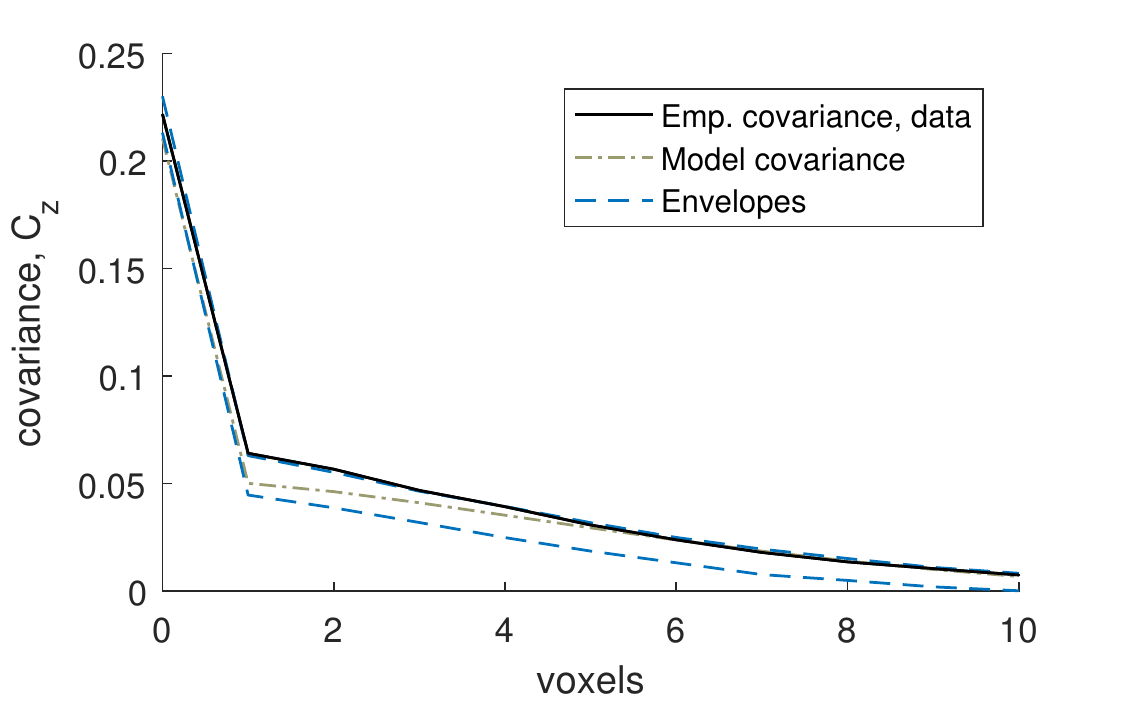}}
	\subcaptionbox{}
	[0.45\textwidth]
	{\includegraphics[width=7cm]{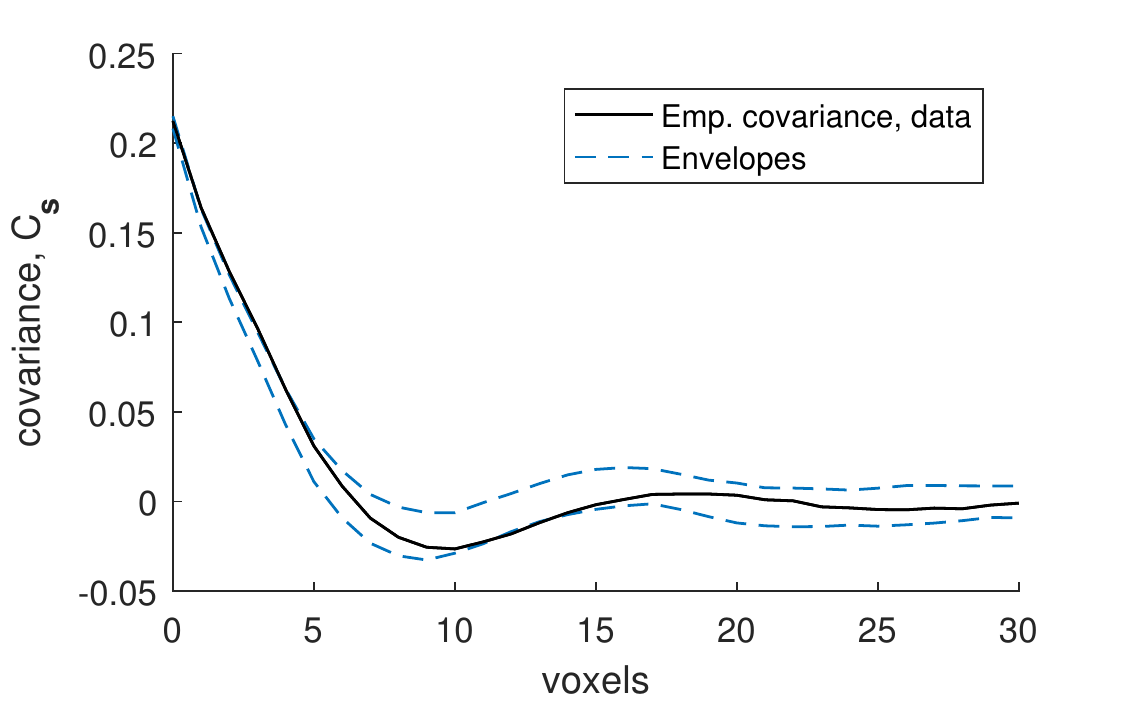}}
	\subcaptionbox{}
	[0.45\textwidth]
	{\includegraphics[width=7cm]{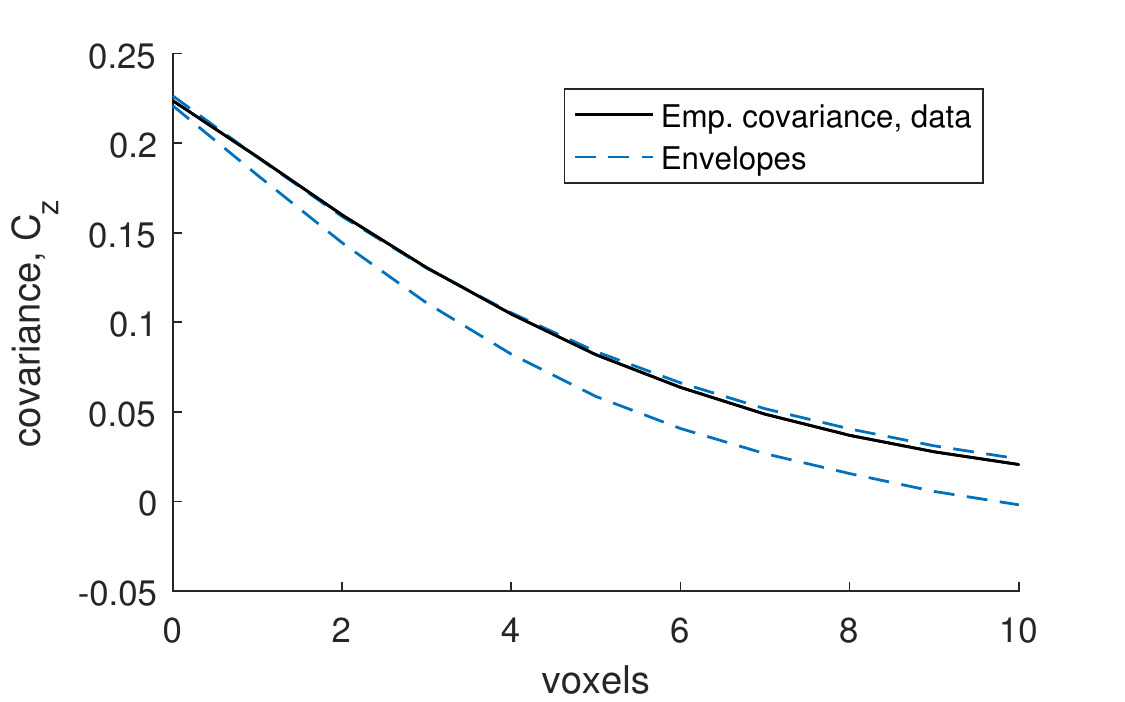}}
	\subcaptionbox{}
	[0.24\textwidth]
	{\includegraphics[width=4.5cm]{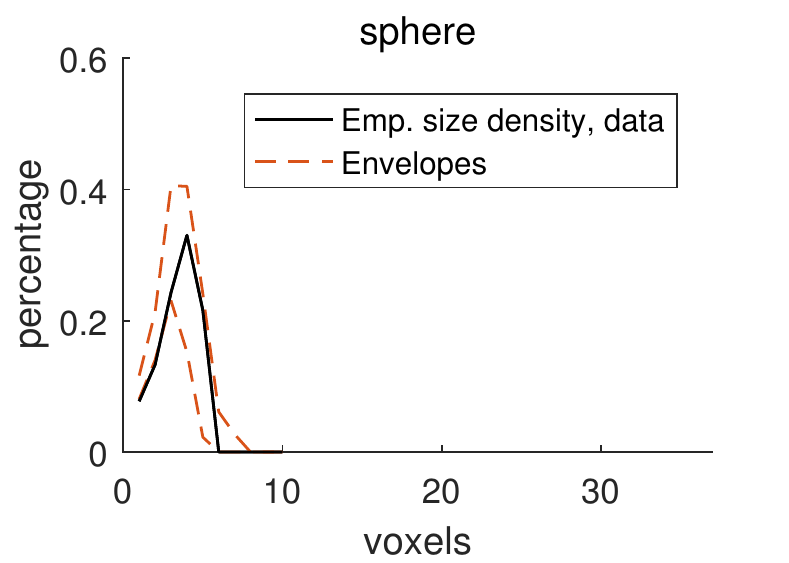}}
	\subcaptionbox{}
	[0.24\textwidth]
	{\includegraphics[width=4.5cm]{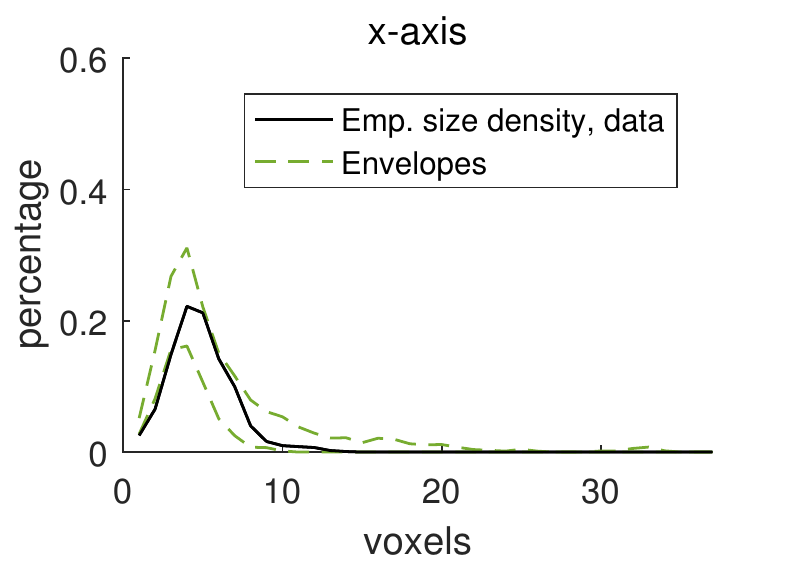}}
	\subcaptionbox{}
	[0.24\textwidth]
	{\includegraphics[width=4.5cm]{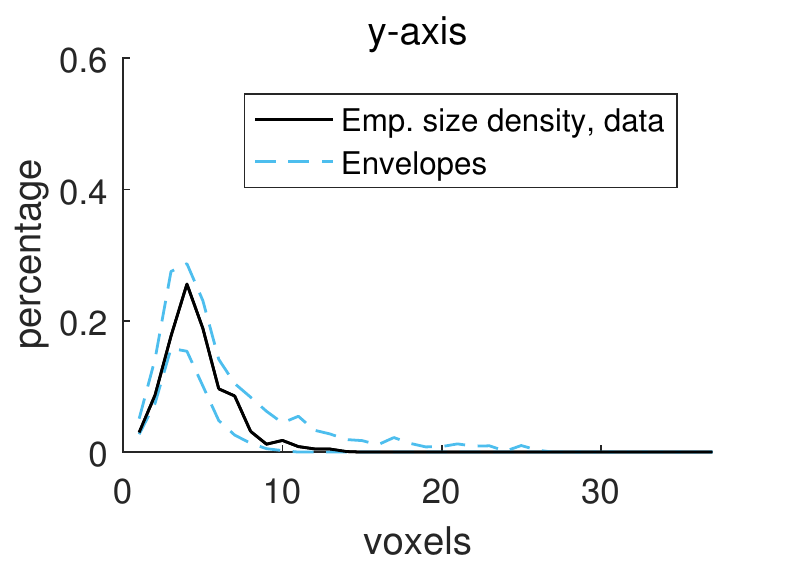}}
	\subcaptionbox{}
	[0.24\textwidth]
	{\includegraphics[width=4.5cm]{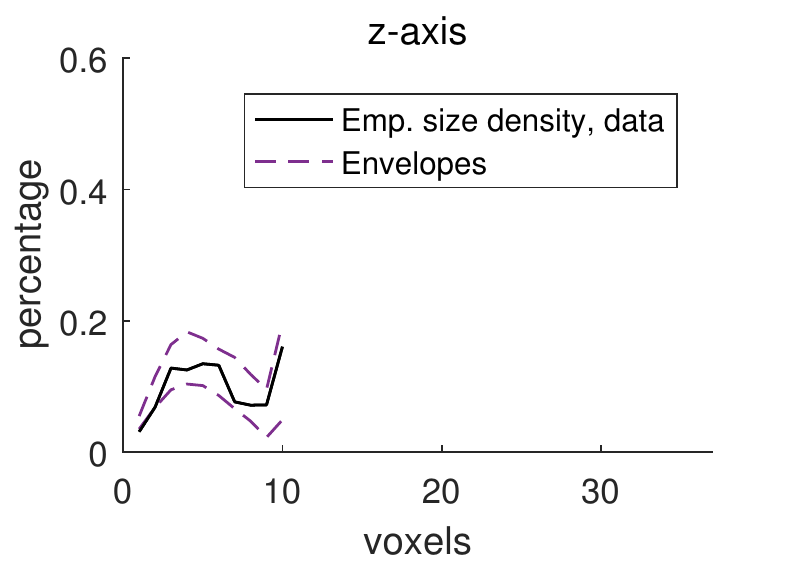}}
	\subcaptionbox{}
	[0.24\textwidth]
	{\includegraphics[width=4.5cm]{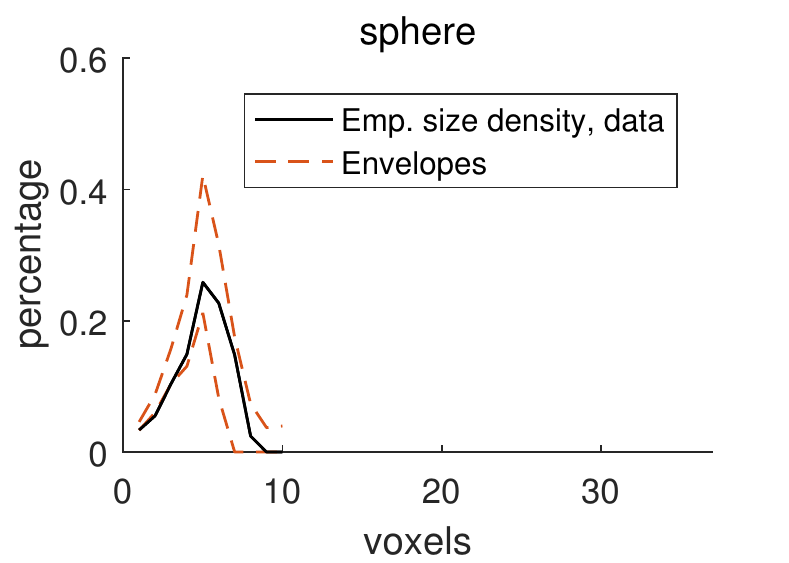}}
	\subcaptionbox{}
	[0.24\textwidth]
	{\includegraphics[width=4.5cm]{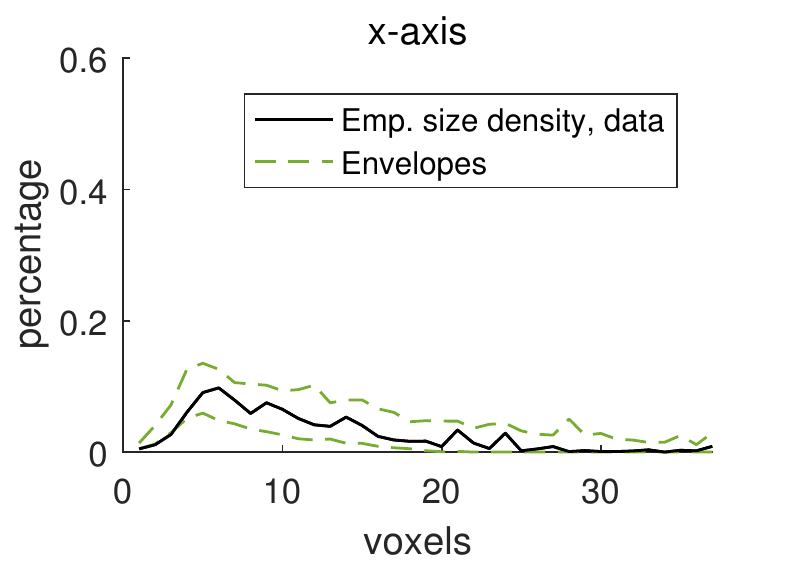}}
	\subcaptionbox{}
	[0.24\textwidth]
	{\includegraphics[width=4.5cm]{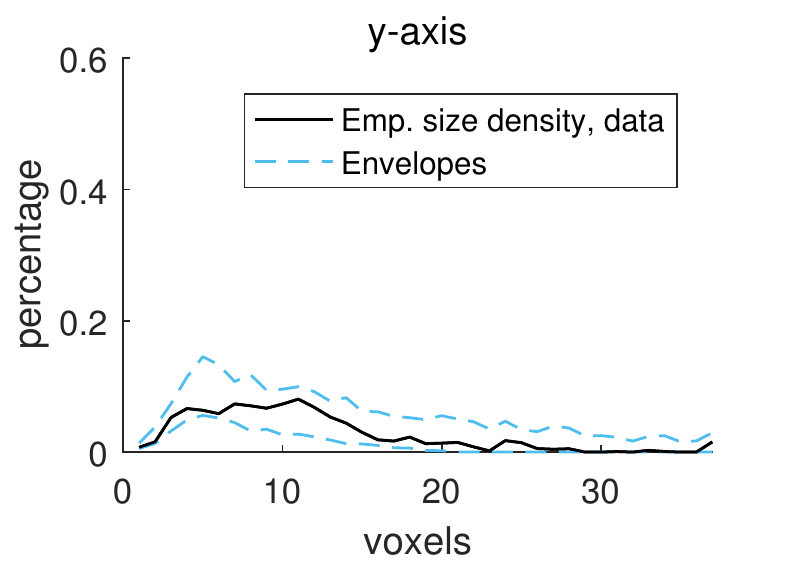}}
	\subcaptionbox{}
	[0.24\textwidth]
	{\includegraphics[width=4.5cm]{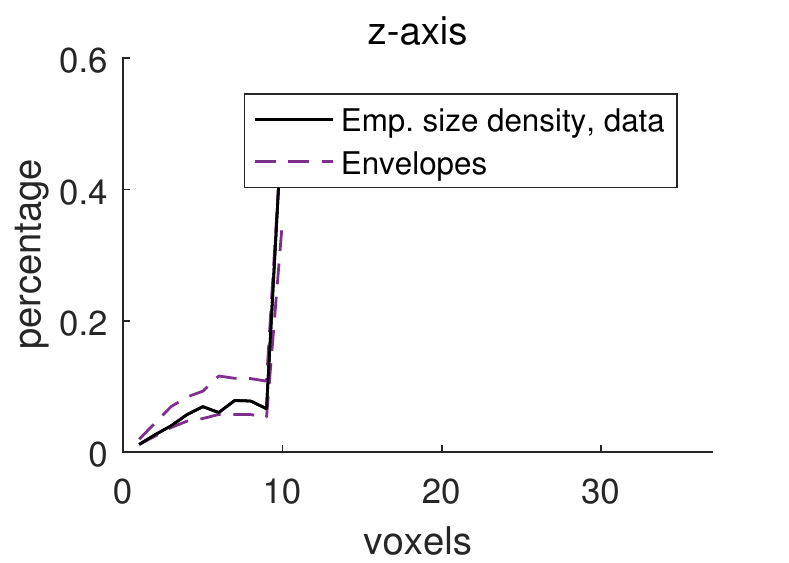}}
	\caption{Results for the $HPC30_2$ sample.  Marginal covariance functions are shown for the noise model (a and b) and the pore model (c and d), together with $95\%$ simultaneous envelopes estimated from $500$ simulations from the corresponding models.  The empirical covariance function $C_{\mv{s}}$ estimated from the binarized CLSM image $\mv{y}$ and the CLSM pore structure $\tildey$ are shown in (a) and (c) respectively.  The corresponding estimates for $C_z$ are shown in (b) and (d). For the noise model, the model covariance function with point-estimates of the parameters is also shown.  Size densities estimated from $\tildey$ on the pore space ((e)--(h)) and the pore matrix ((i)--(l)) are also shown together with $95\%$ simultaneous envelope estimated from $500$ simulations from the pore model. The following structuring elements, with length/radius two, were used:  the unit sphere (e and i), and lines aligned with the x-axis (f and j), the y-axis (g and k), and the z-axis (h and l).}
	\label{fig:summfuncrightp2}
\end{figure}

\begin{figure}[t]
	\centering
	\subcaptionbox{}
	[0.45\textwidth]
	{\includegraphics[width=7cm]{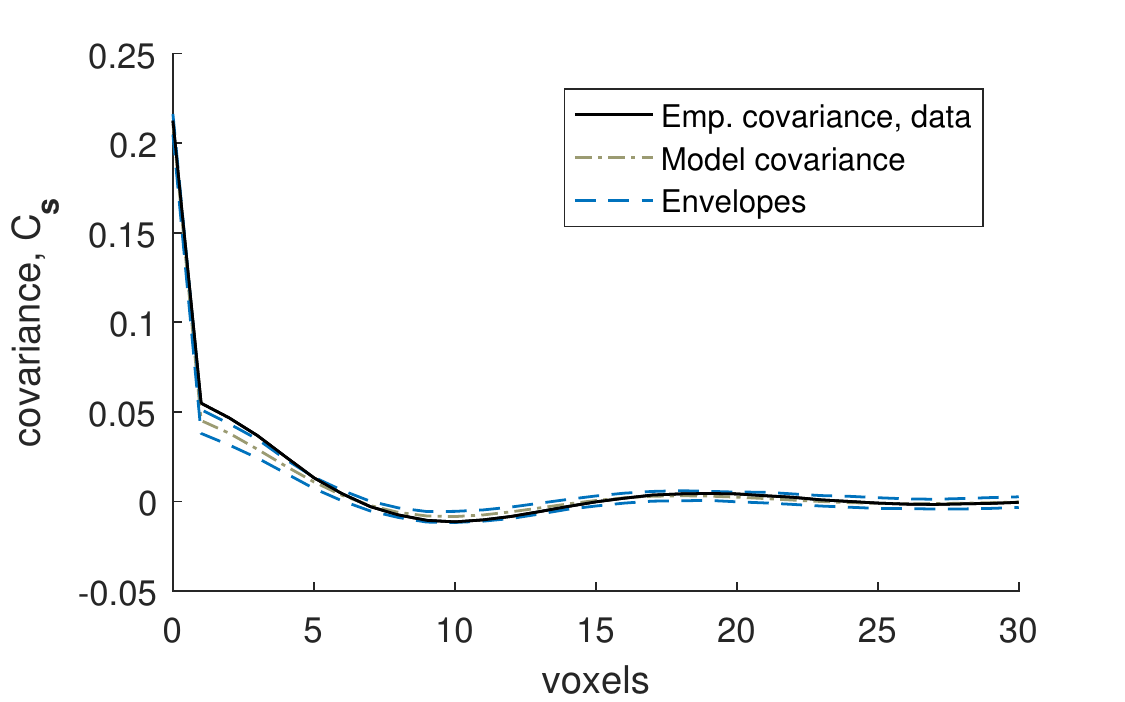}}
	\subcaptionbox{}
	[0.45\textwidth]
	{\includegraphics[width=7cm]{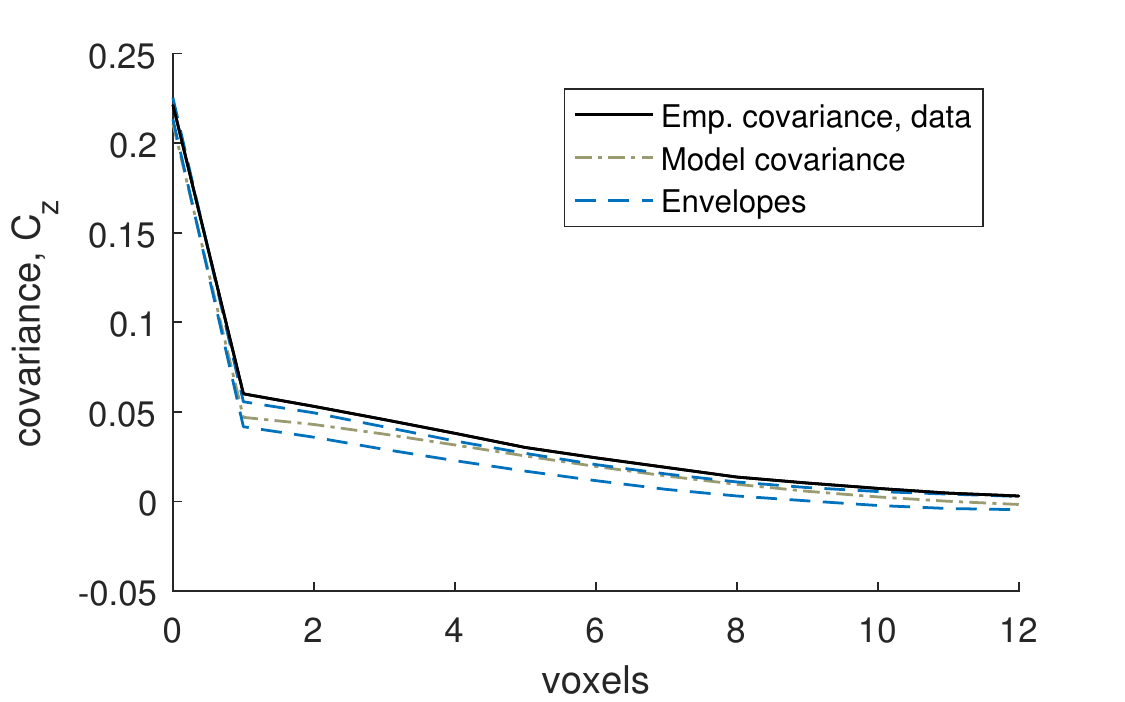}}
	\subcaptionbox{}
	[0.45\textwidth]
	{\includegraphics[width=7cm]{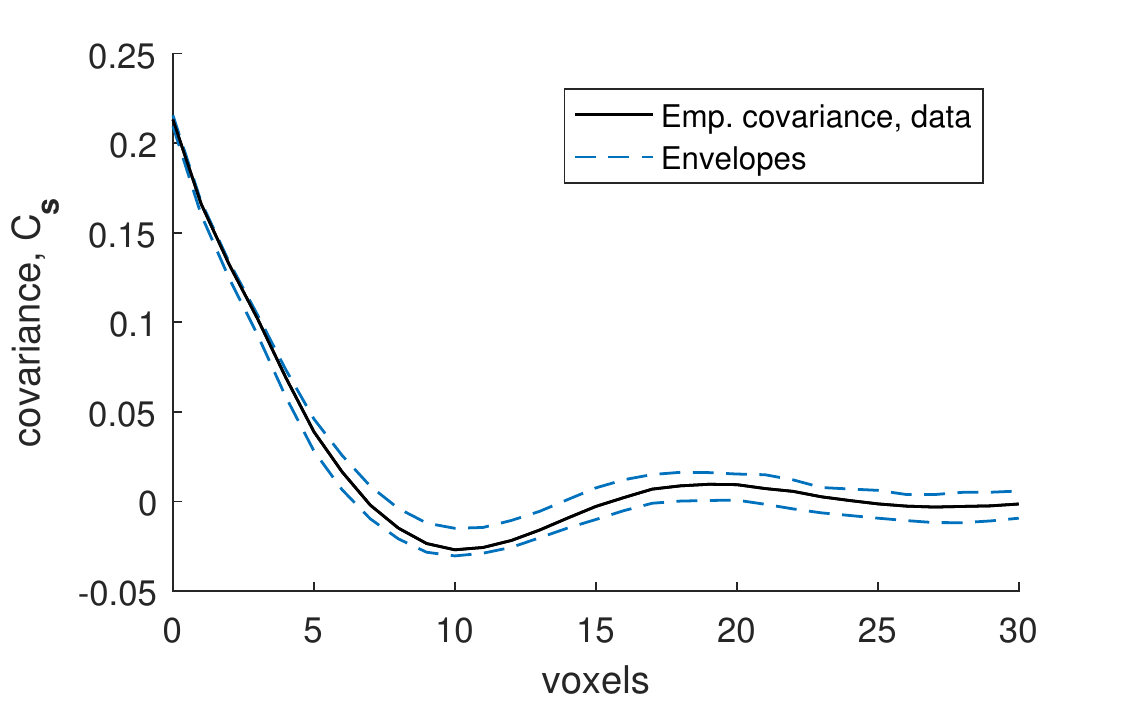}}
	\subcaptionbox{}
	[0.45\textwidth]
	{\includegraphics[width=7cm]{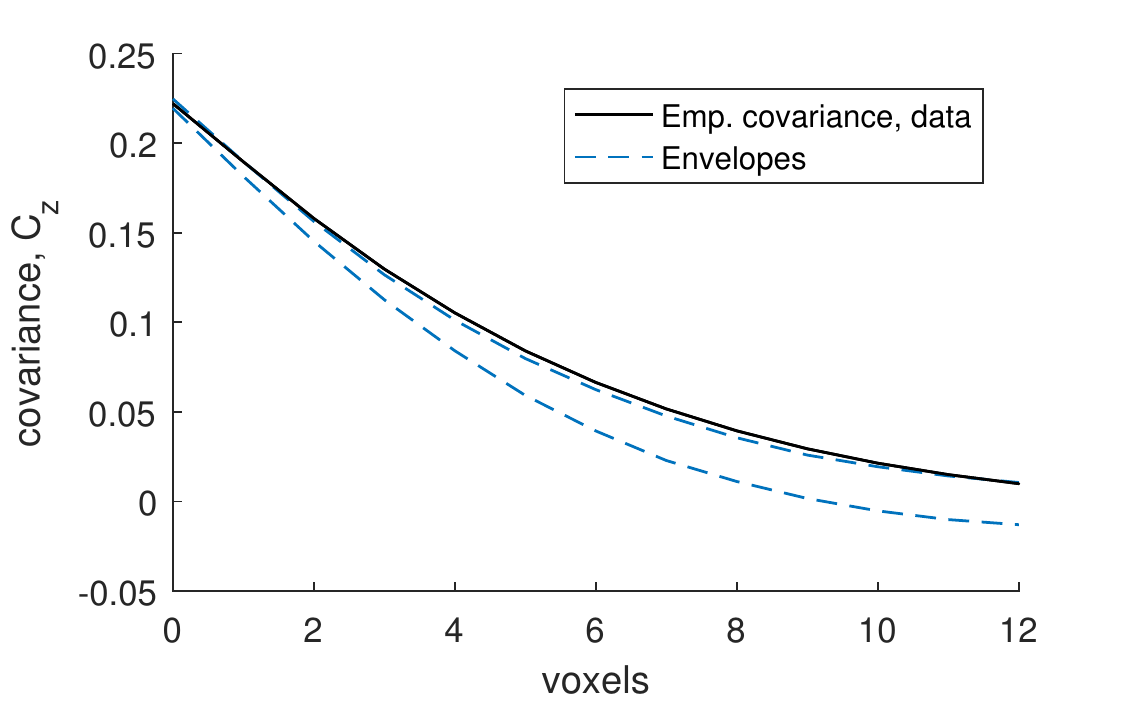}}
	\subcaptionbox{}
	[0.24\textwidth]
	{\includegraphics[width=4.5cm]{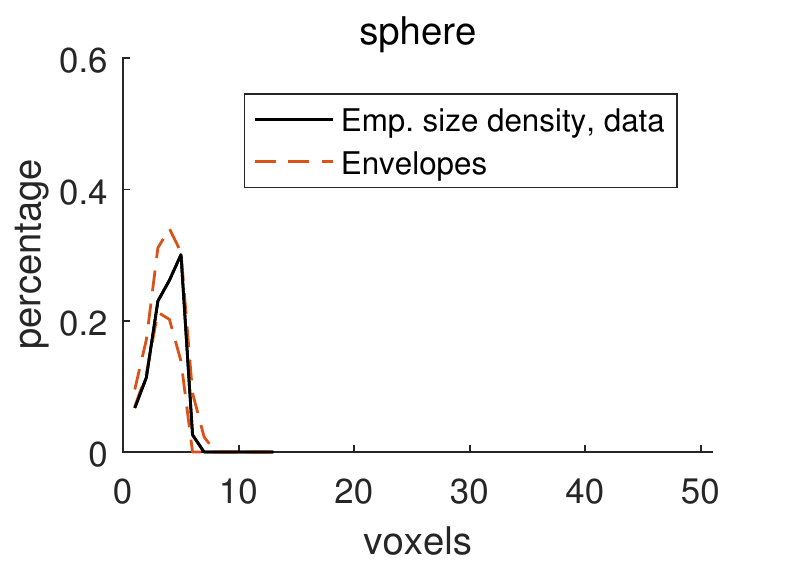}}
	\subcaptionbox{}
	[0.24\textwidth]
	{\includegraphics[width=4.5cm]{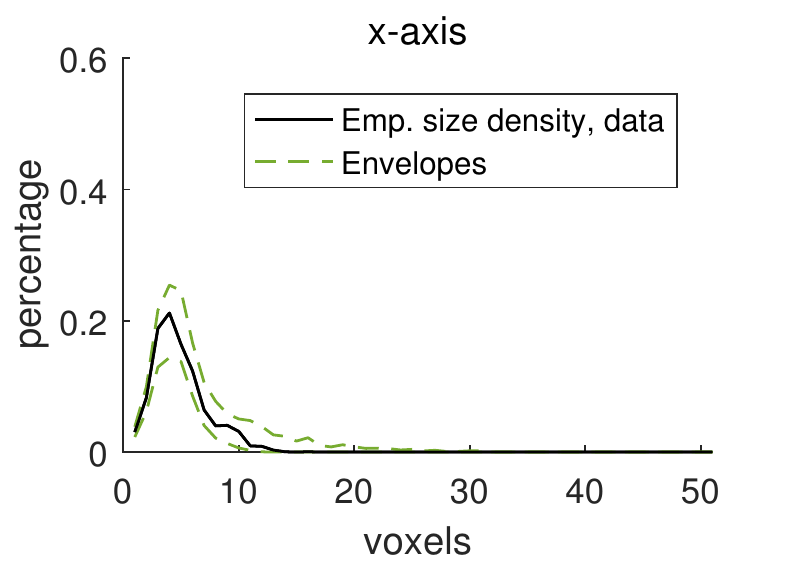}}
	\subcaptionbox{}
	[0.24\textwidth]
	{\includegraphics[width=4.5cm]{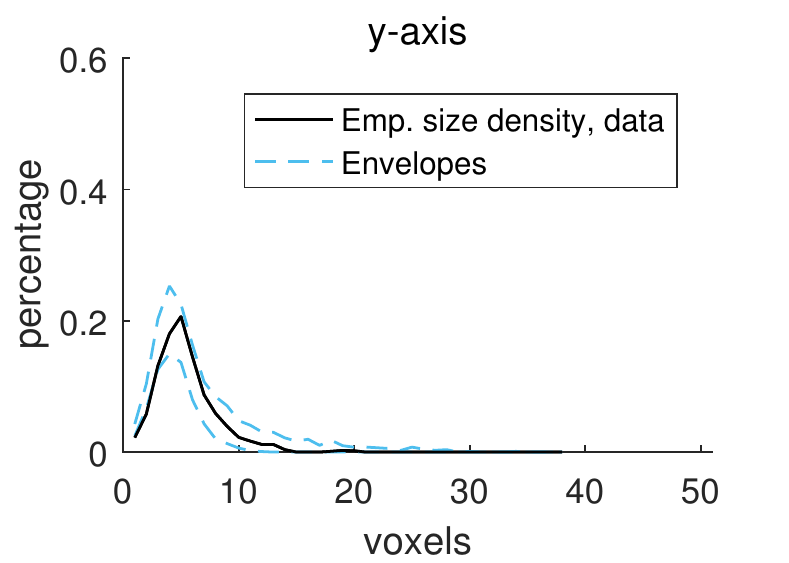}}
	\subcaptionbox{}
	[0.24\textwidth]
	{\includegraphics[width=4.5cm]{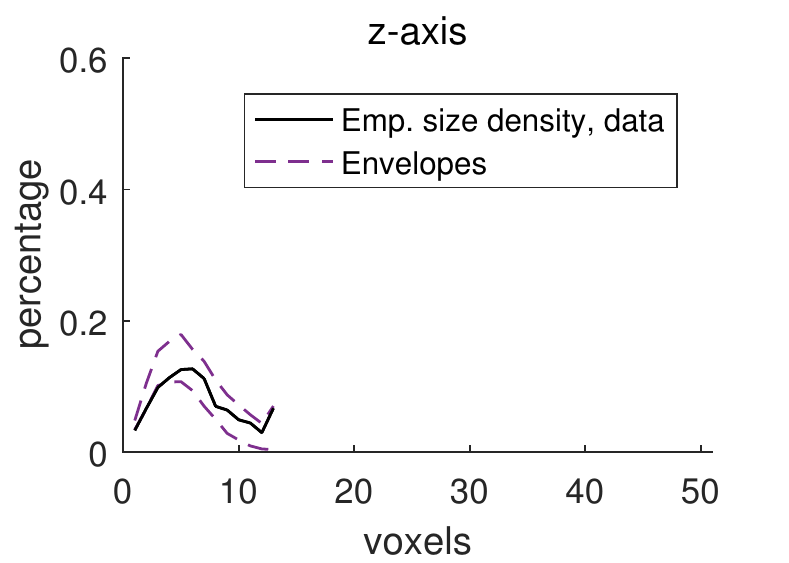}}
	\subcaptionbox{}
	[0.24\textwidth]
	{\includegraphics[width=4.5cm]{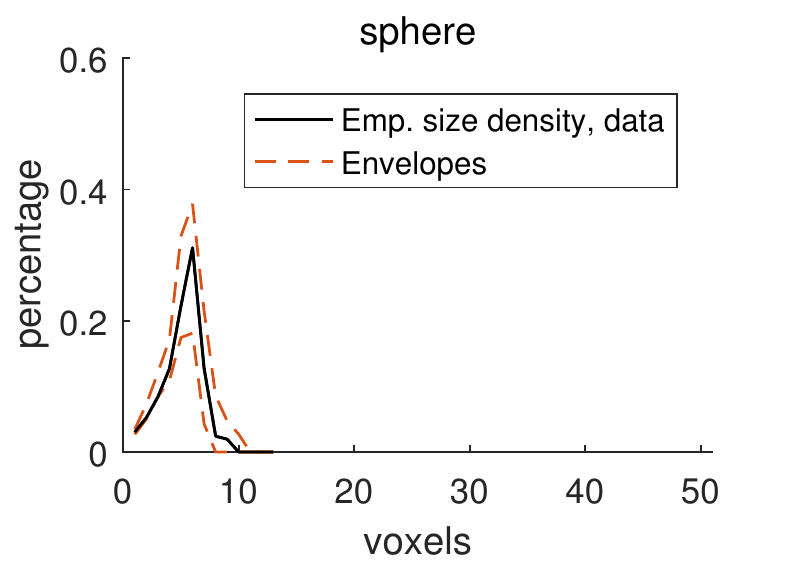}}
	\subcaptionbox{}
	[0.24\textwidth]
	{\includegraphics[width=4.5cm]{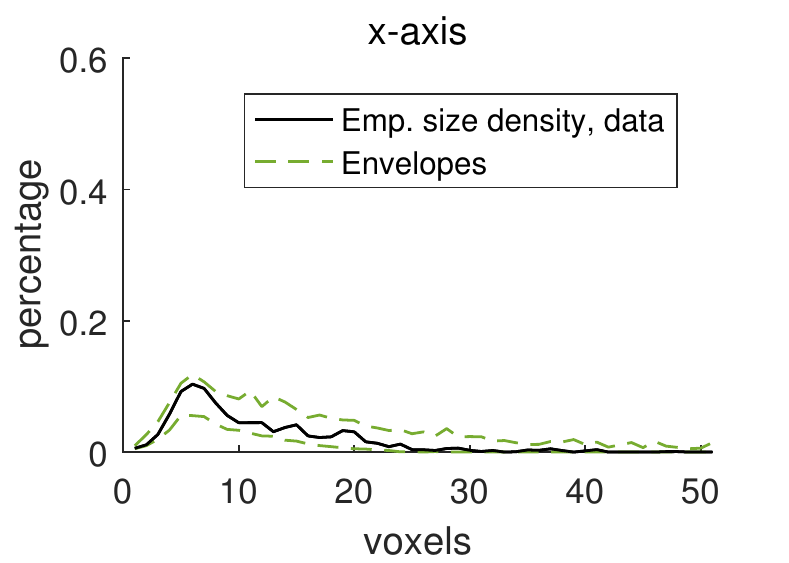}}
	\subcaptionbox{}
	[0.24\textwidth]
	{\includegraphics[width=4.5cm]{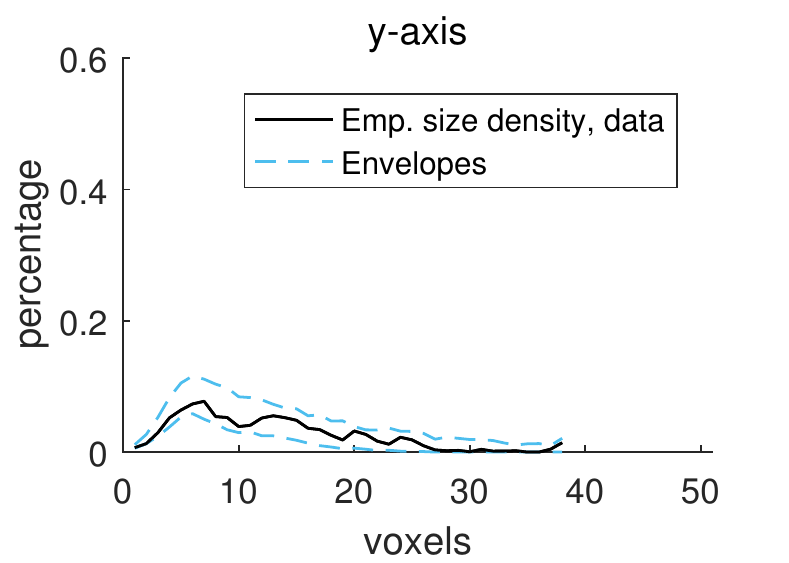}}
	\subcaptionbox{}
	[0.24\textwidth]
	{\includegraphics[width=4.5cm]{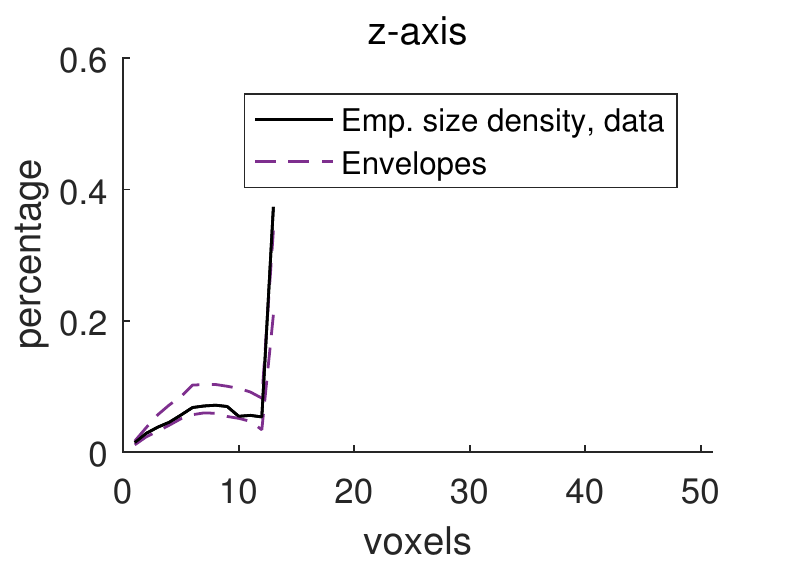}}
	\caption{Results for the $HPC30_3$ sample.  Marginal covariance functions are shown for the noise model (a and b) and the pore model (c and d), together with $95\%$ simultaneous envelopes estimated from $500$ simulations from the corresponding models.  The empirical covariance function $C_{\mv{s}}$ estimated from the binarized CLSM image $\mv{y}$ and the CLSM pore structure $\tildey$ are shown in (a) and (c) respectively.  The corresponding estimates for $C_z$ are shown in (b) and (d). For the noise model, the model covariance function with point-estimates of the parameters is also shown.  Size densities estimated from $\tildey$ on the pore space ((e)--(h)) and the pore matrix ((i)--(l)) are also shown together with $95\%$ simultaneous envelope estimated from $500$ simulations from the pore model. The following structuring elements, with length/radius two, were used:  the unit sphere (e and i), and lines aligned with the x-axis (f and j), the y-axis (g and k), and the z-axis (h and l).}
	\label{fig:summfuncleft11}
\end{figure}

\begin{figure}[H]
	\centering
	\subcaptionbox{}
	[0.45\textwidth]
	{\includegraphics[width=7cm]{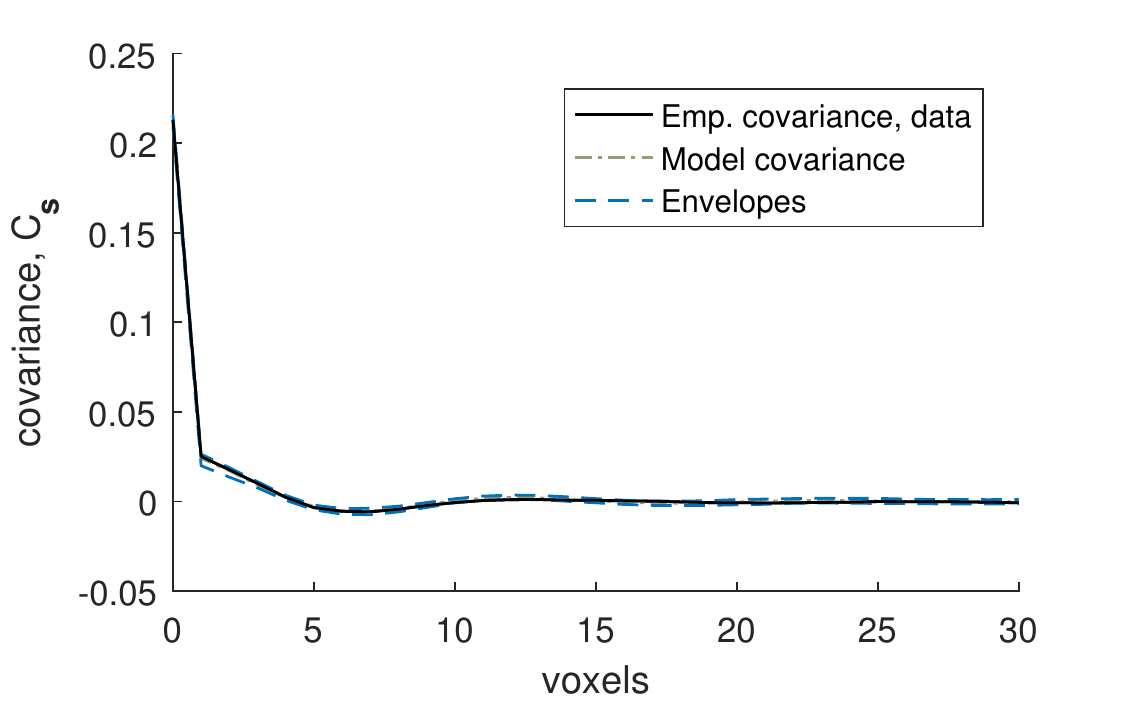}}
	\subcaptionbox{}
	[0.45\textwidth]
	{\includegraphics[width=7cm]{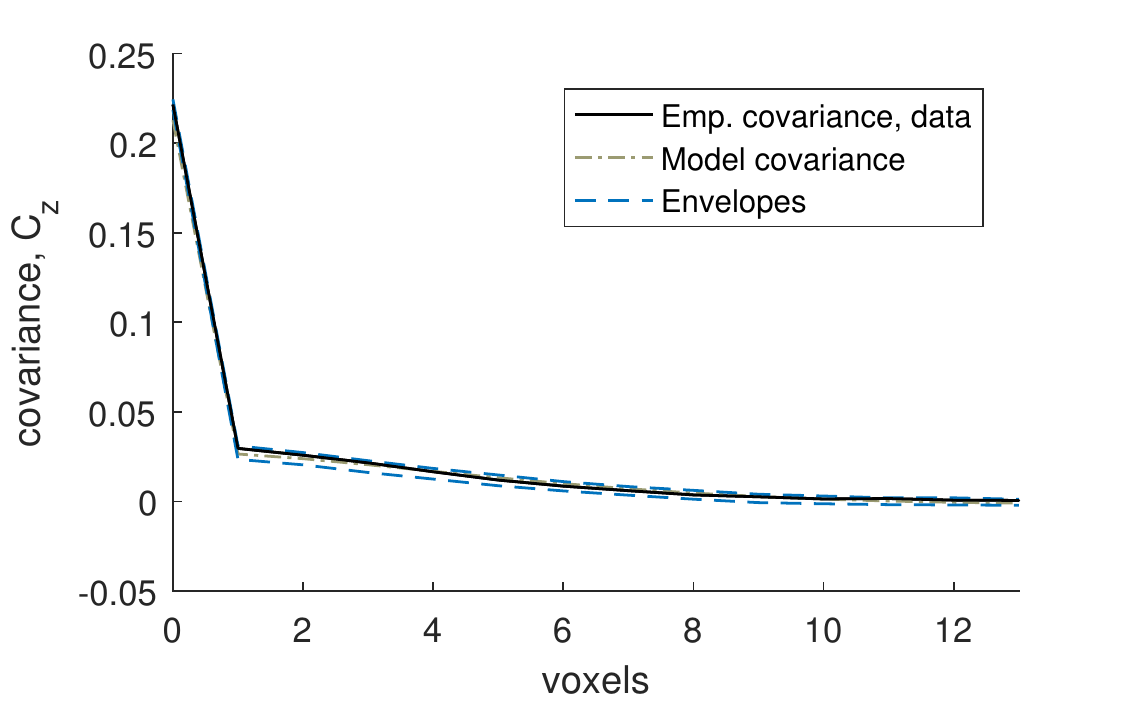}}
	\subcaptionbox{}
	[0.45\textwidth]
	{\includegraphics[width=7cm]{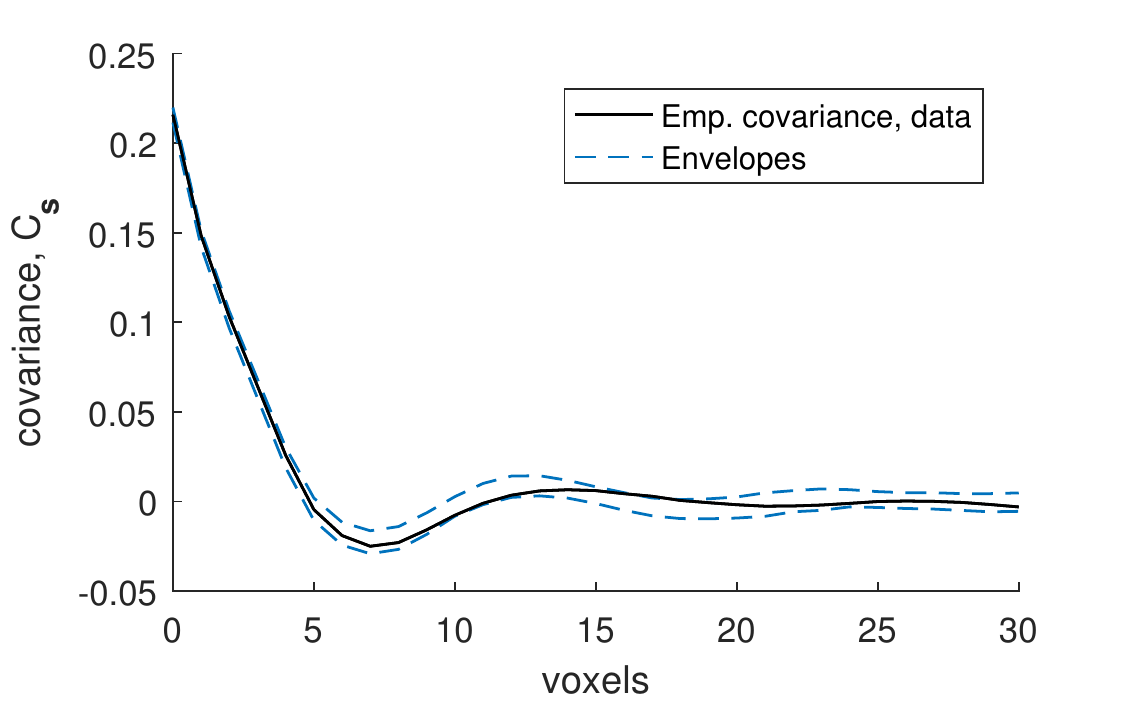}}
	\subcaptionbox{}
	[0.45\textwidth]
	{\includegraphics[width=7cm]{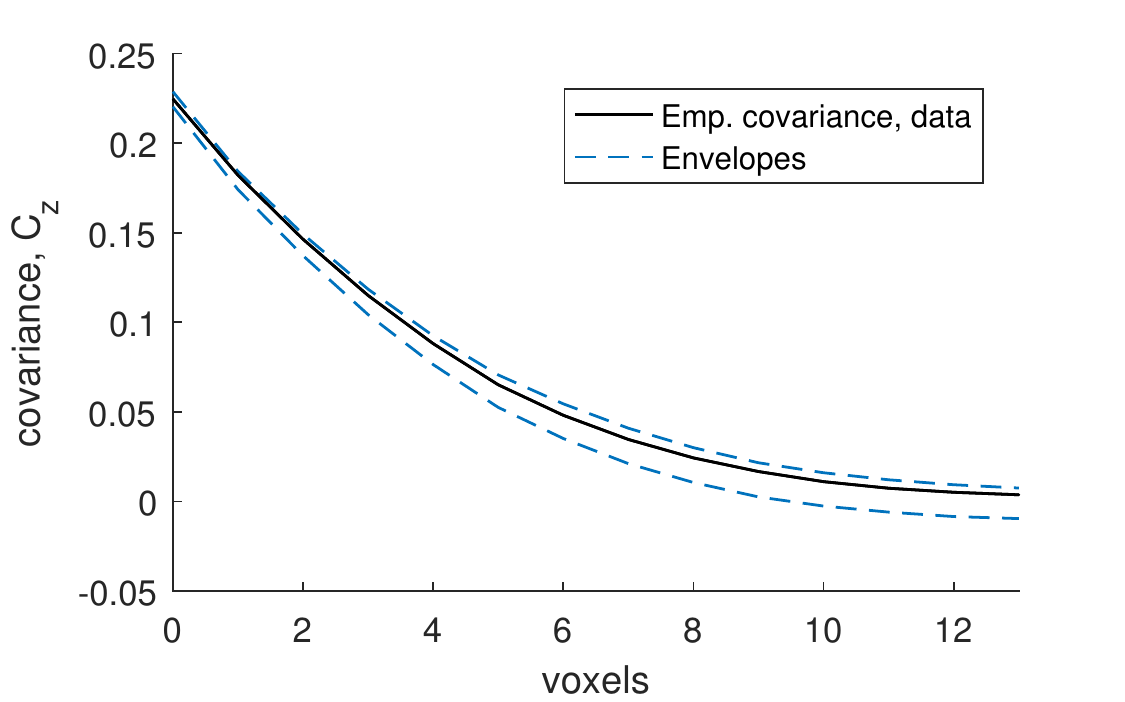}}
	\subcaptionbox{}
	[0.24\textwidth]
	{\includegraphics[width=4.5cm]{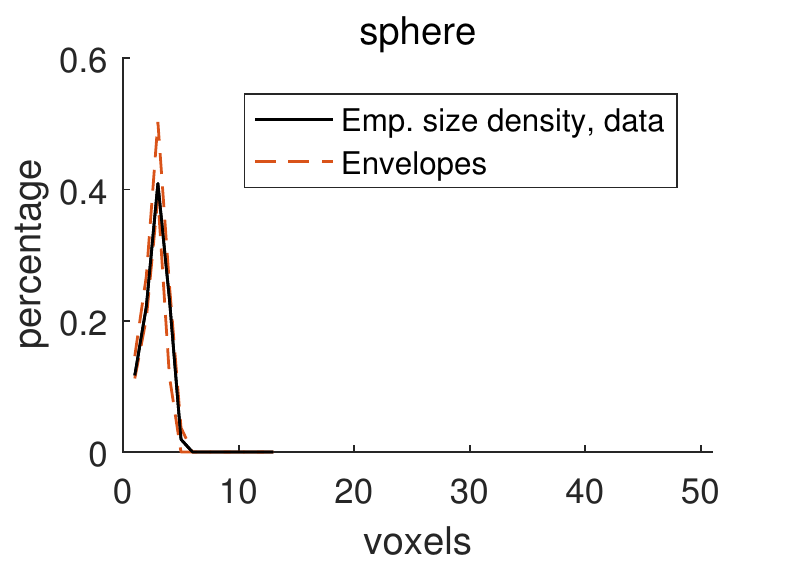}}
	\subcaptionbox{}
	[0.24\textwidth]
	{\includegraphics[width=4.5cm]{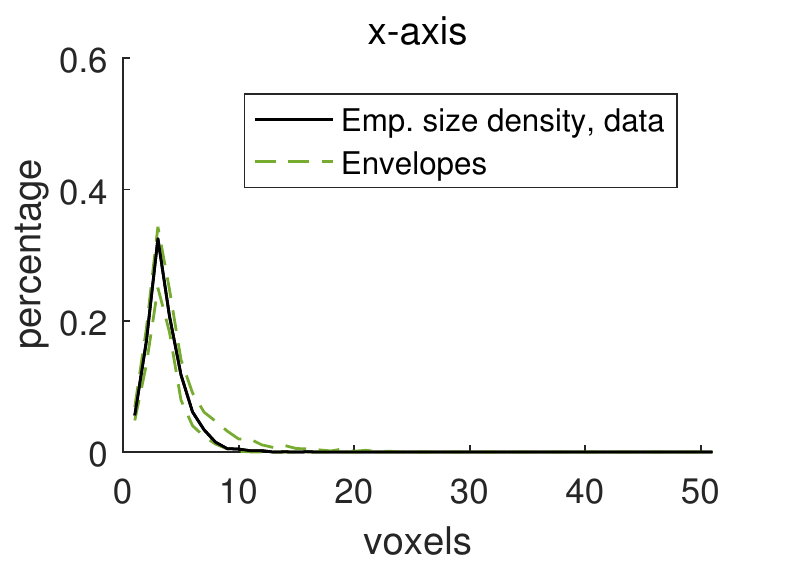}}
	\subcaptionbox{}
	[0.24\textwidth]
	{\includegraphics[width=4.5cm]{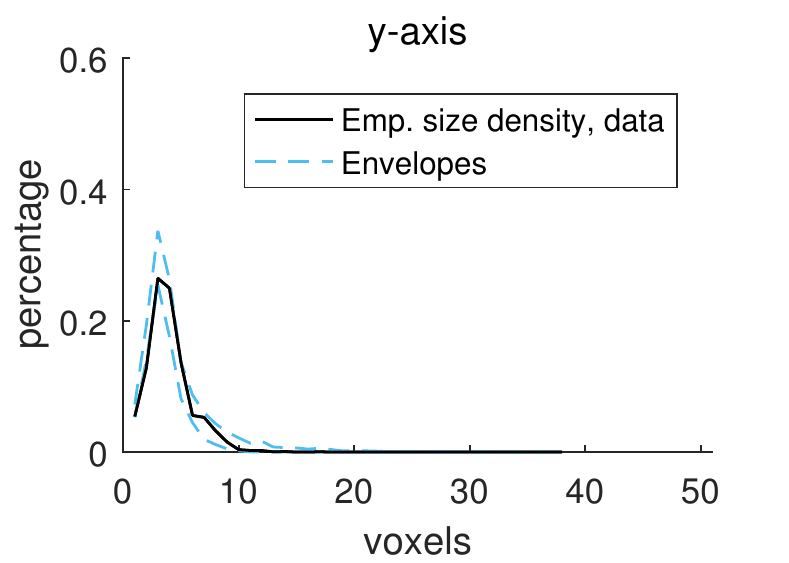}}
	\subcaptionbox{}
	[0.24\textwidth]
	{\includegraphics[width=4.5cm]{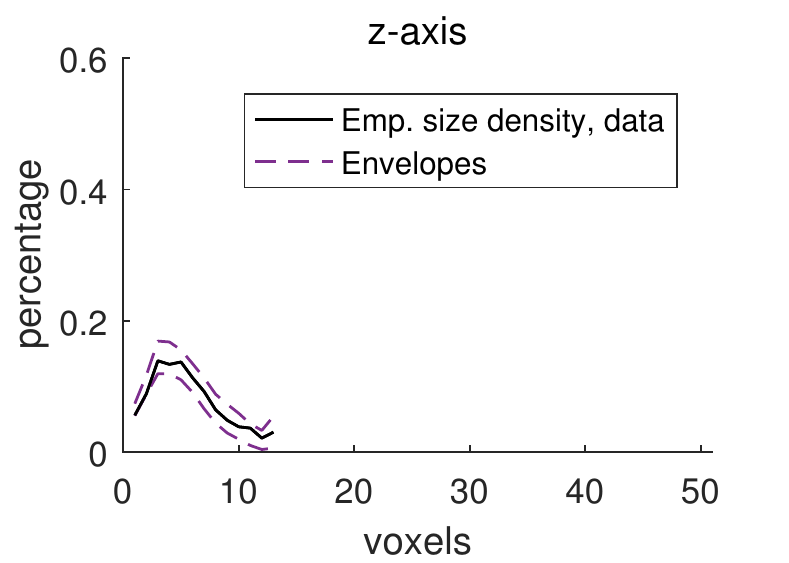}}
	\subcaptionbox{}
	[0.24\textwidth]
	{\includegraphics[width=4.5cm]{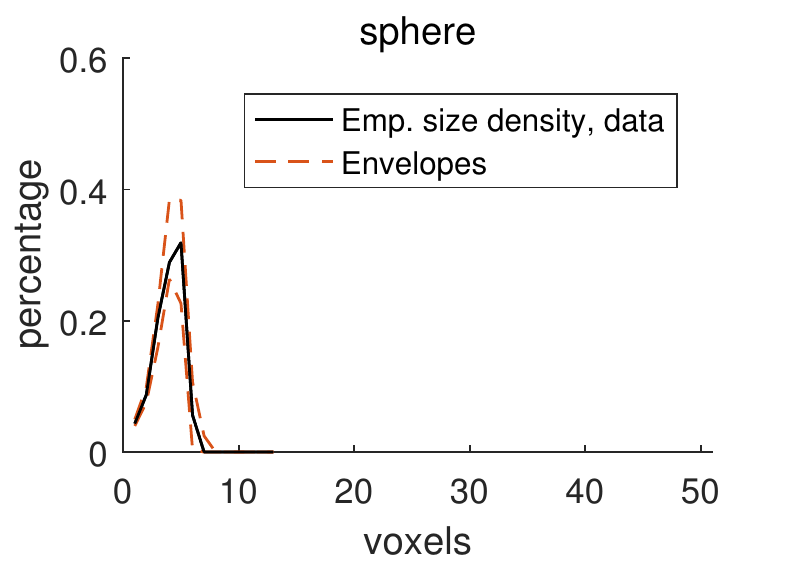}}
	\subcaptionbox{}
	[0.24\textwidth]
	{\includegraphics[width=4.5cm]{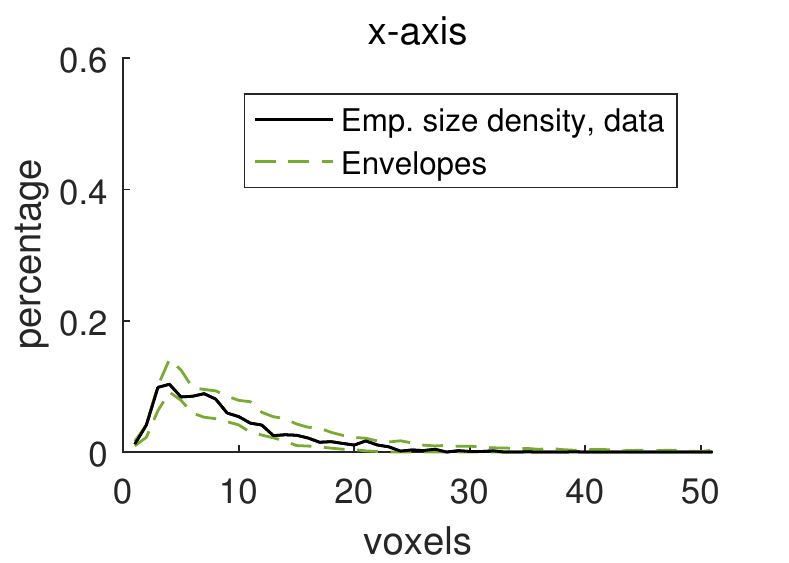}}
	\subcaptionbox{}
	[0.24\textwidth]
	{\includegraphics[width=4.5cm]{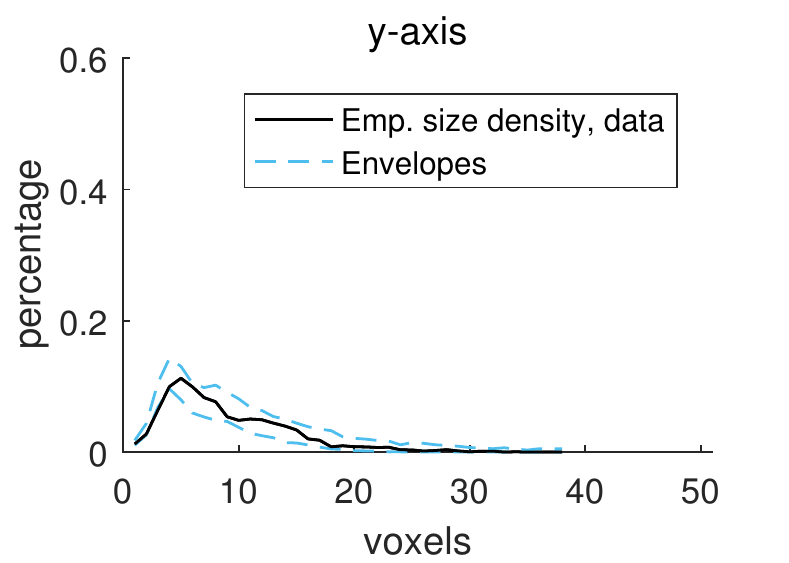}}
	\subcaptionbox{}
	[0.24\textwidth]
	{\includegraphics[width=4.5cm]{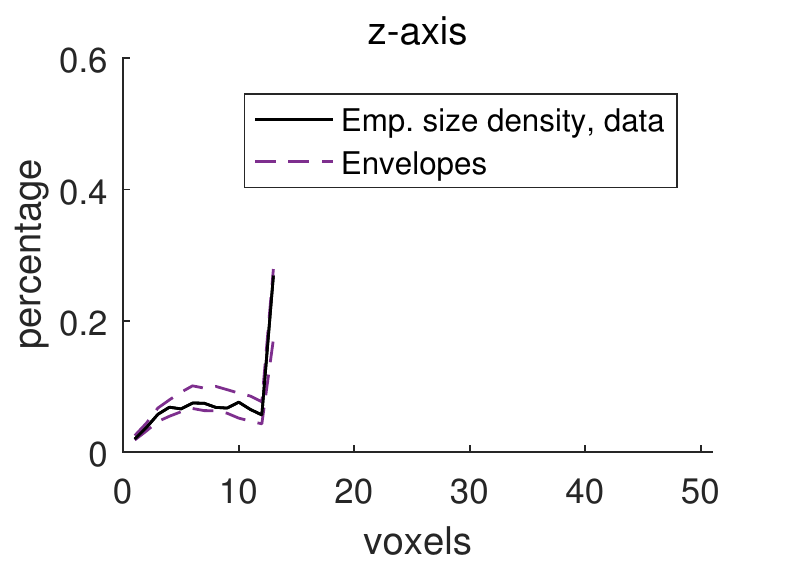}}
	\caption{Results for the $HPC30_4$ sample.  Marginal covariance functions are shown for the noise model (a and b) and the pore model (c and d), together with $95\%$ simultaneous envelopes estimated from $500$ simulations from the corresponding models.  The empirical covariance function $C_{\mv{s}}$ estimated from the binarized CLSM image $\mv{y}$ and the CLSM pore structure $\tildey$ are shown in (a) and (c) respectively.  The corresponding estimates for $C_z$ are shown in (b) and (d). For the noise model, the model covariance function with point-estimates of the parameters is also shown.  Size densities estimated from $\tildey$ on the pore space ((e)--(h)) and the pore matrix ((i)--(l)) are also shown together with $95\%$ simultaneous envelope estimated from $500$ simulations from the pore model. The following structuring elements, with length/radius two, were used:  the unit sphere (e and i), and lines aligned with the x-axis (f and j), the y-axis (g and k), and the z-axis (h and l).}
	\label{fig:summfuncleft21}
\end{figure}

\begin{figure}[H]
	\centering
	\subcaptionbox{}
	[0.45\textwidth]
	{\includegraphics[width=7cm]{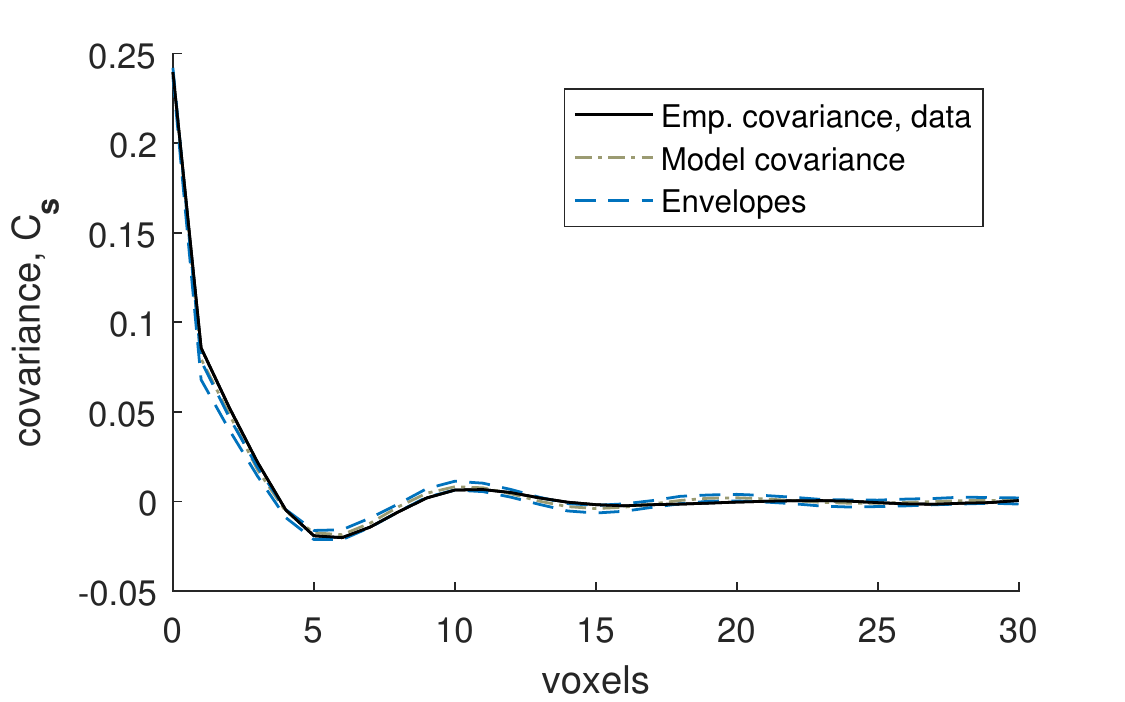}}
	\subcaptionbox{}
	[0.45\textwidth]
	{\includegraphics[width=7cm]{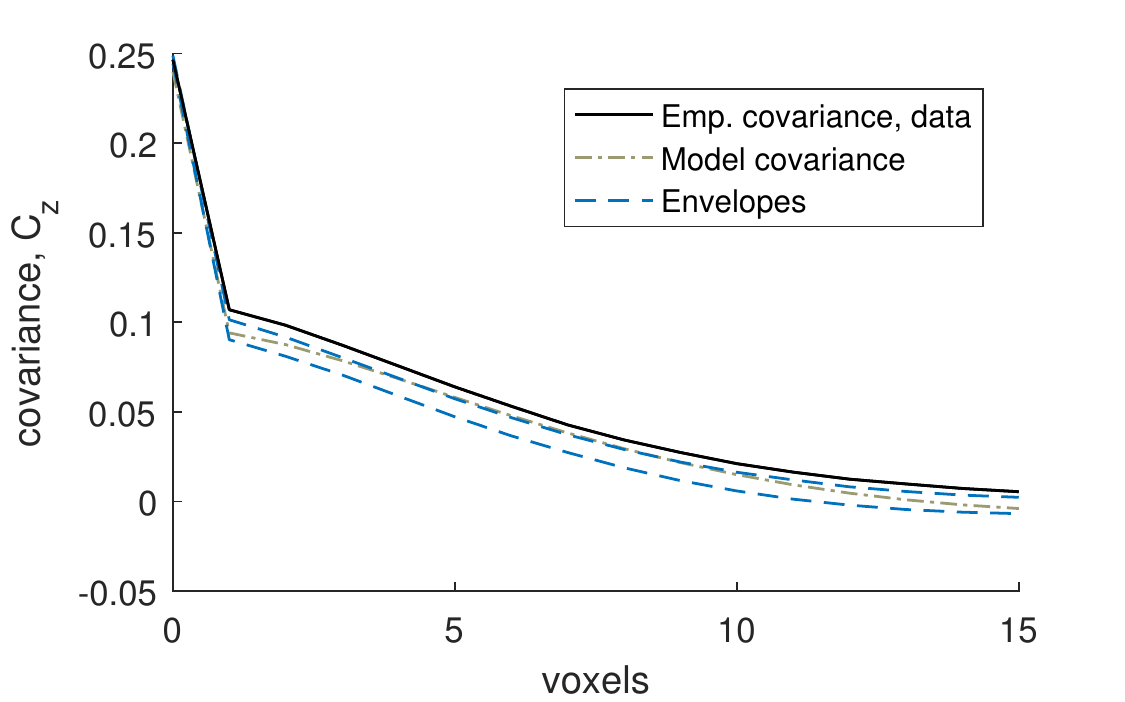}}
	\subcaptionbox{}
	[0.45\textwidth]
	{\includegraphics[width=7cm]{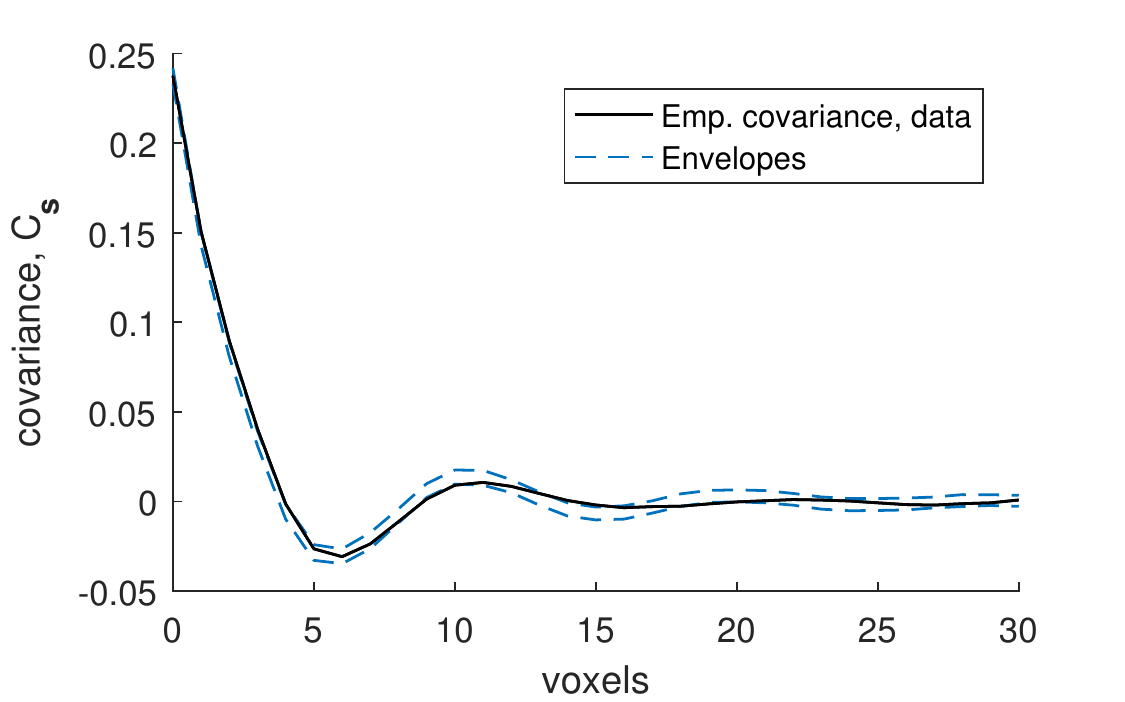}}
	\subcaptionbox{}
	[0.45\textwidth]
	{\includegraphics[width=7cm]{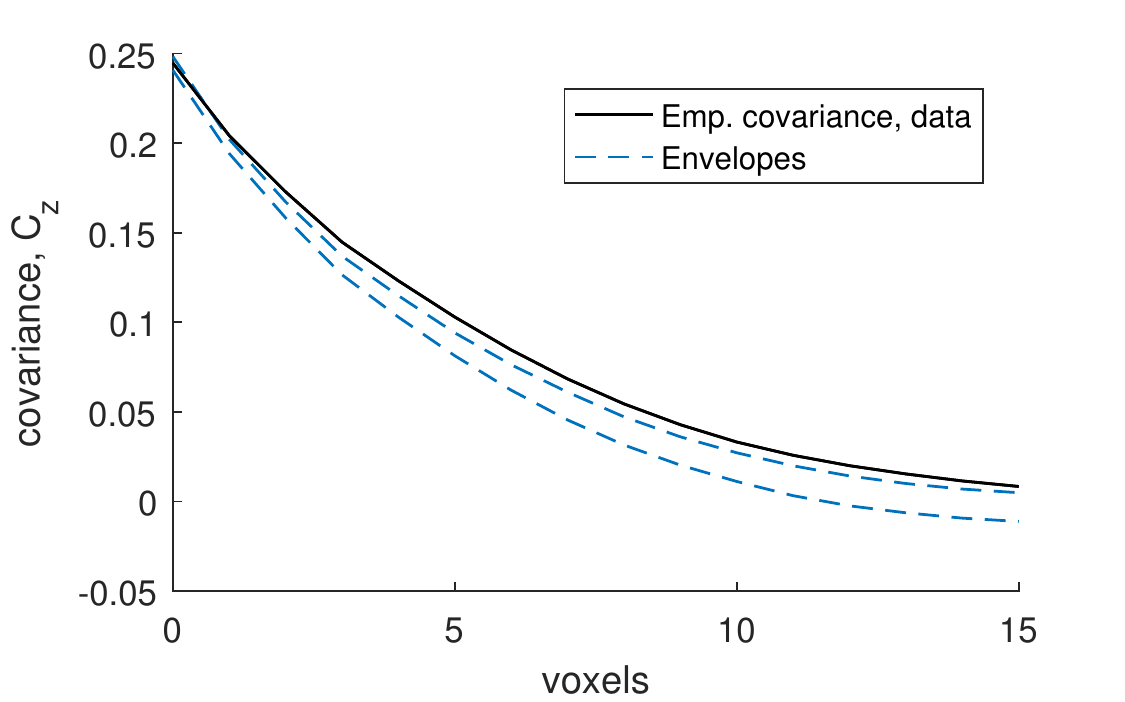}}
	\subcaptionbox{}
	[0.24\textwidth]
	{\includegraphics[width=4.5cm]{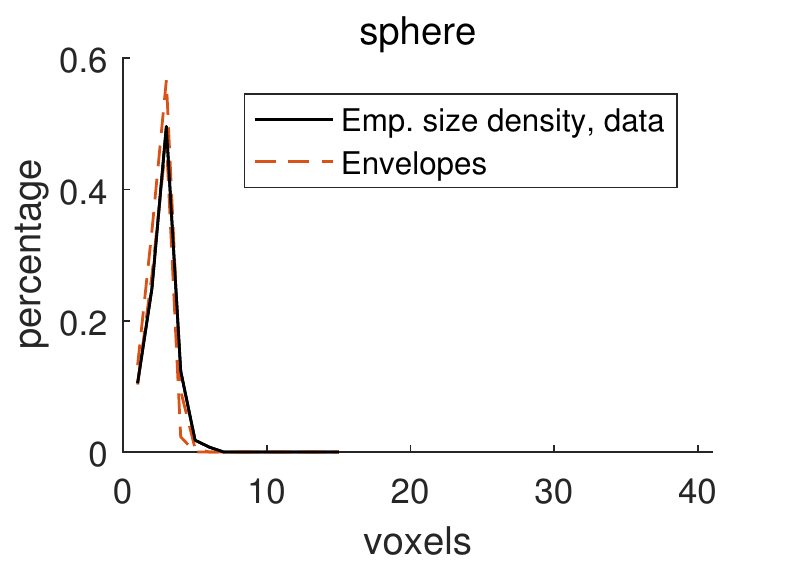}}
	\subcaptionbox{}
	[0.24\textwidth]
	{\includegraphics[width=4.5cm]{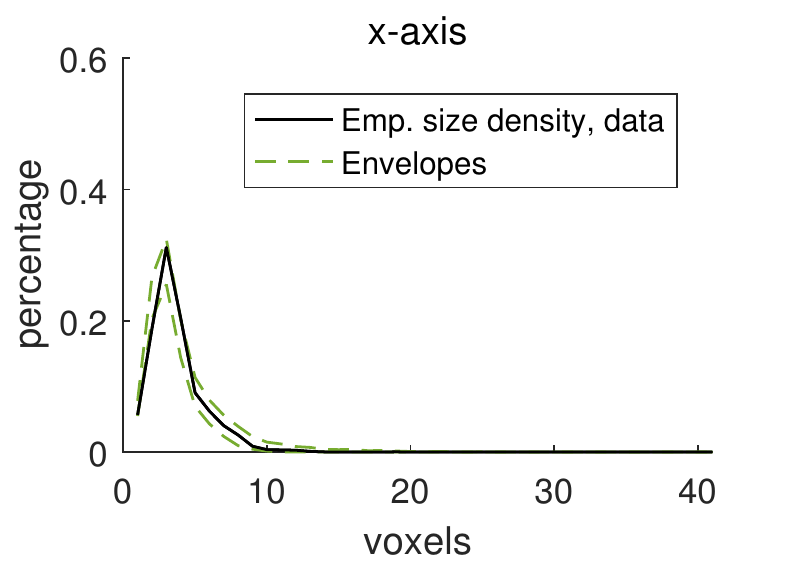}}
	\subcaptionbox{}
	[0.24\textwidth]
	{\includegraphics[width=4.5cm]{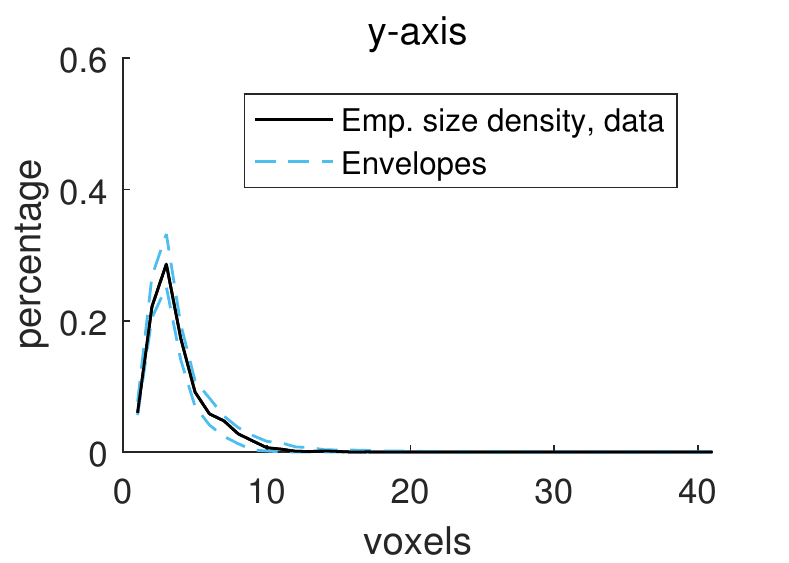}}
	\subcaptionbox{}
	[0.24\textwidth]
	{\includegraphics[width=4.5cm]{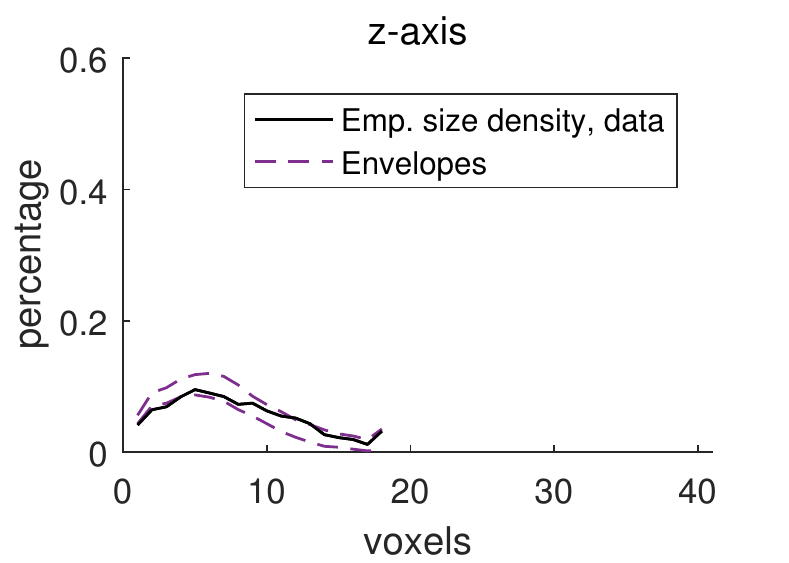}}
	\subcaptionbox{}
	[0.24\textwidth]
	{\includegraphics[width=4.5cm]{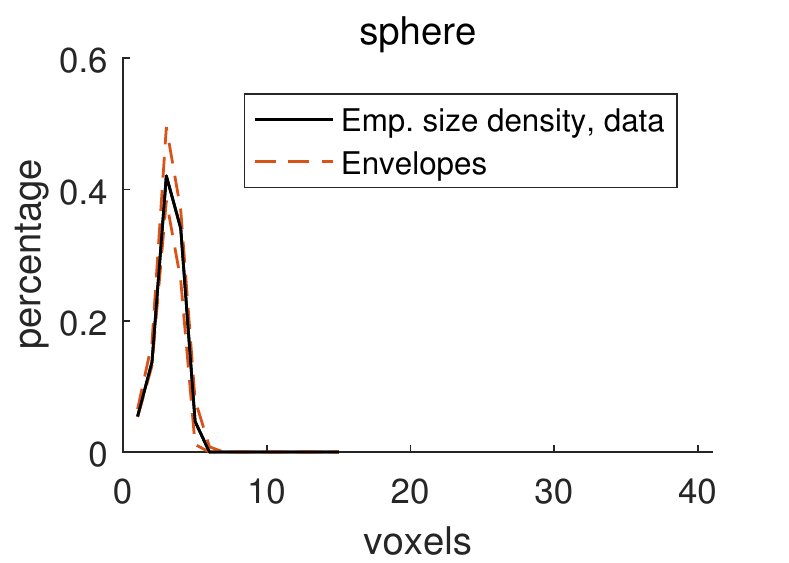}}
	\subcaptionbox{}
	[0.24\textwidth]
	{\includegraphics[width=4.5cm]{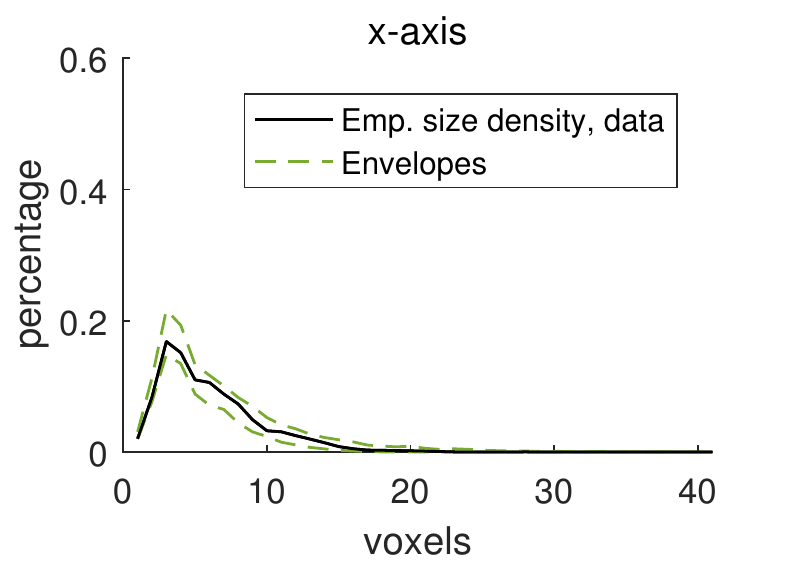}}
	\subcaptionbox{}
	[0.24\textwidth]
	{\includegraphics[width=4.5cm]{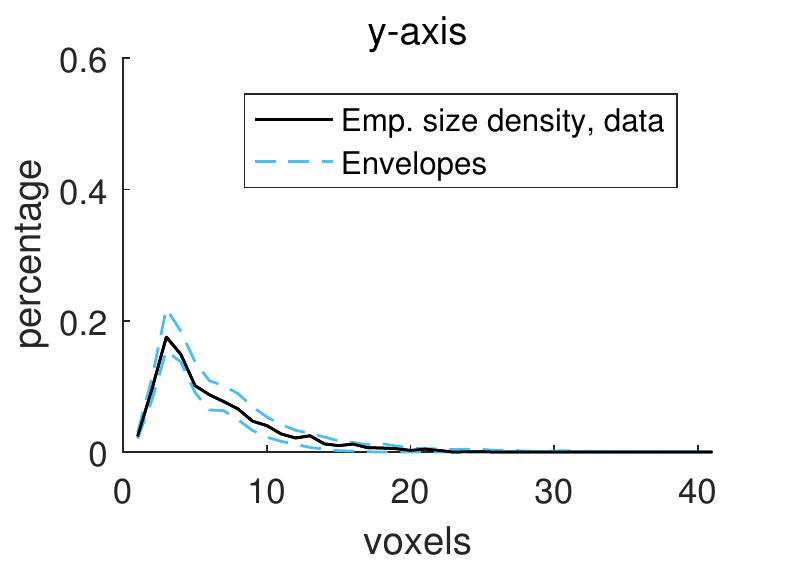}}
	\subcaptionbox{}
	[0.24\textwidth]
	{\includegraphics[width=4.5cm]{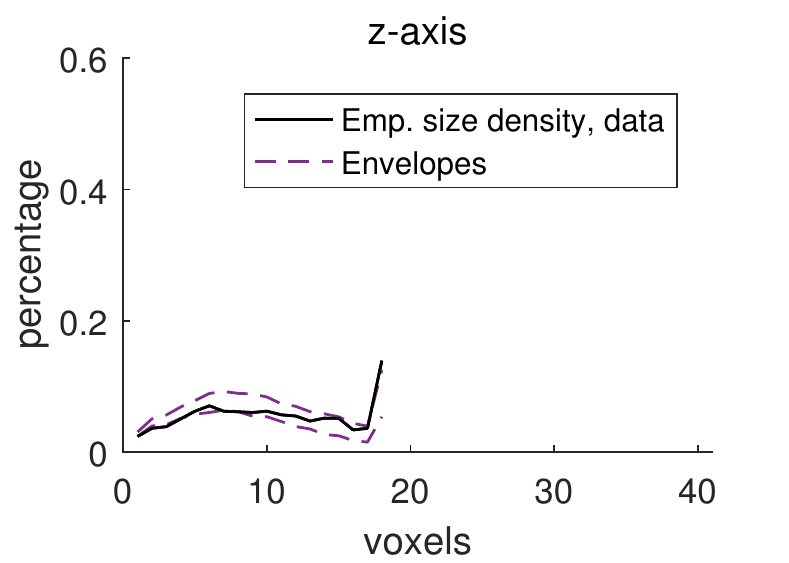}}
	\caption{Results for the $HPC40_1$ sample.  Marginal covariance functions are shown for the noise model (a and b) and the pore model (c and d), together with $95\%$ simultaneous envelopes estimated from $500$ simulations from the corresponding models.  The empirical covariance function $C_{\mv{s}}$ estimated from the binarized CLSM image $\mv{y}$ and the CLSM pore structure $\tildey$ are shown in (a) and (c) respectively.  The corresponding estimates for $C_z$ are shown in (b) and (d). For the noise model, the model covariance function with point-estimates of the parameters is also shown.  Size densities estimated from $\tildey$ on the pore space ((e)--(h)) and the pore matrix ((i)--(l)) are also shown together with $95\%$ simultaneous envelope estimated from $500$ simulations from the pore model. The following structuring elements, with length/radius two, were used:  the unit sphere (e and i), and lines aligned with the x-axis (f and j), the y-axis (g and k), and the z-axis (h and l).}
	\label{fig:summfuncdown1}
\end{figure}

\begin{figure}[H]
	\centering
	\subcaptionbox{}
	[0.45\textwidth]
	{\includegraphics[width=7cm]{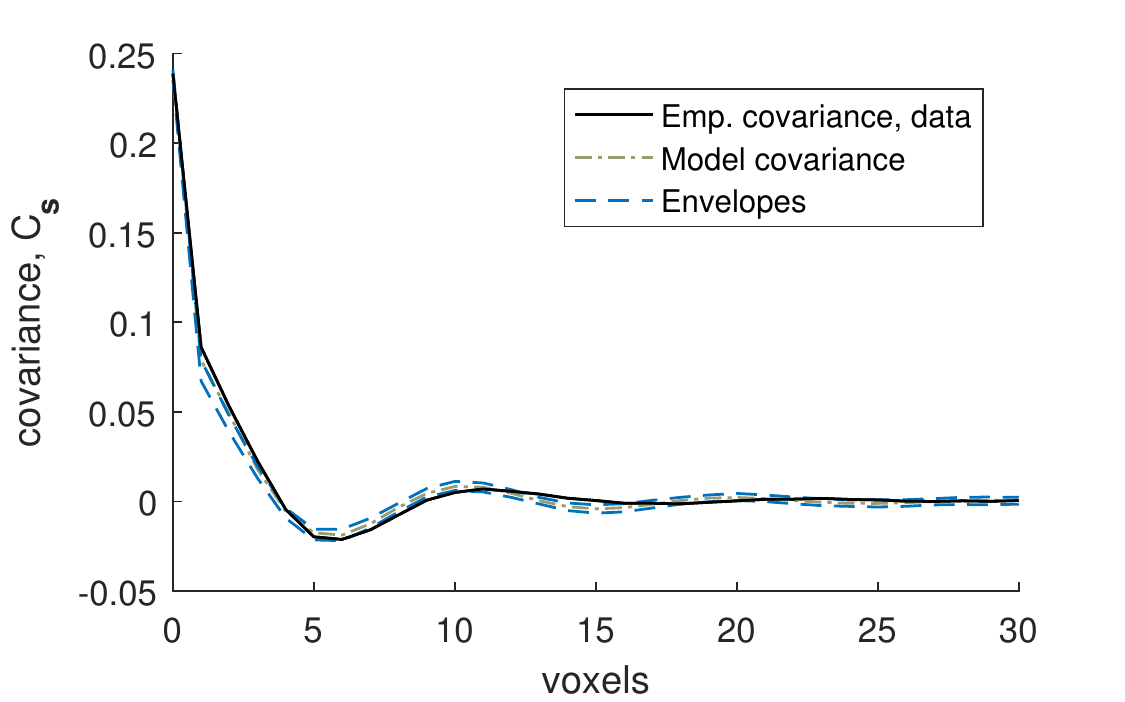}}
	\subcaptionbox{}
	[0.45\textwidth]
	{\includegraphics[width=7cm]{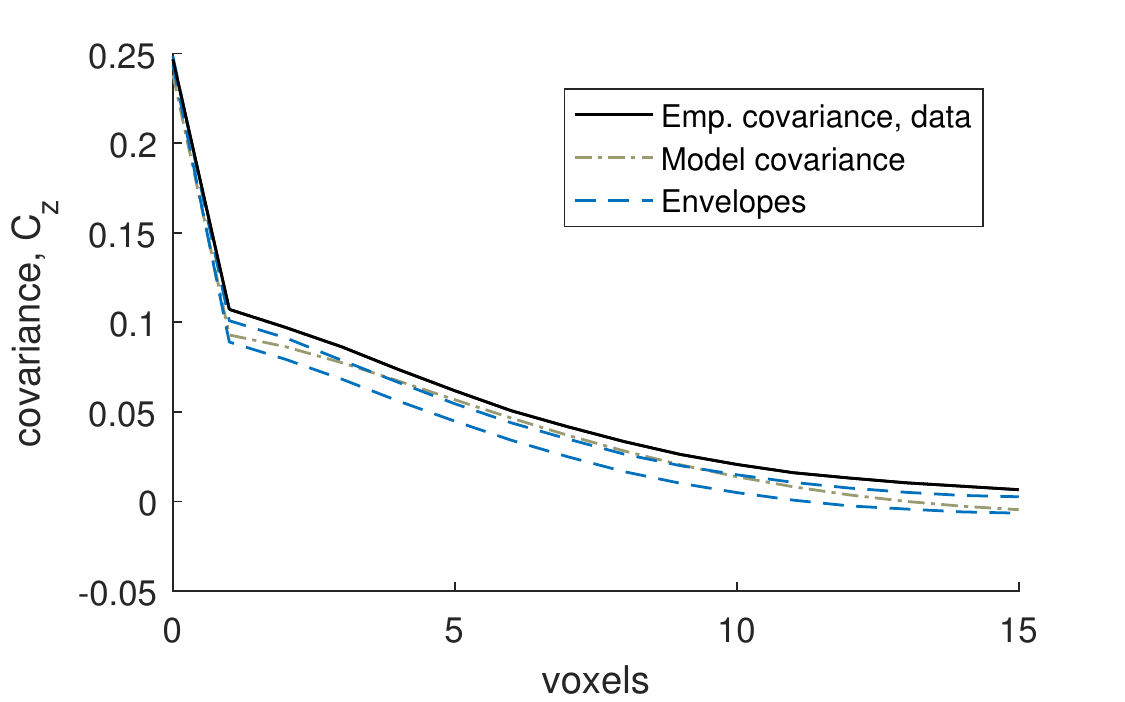}}
	\subcaptionbox{}
	[0.45\textwidth]
	{\includegraphics[width=7cm]{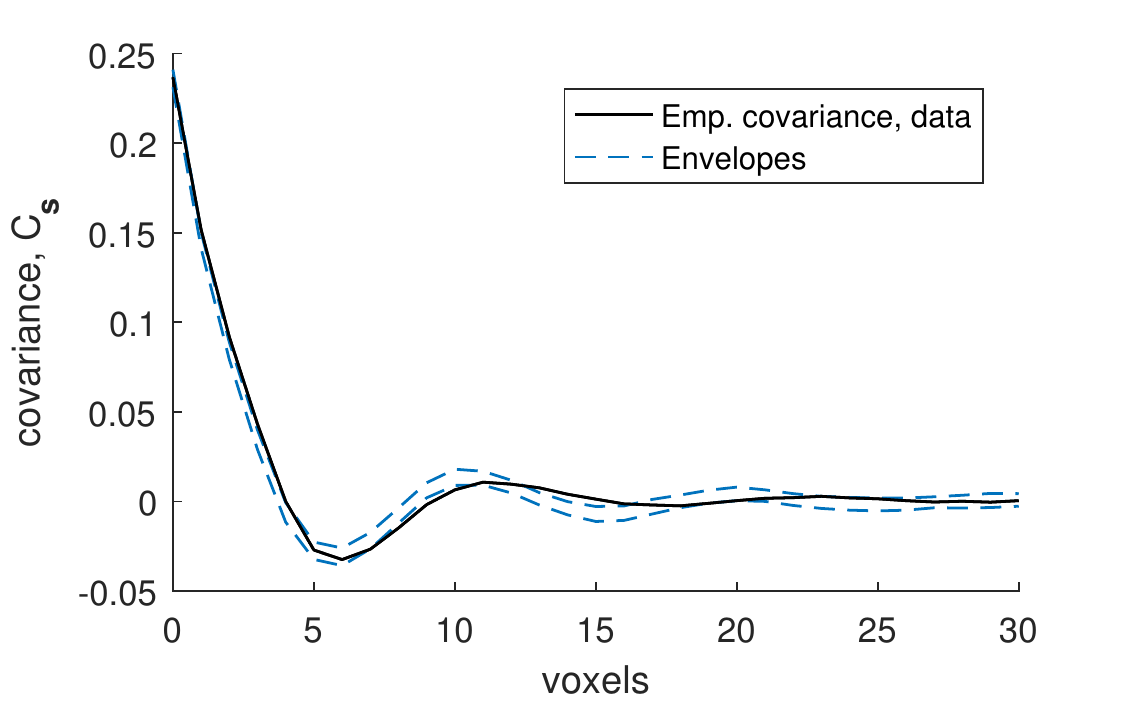}}
	\subcaptionbox{}
	[0.45\textwidth]
	{\includegraphics[width=7cm]{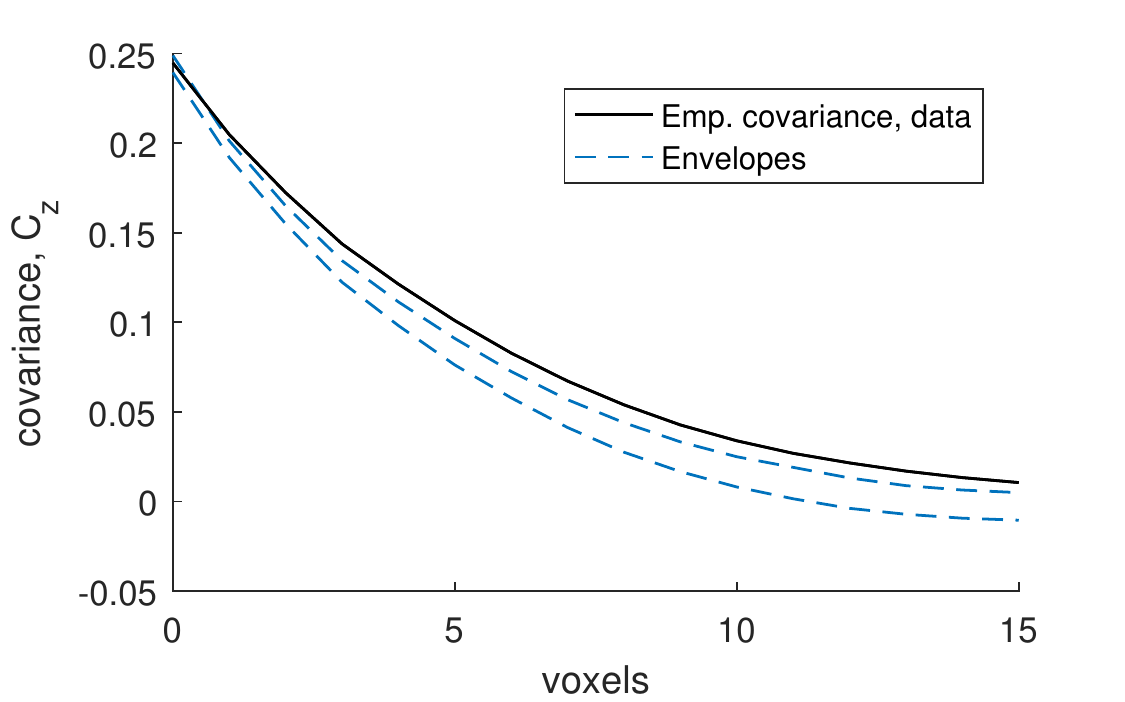}}
	\subcaptionbox{}
	[0.24\textwidth]
	{\includegraphics[width=4.5cm]{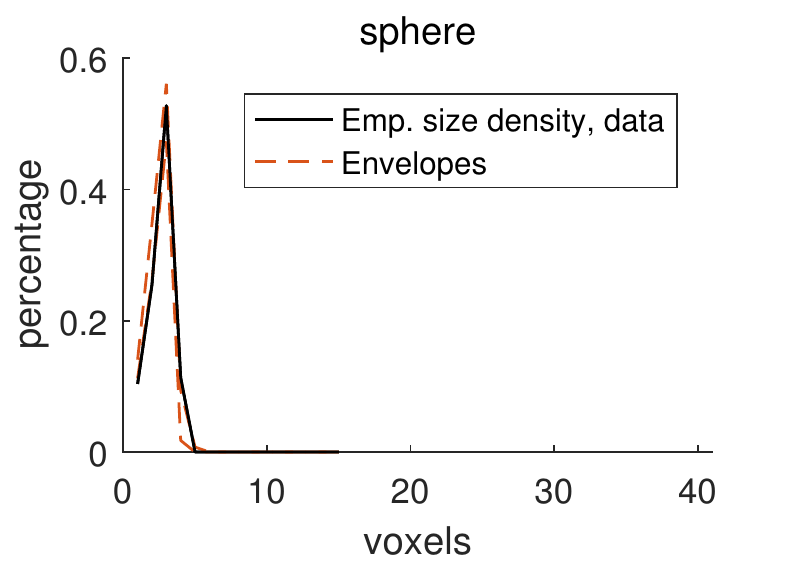}}
	\subcaptionbox{}
	[0.24\textwidth]
	{\includegraphics[width=4.5cm]{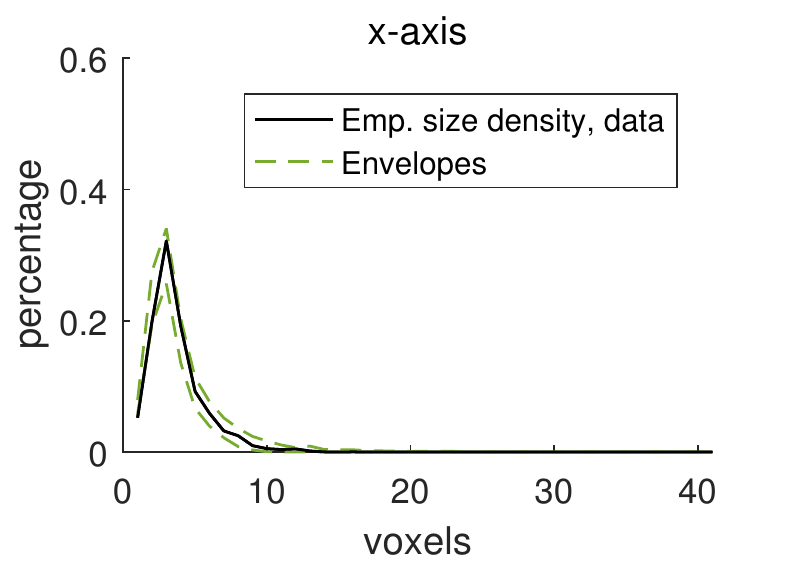}}
	\subcaptionbox{}
	[0.24\textwidth]
	{\includegraphics[width=4.5cm]{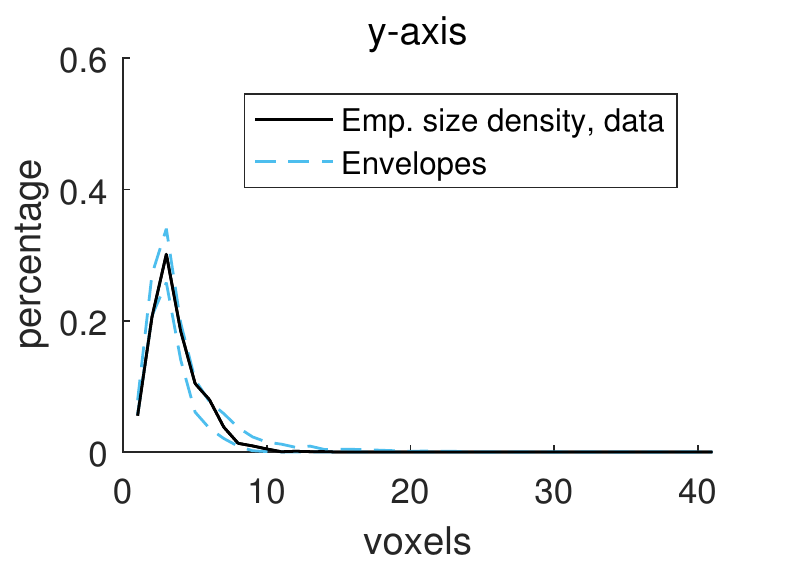}}
	\subcaptionbox{}
	[0.24\textwidth]
	{\includegraphics[width=4.5cm]{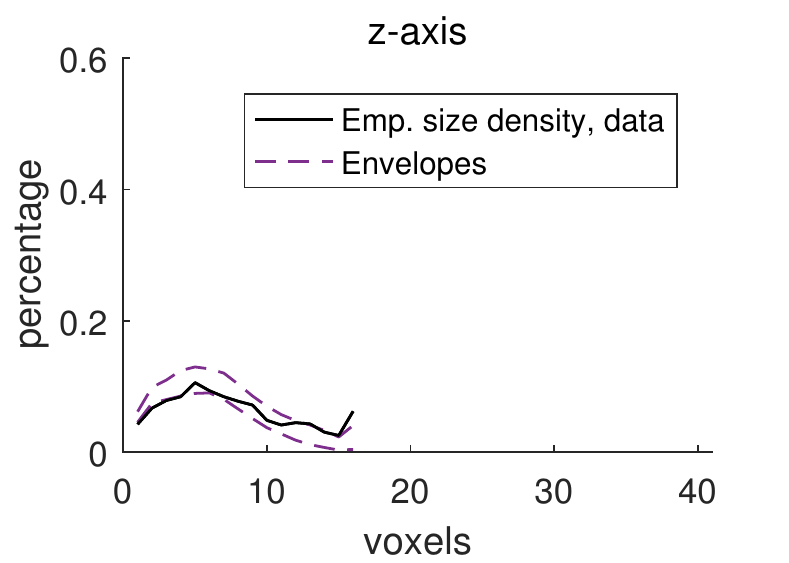}}
	\subcaptionbox{}
	[0.24\textwidth]
	{\includegraphics[width=4.5cm]{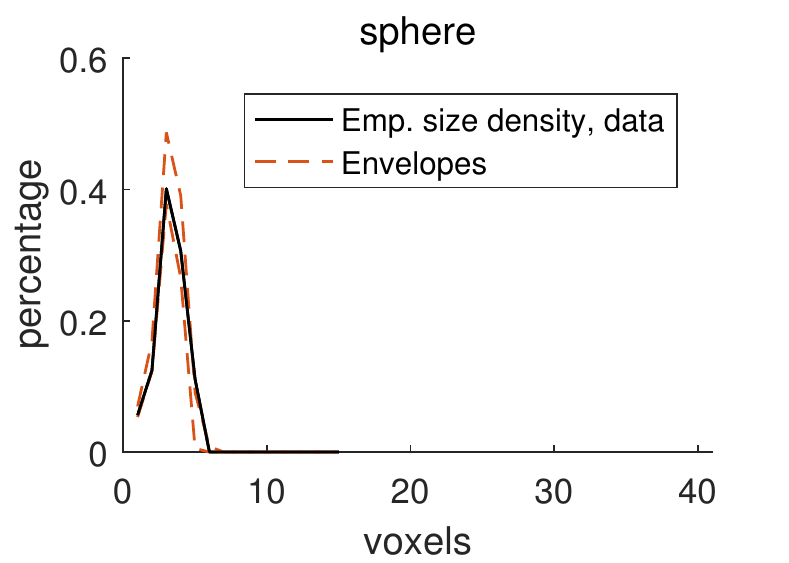}}
	\subcaptionbox{}
	[0.24\textwidth]
	{\includegraphics[width=4.5cm]{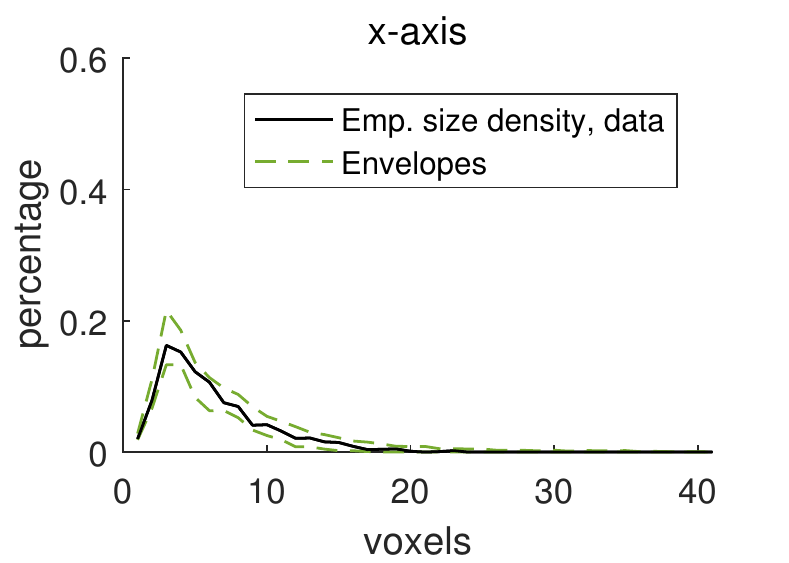}}
	\subcaptionbox{}
	[0.24\textwidth]
	{\includegraphics[width=4.5cm]{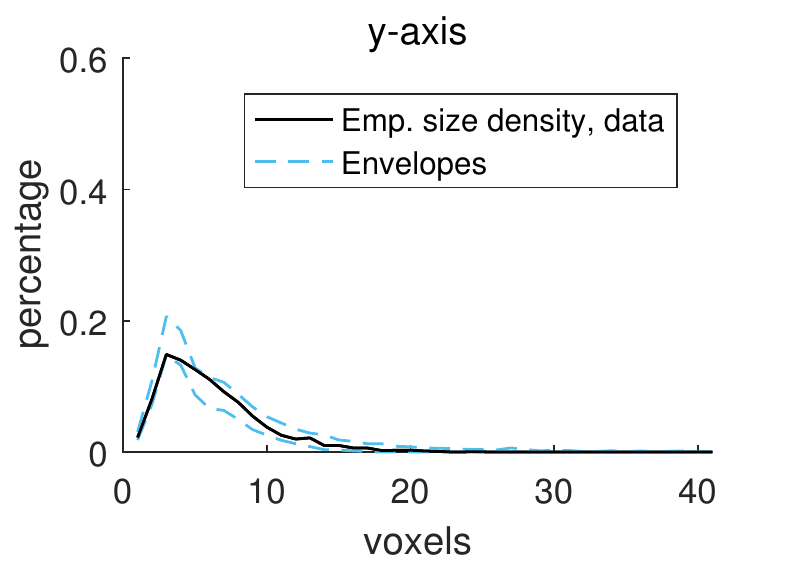}}
	\subcaptionbox{}
	[0.24\textwidth]
	{\includegraphics[width=4.5cm]{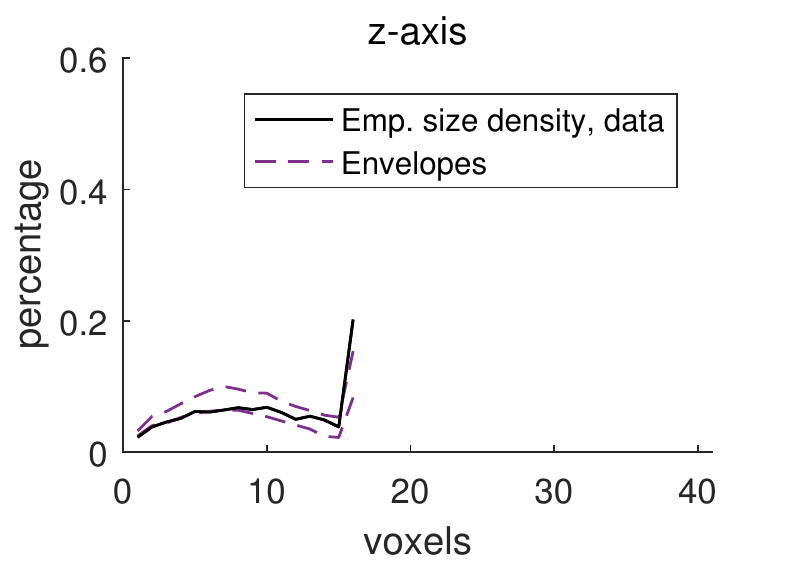}}
	\caption{Results for the $HPC40_2$ sample.  Marginal covariance functions are shown for the noise model (a and b) and the pore model (c and d), together with $95\%$ simultaneous envelopes estimated from $500$ simulations from the corresponding models.  The empirical covariance function $C_{\mv{s}}$ estimated from the binarized CLSM image $\mv{y}$ and the CLSM pore structure $\tildey$ are shown in (a) and (c) respectively.  The corresponding estimates for $C_z$ are shown in (b) and (d). For the noise model, the model covariance function with point-estimates of the parameters is also shown.  Size densities estimated from $\tildey$ on the pore space ((e)--(h)) and the pore matrix ((i)--(l)) are also shown together with $95\%$ simultaneous envelope estimated from $500$ simulations from the pore model. The following structuring elements, with length/radius two, were used:  the unit sphere (e and i), and lines aligned with the x-axis (f and j), the y-axis (g and k), and the z-axis (h and l).}
	\label{fig:summfuncdown2}
\end{figure}

\begin{figure}[H]
	\centering
	\subcaptionbox{}
	[0.45\textwidth]
	{\includegraphics[width=7cm]{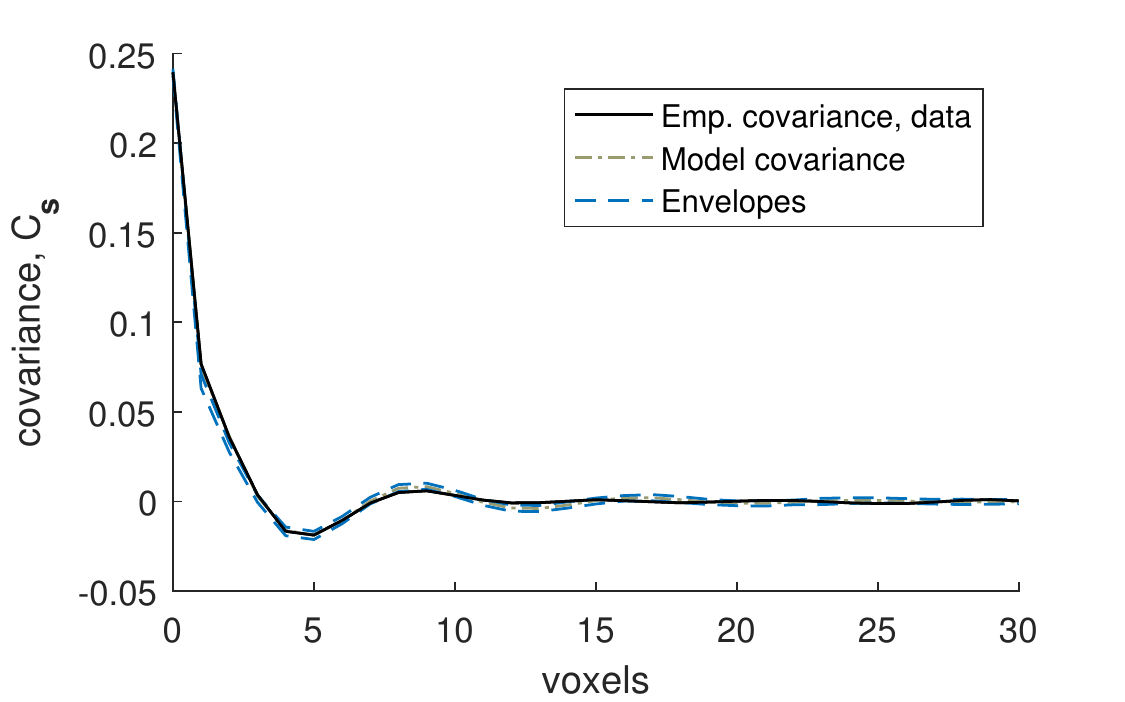}}
	\subcaptionbox{}
	[0.45\textwidth]
	{\includegraphics[width=7cm]{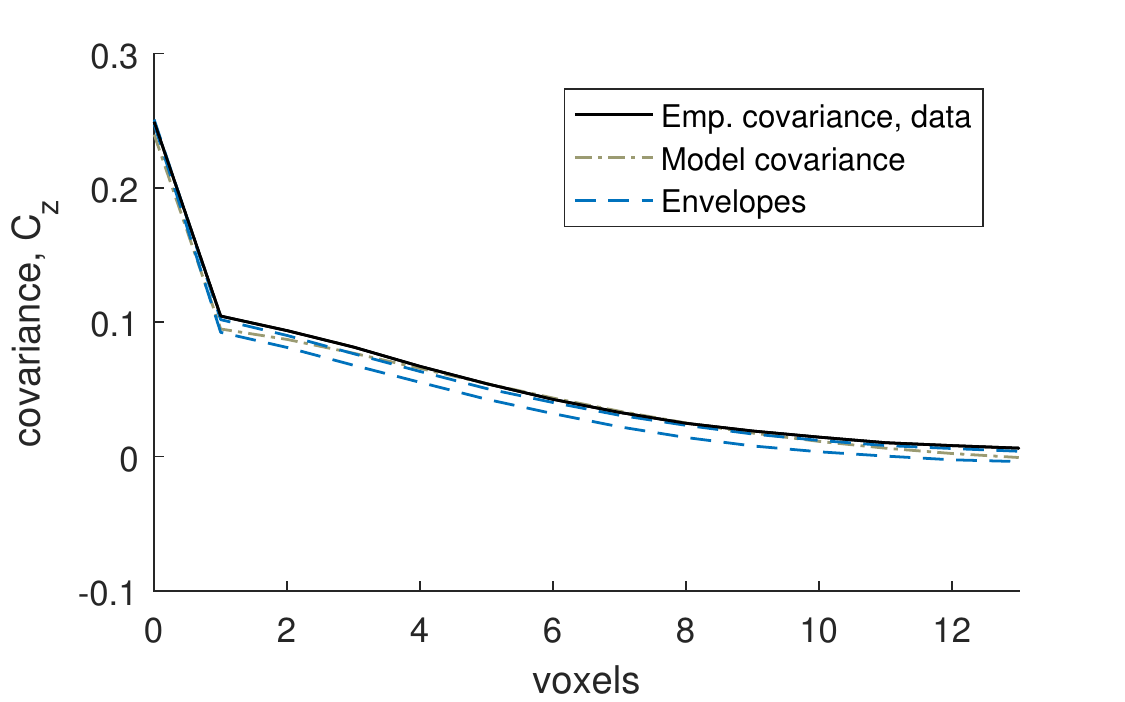}}
	\subcaptionbox{}
	[0.45\textwidth]
	{\includegraphics[width=7cm]{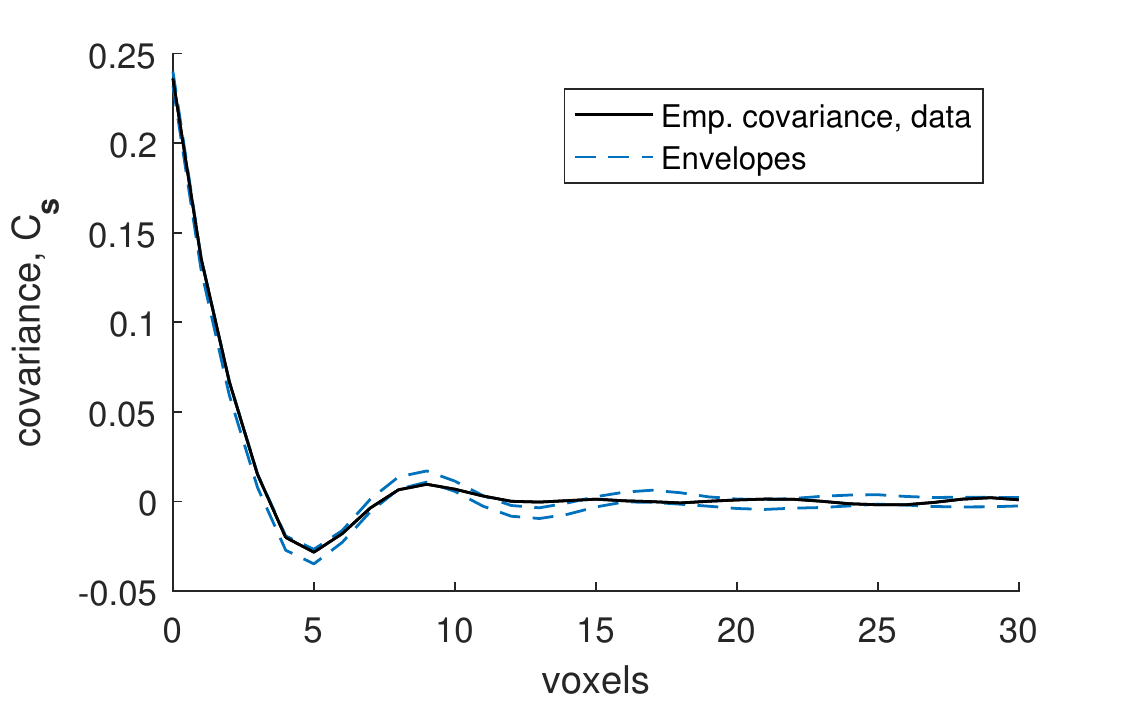}}
	\subcaptionbox{}
	[0.45\textwidth]
	{\includegraphics[width=7cm]{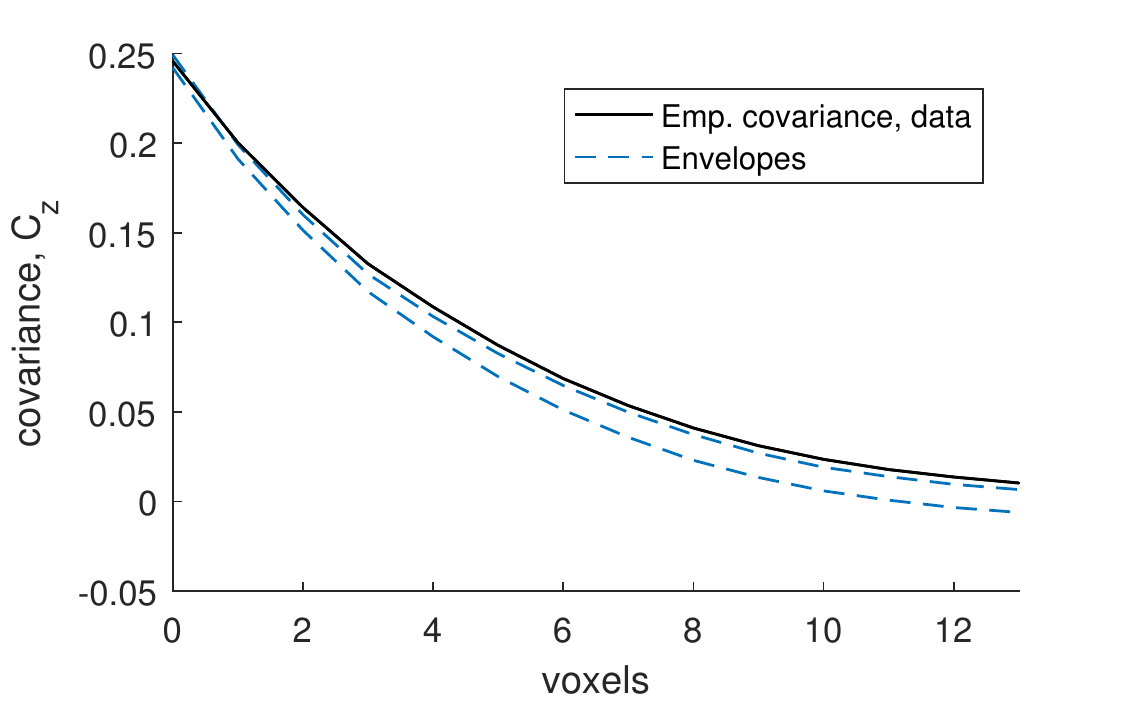}}
	\subcaptionbox{}
	[0.24\textwidth]
	{\includegraphics[width=4.5cm]{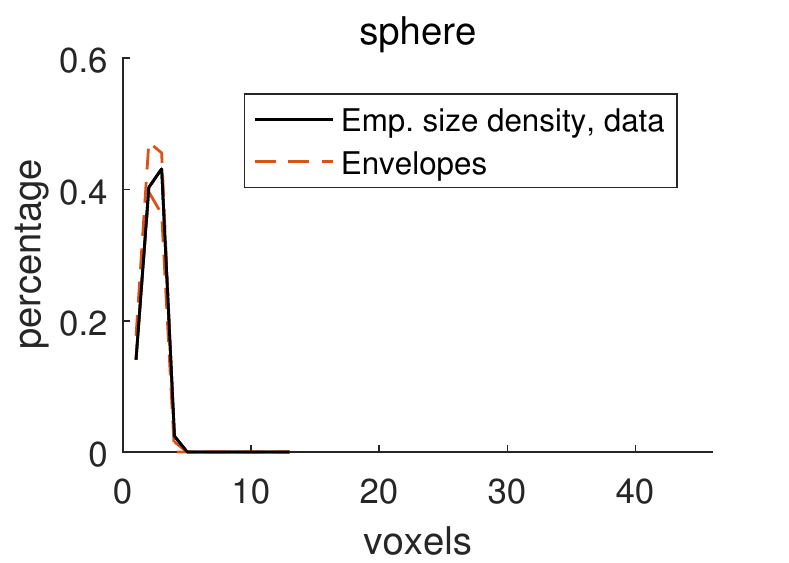}}
	\subcaptionbox{}
	[0.24\textwidth]
	{\includegraphics[width=4.5cm]{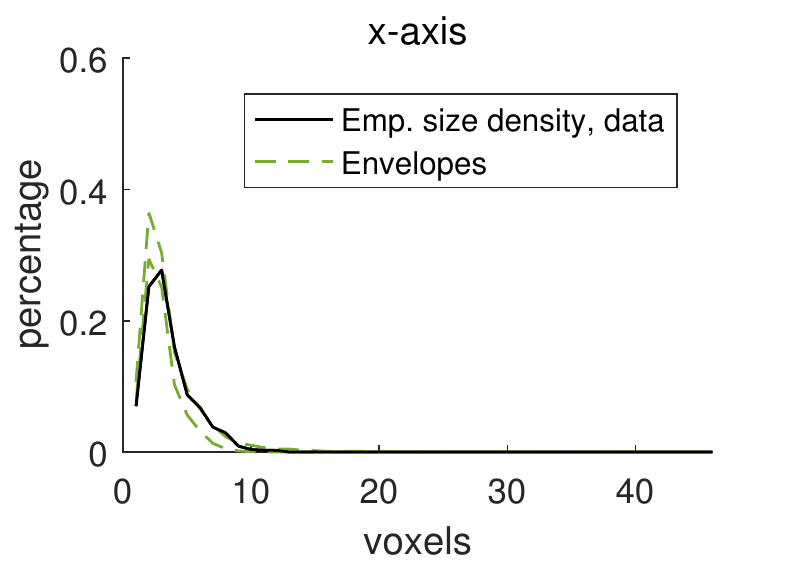}}
	\subcaptionbox{}
	[0.24\textwidth]
	{\includegraphics[width=4.5cm]{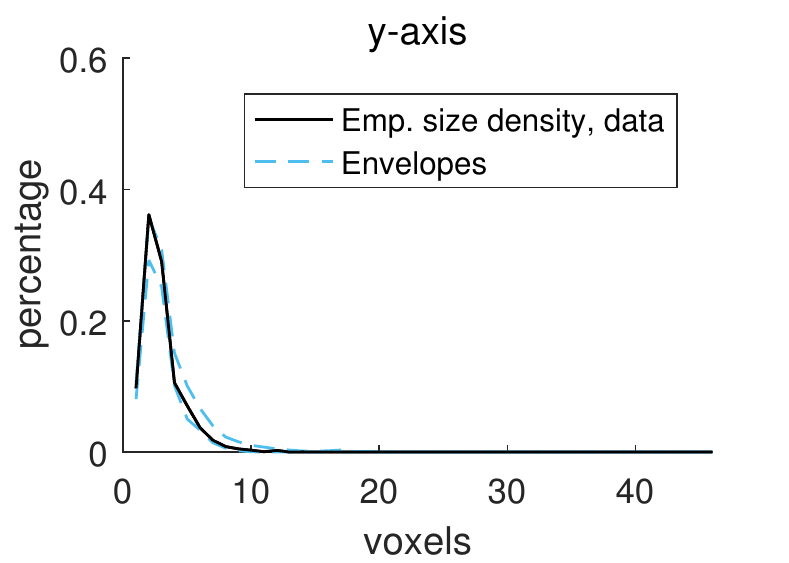}}
	\subcaptionbox{}
	[0.24\textwidth]
	{\includegraphics[width=4.5cm]{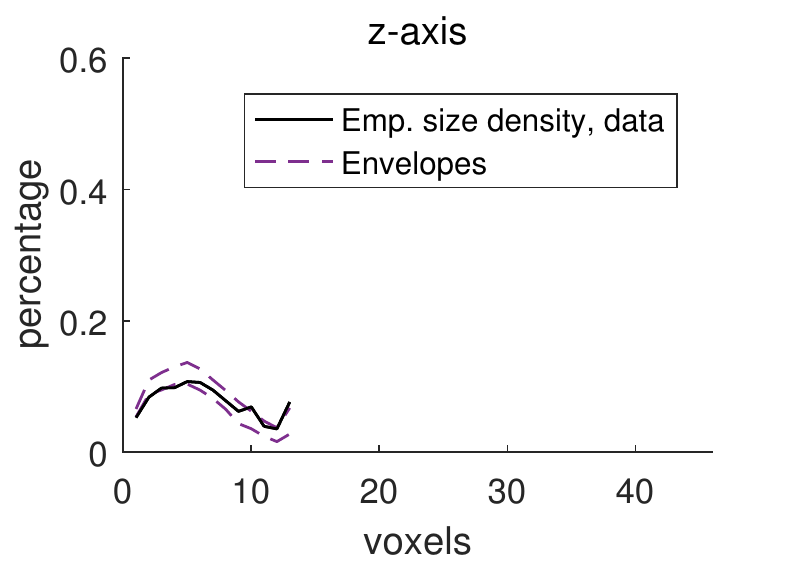}}
	\subcaptionbox{}
	[0.24\textwidth]
	{\includegraphics[width=4.5cm]{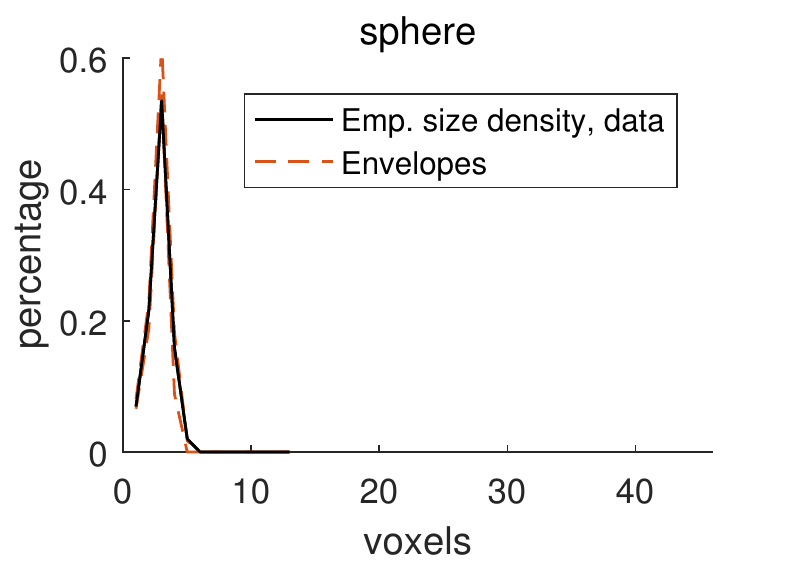}}
	\subcaptionbox{}
	[0.24\textwidth]
	{\includegraphics[width=4.5cm]{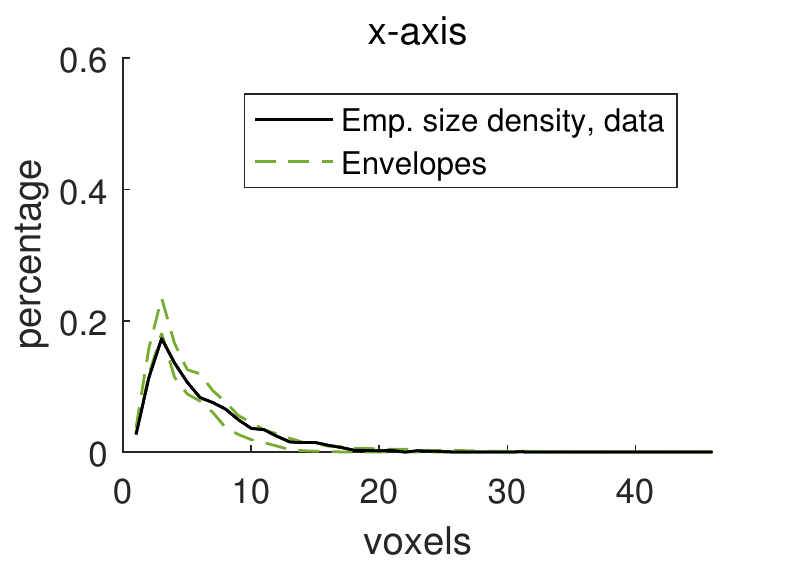}}
	\subcaptionbox{}
	[0.24\textwidth]
	{\includegraphics[width=4.5cm]{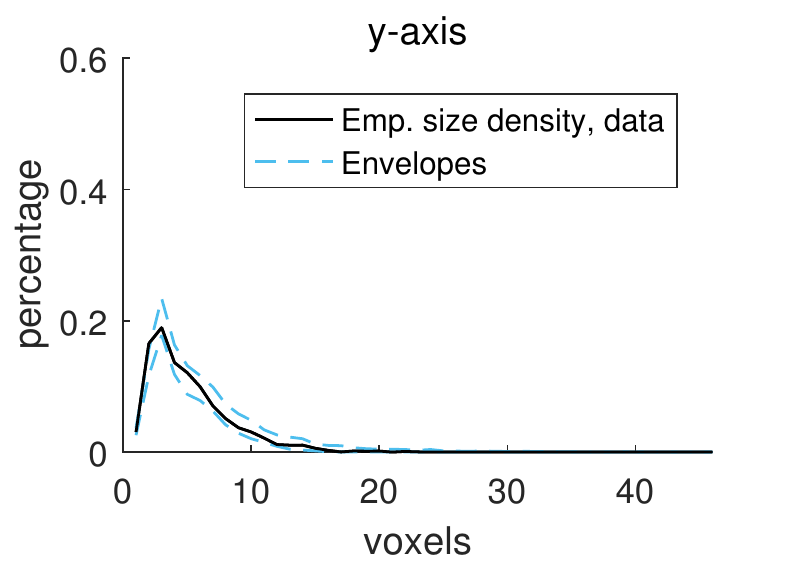}}
	\subcaptionbox{}
	[0.24\textwidth]
	{\includegraphics[width=4.5cm]{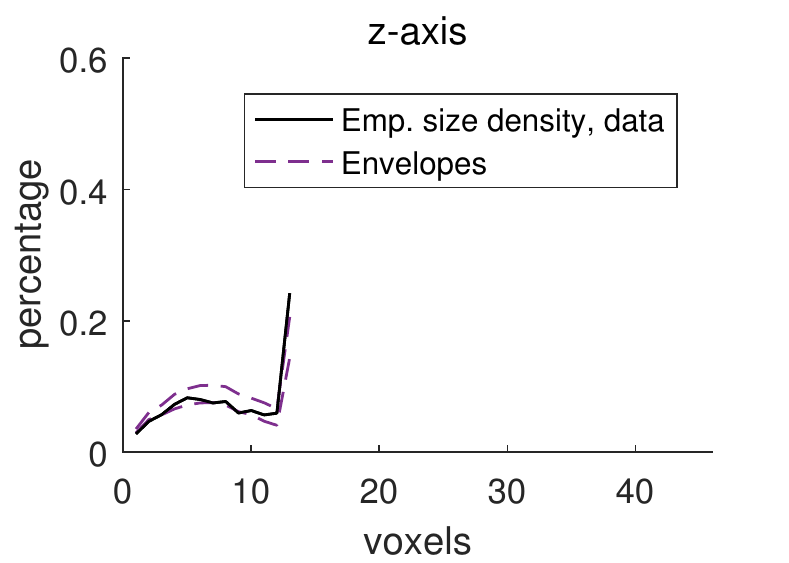}}
	\caption{Results for the $HPC40_3$ sample.  Marginal covariance functions are shown for the noise model (a and b) and the pore model (c and d), together with $95\%$ simultaneous envelopes estimated from $500$ simulations from the corresponding models.  The empirical covariance function $C_{\mv{s}}$ estimated from the binarized CLSM image $\mv{y}$ and the CLSM pore structure $\tildey$ are shown in (a) and (c) respectively.  The corresponding estimates for $C_z$ are shown in (b) and (d). For the noise model, the model covariance function with point-estimates of the parameters is also shown.  Size densities estimated from $\tildey$ on the pore space ((e)--(h)) and the pore matrix ((i)--(l)) are also shown together with $95\%$ simultaneous envelope estimated from $500$ simulations from the pore model. The following structuring elements, with length/radius two, were used:  the unit sphere (e and i), and lines aligned with the x-axis (f and j), the y-axis (g and k), and the z-axis (h and l).}
	\label{fig:summfuncup1}
\end{figure}

\begin{figure}[H]
	\centering
	\subcaptionbox{}
	[0.45\textwidth]
	{\includegraphics[width=7cm]{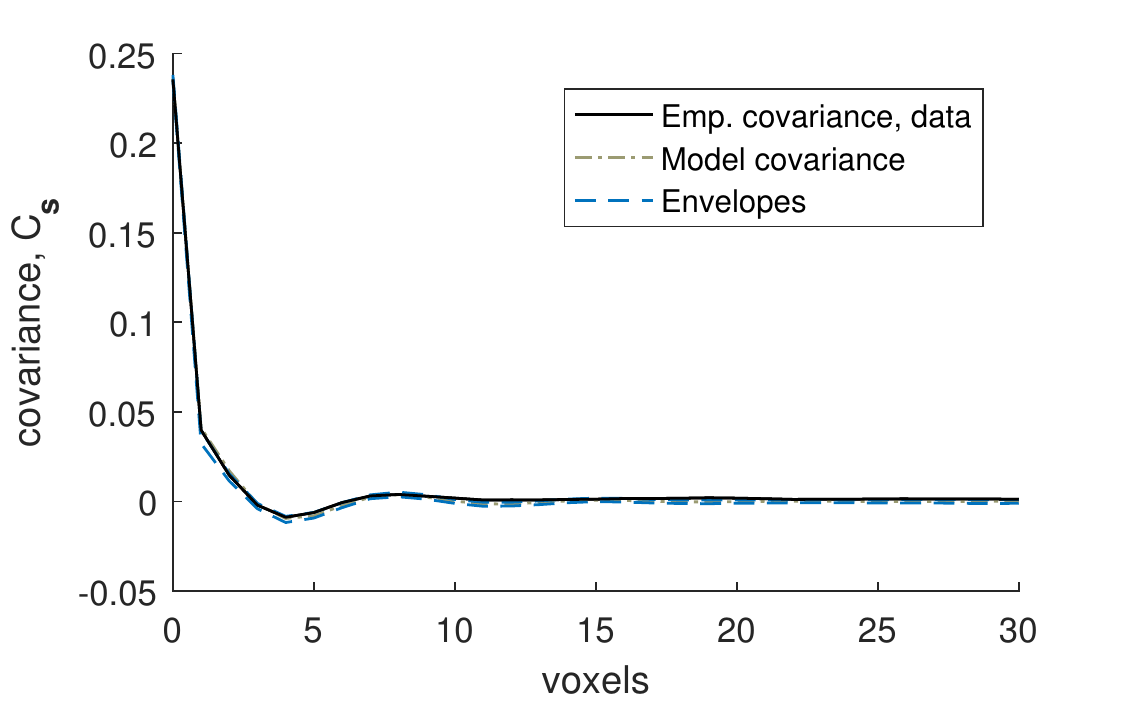}}
	\subcaptionbox{}
	[0.45\textwidth]
	{\includegraphics[width=7cm]{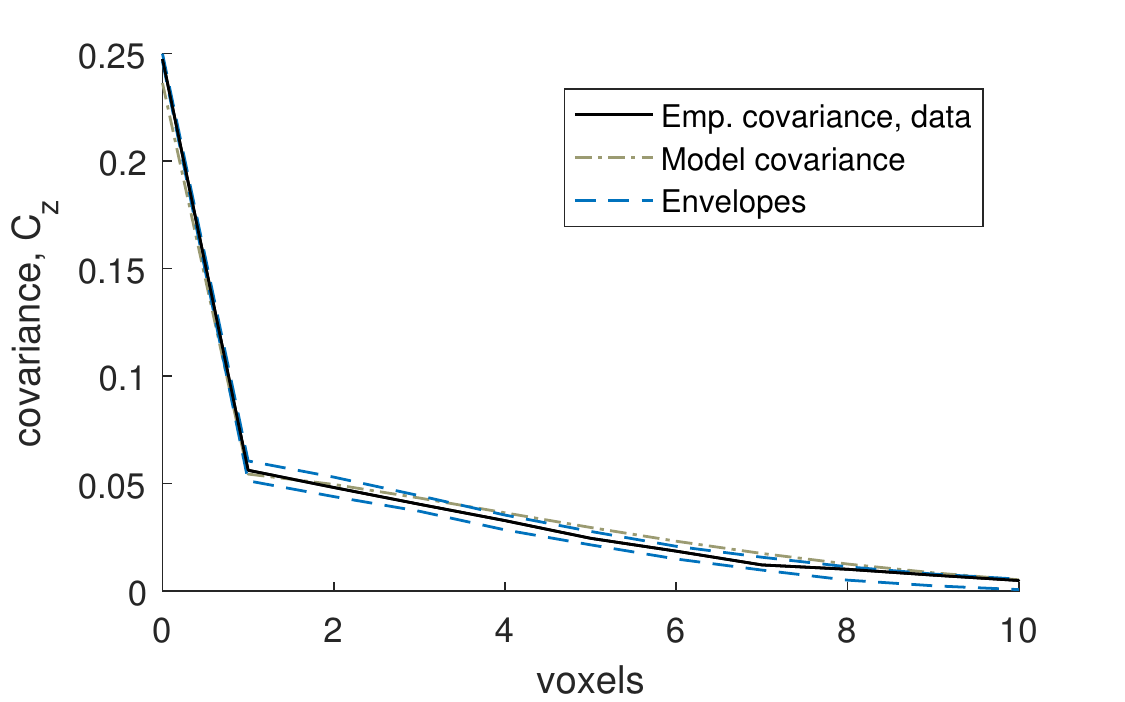}}
	\subcaptionbox{}
	[0.45\textwidth]
	{\includegraphics[width=7cm]{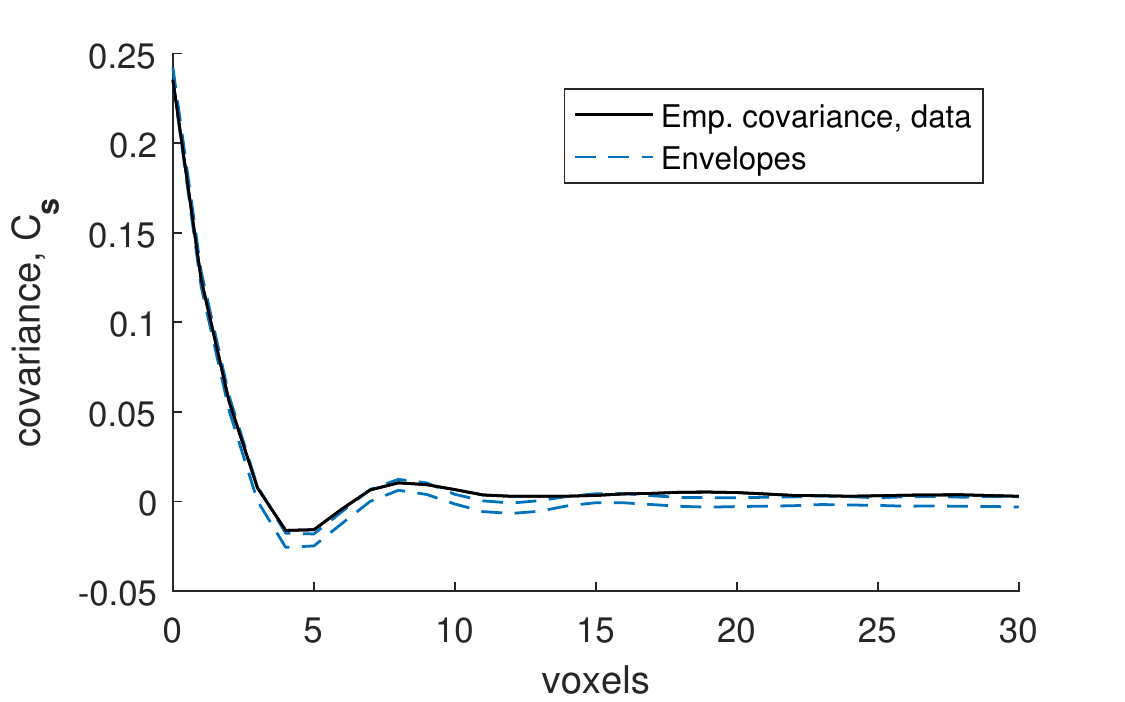}}
	\subcaptionbox{}
	[0.45\textwidth]
	{\includegraphics[width=7cm]{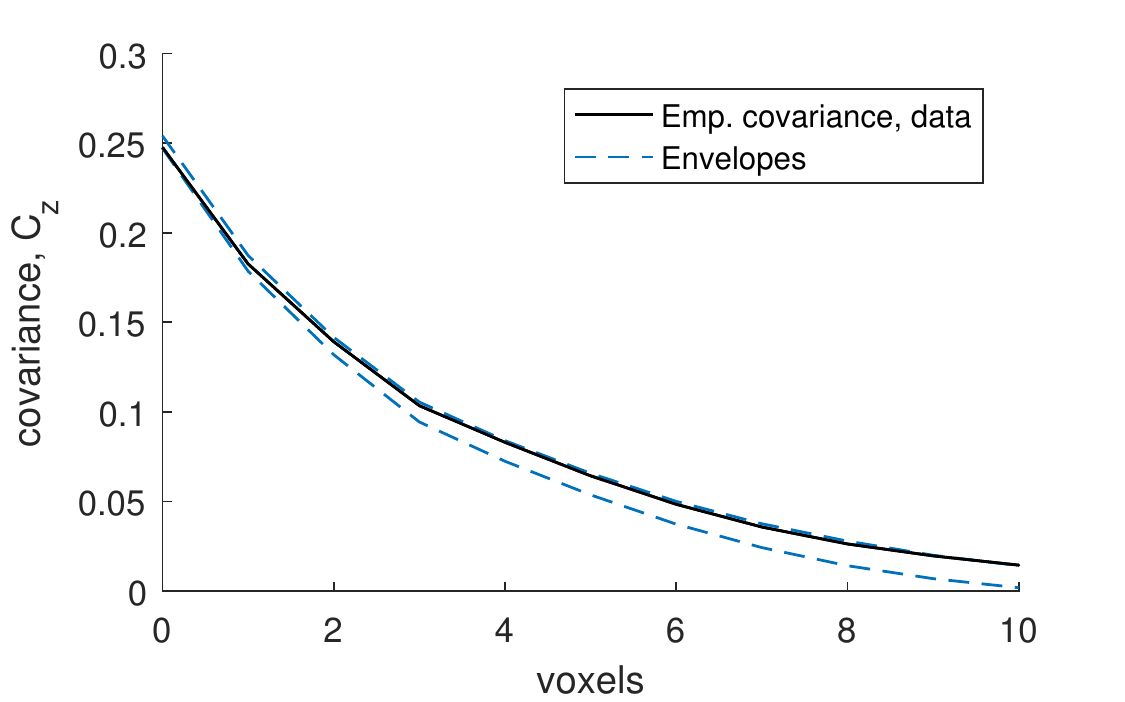}}
	\subcaptionbox{}
	[0.24\textwidth]
	{\includegraphics[width=4.5cm]{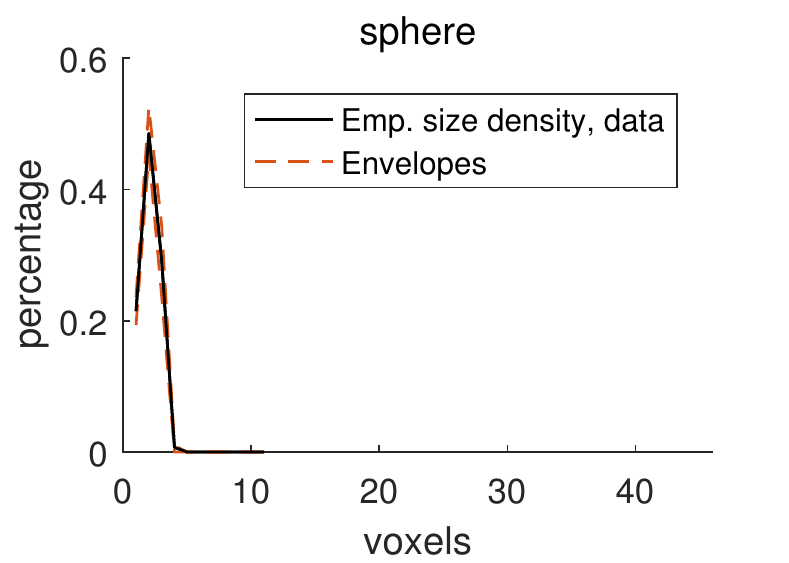}}
	\subcaptionbox{}
	[0.24\textwidth]
	{\includegraphics[width=4.5cm]{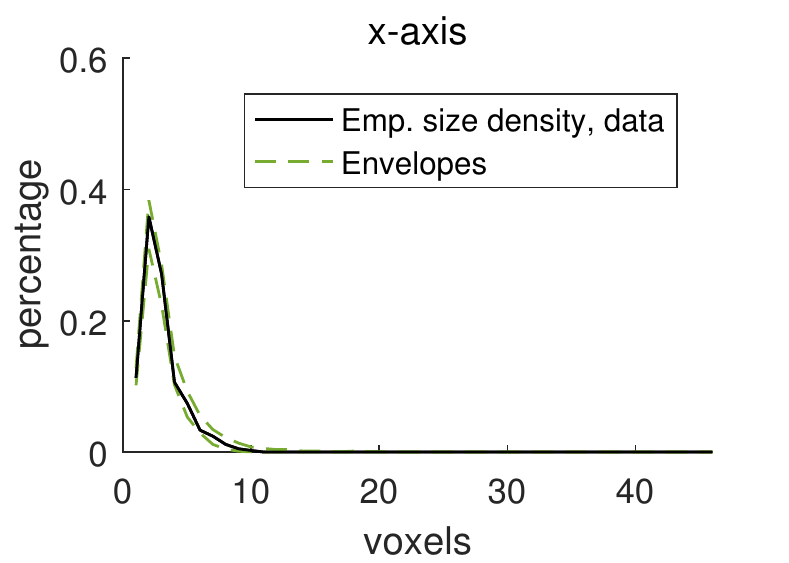}}
	\subcaptionbox{}
	[0.24\textwidth]
	{\includegraphics[width=4.5cm]{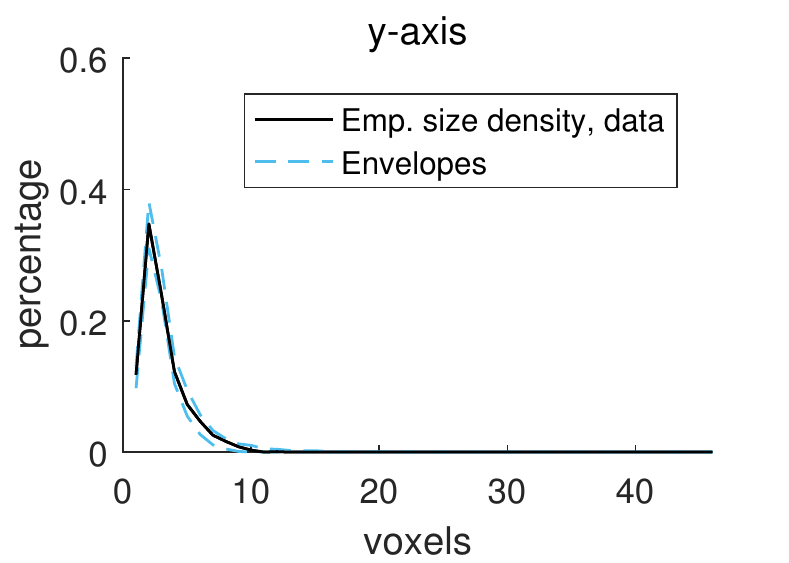}}
	\subcaptionbox{}
	[0.24\textwidth]
	{\includegraphics[width=4.5cm]{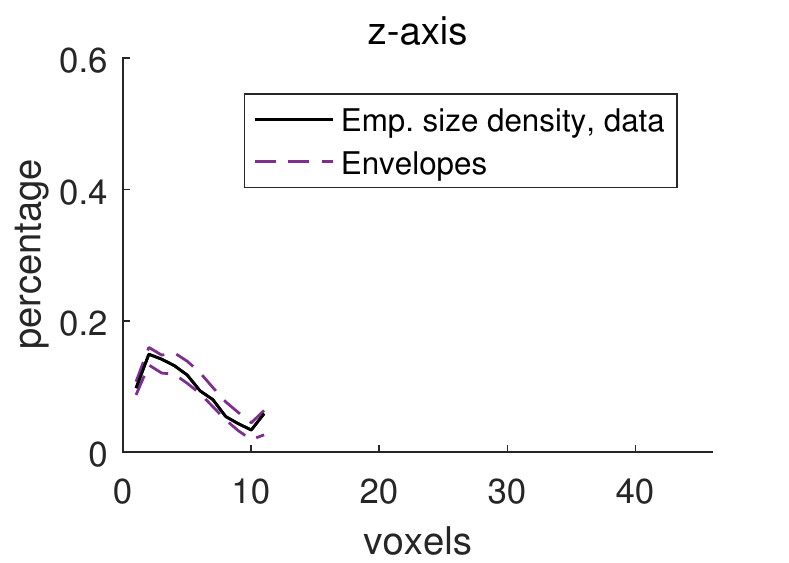}}
	\subcaptionbox{}
	[0.24\textwidth]
	{\includegraphics[width=4.5cm]{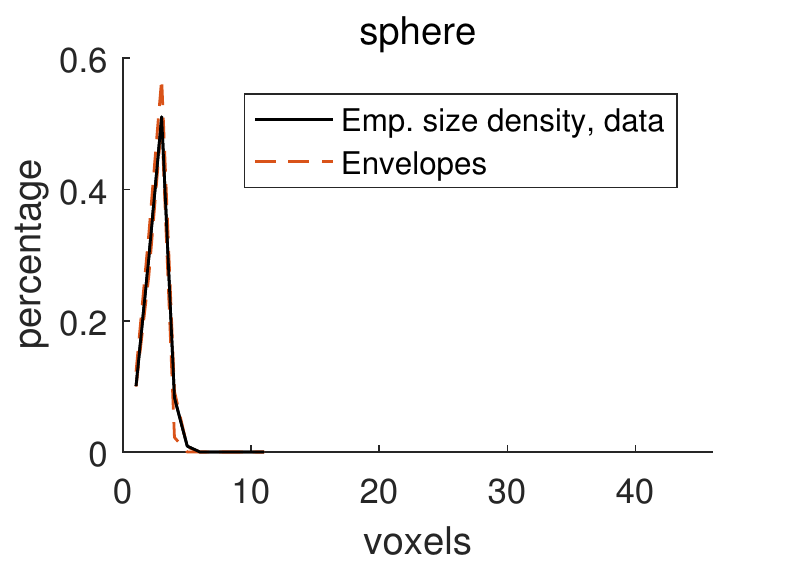}}
	\subcaptionbox{}
	[0.24\textwidth]
	{\includegraphics[width=4.5cm]{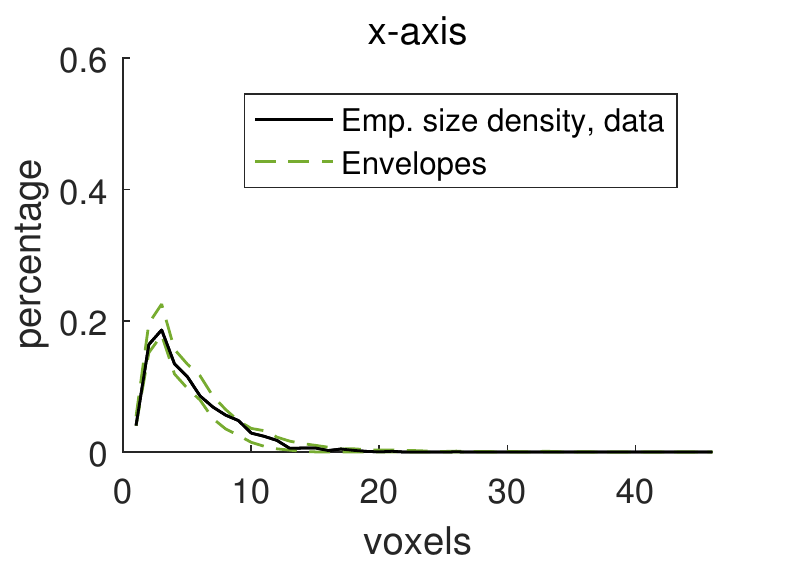}}
	\subcaptionbox{}
	[0.24\textwidth]
	{\includegraphics[width=4.5cm]{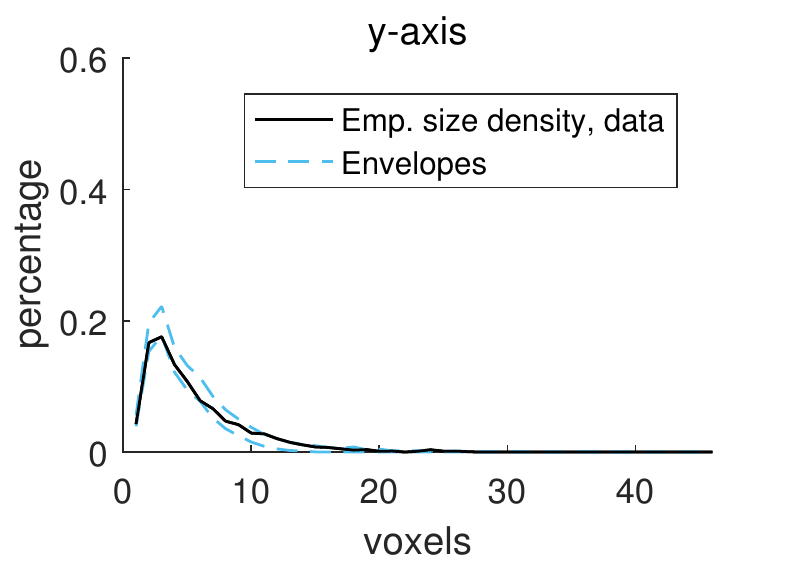}}
	\subcaptionbox{}
	[0.24\textwidth]
	{\includegraphics[width=4.5cm]{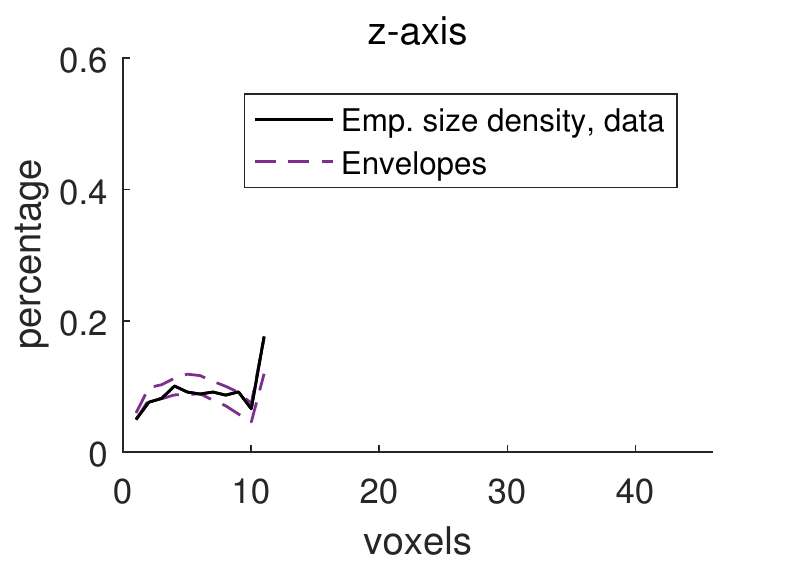}}
	\caption{Results for the $HPC40_4$ sample.  Marginal covariance functions are shown for the noise model (a and b) and the pore model (c and d), together with $95\%$ simultaneous envelopes estimated from $500$ simulations from the corresponding models.  The empirical covariance function $C_{\mv{s}}$ estimated from the binarized CLSM image $\mv{y}$ and the CLSM pore structure $\tildey$ are shown in (a) and (c) respectively.  The corresponding estimates for $C_z$ are shown in (b) and (d). For the noise model, the model covariance function with point-estimates of the parameters is also shown.  Size densities estimated from $\tildey$ on the pore space ((e)--(h)) and the pore matrix ((i)--(l)) are also shown together with $95\%$ simultaneous envelope estimated from $500$ simulations from the pore model. The following structuring elements, with length/radius two, were used:  the unit sphere (e and i), and lines aligned with the x-axis (f and j), the y-axis (g and k), and the z-axis (h and l).}
	\label{fig:summfuncup2}
\end{figure}

The covariance functions and size distributions (Figures \ref{fig:summfuncrightp2}--\ref{fig:summfuncup2}) for the remaining samples show that the model provides a relatively good fit. It is clear that the signal to noise ratio of the covariance functions corresponding to the noise model \noisemodel{} is underestimated (Panels (a)--(b)). However, as was the case for sample $HPC30_1$,  this underestimation does not have a big effect on the pore model.

The marginal covariance functions in the $\mv{s}$-plane, $C_{\mv{s}}$, for both the noise and pore model, fit better at smaller distances (Panels (a) and (c)). For half of the samples, the ranges of the estimated covariances $\hat{C}_{\mv{s}}$ are slightly underestimated, although the shapes of $\hat{C}_{\mv{s}}$ are consistent with the model envelopes (Figures \ref{fig:summfuncrightp2}, \ref{fig:summfuncleft21}--\ref{fig:summfuncdown2}, Panels (a) and (c)). A possible explanation comes from the fact that even though the samples were kept small in the $z$-direction, the pore sizes decrease slightly along the $z$-axis from the bottom to the top of each sample. The estimated range parameters could then have been chosen to better fit the part of the sample with smaller pore sizes.  For samples with high estimated oscillation parameters $\theta_{\mv{s}}$, the smaller peaks and troughs of the covariances $\hat{C}_{\mv{s}}$ do not seem to be as pronounced as those of the model envelopes (Figures \ref{fig:summfuncleft21}--\ref{fig:summfuncup1}, Panels (a) and (c)). This might be because the peaks and troughs shift and therefore cancel out due to the decreasing pore size along the $z$-axis. The size distributions lie within their envelopes to a higher extent than the covariance functions (Panels (e)--(l)). Thus the pore shapes of stochastic simulations from the fitted pore models correspond well to the pore shapes of the pore structures obtained from the CLSM samples.


\newpage

\end{appendix}

\end{document}